\documentclass{JHEP3}
\preprint{TTK-10-41}
\pdfoutput=1
\newcommand\myarraystretch{1.5}
\usepackage{graphicx}
\newcommand{\texorpdfstring}[2]{#1}
\usepackage{amssymb}
\usepackage{amsmath}
\newcommand{\bea}{\begin{eqnarray}}
\newcommand{\eea}{\end{eqnarray}}

\newcommand{\be}{\begin{equation}}
\newcommand{\ee}{\end{equation}}
\newcommand{\beqy}{\begin{eqnarray}}
\newcommand{\eeqy}{\end{eqnarray}}

\newcommand{\al}{\alpha}

\newcommand{\ga}{\gamma}

\newcommand{\te}{\theta}

\newcommand{\de}{\delta}
\newcommand{\De}{\Delta}

\newcommand{\e}{\epsilon}

\newcommand{\Om}{\Omega}

\newcommand{\La}{\Lambda}

\newcommand{\bM}{\tilde{M}}

\newcommand{\Lb}{\bar{L}}
\newcommand{\rb}{\bar{r}}

\newcommand{\cO}{{\cal O}}

\newcommand{\ra}{\rightarrow}
\newcommand{\Ra}{\Rightarrow}
\newcommand{\im}{\Longleftrightarrow}

\newcommand{\LF}{\left(}
\newcommand{\RF}{\right)}
\newcommand{\LT}{\left[}
\newcommand{\RT}{\right]}
\newcommand{\ie}{{\it i.e.\ }}

\newcommand{\mx}{\mbox}
\newcommand{\mt}{\rm}
\newcommand{\mtx}{{\mt{max}}}
\newcommand{\mto}{{\mt{out}}}

\newcommand{\eff}{{\mt{eff}}}

\title{Testing the Void against Cosmological data: fitting CMB, BAO, SN and $H_0$}

\author{Tirthabir Biswas \\ Department of Physics,\\
Saint Cloud State University,\\
Wick Science Building, Saint Cloud, MN 56301, U.S.A\\ 
and\\
Department of Physics\\
Loyola University\\
6363 St. Charles Avenue, Campus Box 92\\
New Orleans, LA 70118, USA\\
\email{tbiswas@loyno.edu}}
\author{Alessio Notari\\Institut f\"ur Theoretische Physik \\
Universit\"at Heidelberg \\
Philosophenweg 16, D-69120 Heidelberg, Germany.\\
\email{notari@thphys.uni-heidelberg.de}
}
\author{Wessel Valkenburg \\ Institut f\"ur Theoretische Teilchenphysik und Kosmologie,\\
RWTH Aachen University,\\
D-52056 Aachen, Germany \\
\email{wessel.valkenburg@physik.rwth-aachen.de}
}

\abstract{
In this paper,  instead of invoking Dark Energy, we try and fit various cosmological observations with a large Gpc scale  under-dense region (Void) which is modeled by a Lema\^itre-Tolman-Bondi metric that at large distances becomes a homogeneous FLRW metric.
We improve on previous analyses by allowing for nonzero overall curvature, accurately computing the distance to the last-scattering surface  and the observed scale of the Baryon Acoustic peaks, and investigating important effects that could arise  from having nontrivial Void density profiles. We mainly focus on the WMAP 7-yr data (TT and TE), Supernova data (SDSS SN), Hubble constant measurements (HST) and Baryon Acoustic Oscillation data (SDSS and LRG). We find that the inclusion of a nonzero overall curvature drastically improves the goodness of fit of the Void model,
 bringing it very close to that of a homogeneous universe containing Dark Energy,
while by varying the profile one can increase the value of the local Hubble parameter which has been a challenge for these models.
We also try to gauge how well our model can fit the large-scale-structure data, but a comprehensive analysis will require the knowledge of perturbations on LTB metrics. 
The model is consistent with the CMB dipole if the observer is 
about $15$ Mpc off the centre of the Void. Remarkably, such an off-center position may be able to account for the recent anomalous measurements of a large bulk flow from kSZ data. Finally we provide several analytical approximations in different regimes for the LTB metric, and a numerical module for {\sc cosmomc},
 thus allowing for a MCMC exploration of the full parameter space.}

\begin{document}



\section{Introduction}
Our aim in this paper is to check whether spherically symmetric inhomogeneous models of our universe can represent a genuine alternative to the Dark Energy paradigm.
Dark Energy is one of the biggest mysteries of modern cosmology and of fundamental physics in general. Despite the fact that it is assumed to constitute about $75\%$ of the energy budget of the Universe, we do not know anything yet of its properties except that it dominates at late time and that it has an effective equation of state close to $-1$. We do not know if dark energy is only a cosmological constant or if it has dynamical properties, and we do not know why it becomes important in the cosmological evolution close to our present epoch.
A different approach may be taken where we assume  that Dark Energy is absent or  is negligible, and that we do not live at any special {\it time}, but  rather at a special {\it space} point. Namely that  we live very close to the  center of a very large local underdense region of the Universe~\cite{PhysRevD.45.3512,1995ApJ...453...17M,Tomita:1999qn,Tomita:2001gh,Tomita:2002td,Moffat:2005yx,Moffat:2005ii,Mansouri:2005rf, Alnes:2005rw, Chung:2006xh, Enqvist:2006cg,Enqvist:2007vb, Biswas:2007gi,ABNV,GarciaBellido:2008nz,Zibin:2008vk,February:2009pv,Romano:2009ej}. This is of course a very radical possibility, which goes against the so-called Copernican Principle, which states that we do not occupy any special place in the universe\footnote{Attempts to generalize such a void to a configuration in which the Copernican Principle is not (or less) violated, {\em e.g.} the so-called Swiss-Cheese universe, so far have failed to agree with observations~\cite{Marra:2007pm,Mattsson:2007tj,Biswas:2007gi,Vanderveld:2008vi,Valkenburg:2009iw}.}, and one may argue in terms of fine-tunings\footnote{An interesting explanation coming from inflationary cosmology for having such a configuration as a generic prediction has been studied in~\cite{Linde:1994gy}.} to establish which of the two possibilities is more unlikely; we simply take the approach here that observations, rather than postulates about what are the structures present in our Universe and what is our location in the Universe,  should be able to distinguish between the two models. Alternatively, we may think of this as an opportunity to test the Copernican Principle on the basis of recent observational data.

We have in mind a Universe described by a Lema\^itre Tolman Bondi (LTB) metric (which describes an inhomogeneous region, centered around the observer) matched, at very large distances from the observer, to an external Friedmann Lema\^itre Robertson Walker homogeneous and isotropic metric. We study different LTB profiles,
but we always require in general the inhomogeneous region to be underdense close to the observer (we call this a local Void). Sometimes there will also be
  a shell-like overdense structure near the boundary of this region (a ``compensated" Void).
We consider, as a zeroeth order approximation, the observer to be at the center, although  we do study some aspects of having an off-center observer.

While it is well established that a local Void can mimic an accelerated expansion~\cite{Tomita:1999qn,Tomita:2001gh,Tomita:2002td,Moffat:2005yx,Moffat:2005ii,Mansouri:2005rf, Alnes:2005rw, Chung:2006xh,Biswas:2007gi,ABNV,Romano:2009mr},  whether such a Void can successfully reproduce all current cosmological data is still a matter of debate.
Most work on the void models have focused on reproducing the shape of the $\La$CDM  luminosity distance ($D_L$) versus redshift ($z$) curve, in order to fit the  type Ia Supernova data, but a few studies~\cite{ABNV,GarciaBellido:2008nz,Zibin:2008vk} have included other data such as Cosmic Microwave Background (CMB) and Baryon Acoustic Oscillations scale (BAO).
In the present paper we try to perform a much more comprehensive analysis.
First, we combine fits of several cosmological observations: the CMB measured by the WMAP 7-year data, recent Supernova data (the SDSS SN~\cite{Kessler:2009ys}, which are a collection of SDSS, ESSENCE, SNLS, HST), the Hubble constant measurements (we use here the HST values~\cite{Sandage:2006cv}), and the measurements of the BAO scale (BAO)~\cite{Percival:2009xn}.  We also try to include the large scale structure data using the SDSS main sample~\cite{Tegmark:2003uf} and the Luminous Red Galaxy (LRG) subset, DR4~\cite{Tegmark:2006az}, although this required some drastic simplifying assumptions.

Second, we enlarge the parameter space with respect to previous analyses. So far, most fits in the literature  have considered a Void embedded exactly in a flat Einstein de Sitter Universe (Eds Void Models). In~\cite{Sollerman:2009yu} voids embedded in a strictly open or flat FLRW universe have been considered and fit to the CMB, SN and BAO, although ignoring the monopole shift that the CMB experiences in a void (as will be explained in Section~\ref{sec:theory_detail}). Here we consider the generic possibility that the outer background FLRW metric has arbitrary spatial curvature. Fitting the CMB  with an asymptotically curved FLRW model (Curved Void Models) allows to fit better the distance to the Last Scattering Surface~\cite{Clifton:2009kx}. For this purposes we develop in this paper an analytic treatment of LTB solutions, including the derivation of the distance-redshift curve, which allows for curvature in the outer region.

Third, we include the effect of the Void on the redshift of the last-scattering surface, which can be encapsulated by an effective change in the CMB temperature monopole. This effect is suppressed for small Voids and it was neglected for example in~\cite{ABNV}, where only Voids of size of about $300 {\rm Mpc}/h$ were considered. However, the effect can become significant for larger voids (of order ${\rm Gpc}$) and it was taken into consideration in~\cite{Zibin:2008vk}.
This effect scales as $(L/r_H)^2$, where $L$ is the size of the Void and $r_H$ is the present horizon. We also derive the physical quantity that is appropriate  for the BAO observations in LTB models. This allowed us to accurately compute the physical scales associated with the BAO peaks in our inhomogenous model and compare it with observations.

Finally, we include several variations on the shape of the Void profile. The simplest void models essentially have two important physical parameters: the size of the void, and the amplitude of density contrast within the void. These models have a relatively constant underdensity at the center, followed by a ``compensating'' overdense shell which is then matched to a homogeneous FLRW background. Initially, mainly for the purpose of illustration we focus on this simple case. Later however we start exploring more nontrivial  radial density profiles.   Two interesting cases emerge: the non-compensated Voids which  go only asymptotically to FLRW, and the ``compensated'' ones which have an additional higher underdensity near the center of the void compared to the surrounding region inside the void. Both these profiles  have  a large effect on the monopole of the CMB which can accommodate for larger values of the measured Hubble constant and/or  locally have a higher apparent acceleration providing us with a better fit to the BAO.

So far, several models have appeared in the literature which mostly differ in the profile and the size of the void. It is useful to  distinguish between three classes of models:
\begin{itemize}
\item[I.]  ``Minimal Void'' models~\cite{ABNV}: These are relatively small in size, extending upto a redshift of $z\sim 0.1$; in this case the main effect is a difference in the expansion rate inside (nearby supernovae, $z<0.1$) and outside (distant supernovae $z>0.1$), which can to some extent  mimic the effect of acceleration. Since the probability of having large voids is exponentially suppressed according to the ``standard'' analysis on growth of structures ~\cite{Hunt:2008wp}, these models have an advantage over the larger void models, but recent studies based on newer supernovae data seem to rule them out. We find that allowing for a background spatial curvature makes these models consistent with supernovae  again, but unfortunately they do not produce a consistent combined fit of SN + CMB. Thus we will mostly devote our attention to the larger voids.
\item[II.] Large ``Compensated'' Void models: These typically extend up to very high redshifts $\sim 0.5-1.5$, and incorporate most of the supernovae; the radial void profile can be used to modify the luminosity distance ($D_L$) versus redshift ($z$) curve all the way up to these high redshifts to fit the supernovae data. As we will see, these models can indeed be consistent with SN + BAO + CMB + HST. However, in order to have a large local Hubble parameter these models require additional features in the Void profile.
\item[III.] Non-Compensated Voids: These match to the background FLRW only asymptotically, \ie it is difficult to say in a sharp way what the Void size is, since they never approach exactly FLRW. These profiles do not necessarily have a compensating overdense shell. It turns out that the monopole correction to the CMB temperature in such cases helps fit the CMB data significantly. This in turn ensures that much larger local values of the Hubble parameter are possible.
\end{itemize}
When comparing these models with data the three classes have to be dealt with slightly differently.
For example, the ``Minimal Void'' case is well under control analytically using an expansion for $L/r_H \ll 1$, which is basically a Newtonian approximation~\cite{Biswas:2007gi,ABNV}. Moreover, all datasets which are located outside this Void (for example the LRG or CMB) can just be analyzed within the standard FLRW framework. Finally, when studying effects on the $D_L-z$ curve in the outer homogenous region, the corrections can be neglected.

The Large Void cases (II \& III) instead require more care. In particular, if we want to employ an analytical approximation, higher order post-newtonian terms in $L/r_H$ must be included. Besides, the distance curve in the outer (nearly) homogenous region receives non-negligible corrections (which have to be taken into account for example when fitting the CMB). Finally the data analysis poses non-trivial problems: since many datasets (such as the BAO and the large scale-structure) are inside the Void, the usual treatments have to be modified accordingly. Finally, for the non-compensated case, since the FLRW metric is approached only asymptotically, we have to define what we mean by the background FLRW metric. It turns out  to be the ``effective'' FLRW model describing the evolution of the last scattering surface.

While fitting the Void models with the major cosmological data-sets is essential in determining the viability of these models,  it is also important to find distinguishing features of these scenarios which are unique to a spherically symmetric model.
In this regard,  a crucial point is  that the observer will in general  be displaced from the exact center of the LTB metric. This has several observational consequences, such as anisotropic expansion, the presence of a large dipole in the CMB, large coherent peculiar velocities (``bulk flows"), etc.
 We discuss the constraint coming from the CMB dipole on how far away ``we'' can be located from the center of the Void. This is important in estimating the amount of fine-tuning involved in such scenarios. The Void also naturally predicts that we will observe an ``apparent'' bulk flow of all the objects within the void. Furthermore the ``bulk velocity'' should be aligned with the CMB dipole. Remarkably there are now indications of precisely such ``Dark Flows'' of matter ~\cite{Kashlinsky:2008us} around us.
It is also worth mentioning that~\cite{GarciaBellido:2008gd} has explored a very interesting effect which is present even when the observer is at the centre: the kinetic Sunyaev-Zeldovich effect,  which can put upper bounds on the size of very large Voids.

When comparing goodness of fit of the different models at stake, we choose to focus on the effective quantity $\chi_{\rm eff}^2 \equiv -2 \ln \mathcal{L}$, where $\mathcal{L}$ is the likelihood. We do not perform a Bayesian Evidence (BE) comparison, because the calculation of the BE in the case of the Void scenario is not straightforward. We explain this in Section~\ref{divergences}. Secondly, since we do find models with a $\chi_{\rm eff}^2$ comparable to that of $\Lambda$CDM (at best a difference in $\chi_{\rm eff}^2$ of $\sim 2$ on $3406$ degrees of freedom in the data) but do not find models with exactly the same or a lower value of $\chi_{\rm eff}^2$, we do not expect to find significant outcomes of a BE calculation, which would anyway still favour $\Lambda$CDM.

The paper is organized as follows: in section~\ref{sec:theory_intro} we present the theoretical framework to study void models and some general analytical results. In section~\ref{sec:theory_detail} we discuss qualitatively how to compare the void models with data. In section~\ref{sec:technicalresults} we include a more technical discussion, which some of the readers may want to skip,
 on how to perform such an analysis, including the set of parameters that are used in the analysis. In section~\ref{sec:num_analysis} we describe how our numerical code works\footnote{We implemented our code as a module in {\sc cosmomc}~\cite{Lewis:2002ah}. We will publicly release our module at \href{http://web.physik.rwth-aachen.de/download/valkenburg/}{http://web.physik.rwth-aachen.de/download/valkenburg/}  }. In section~\ref{risult} we describe our main results for the simple void profiles which help the  understanding of how the presence of the void affects the different important observational quantities, and also serve as an illustration of the numerical prescription we are following.
Next  in section~\ref{sec:exotic},  we explore more elaborated compensated profiles and non-compensated profiles and how it improves the various fits. We also include  some additional CMB and HST data. In section~\ref{sec:otherpreds} we comment on other cosmological predictions. We conclude in section~\ref{sec:conclusions}, summarizing our findings, and pointing out some unique distinguishing signatures of these models.

In the appendices we present a number of technical results: in Appendix~\ref{approximations} analytical approximations of LTB metrics and geodesics in different regimes,  in Appendix~\ref{app:t_ast_CMB}  the correction to the CMB temperature in the general case which includes curvature in the outer FLRW region and in Appendix~\ref{app:snchi2} a discussion about the likelihood of multivalued functions, which is needed in some part of the parameter space in our Void models. Finally in Appendix~\ref{figurestables}  we include figures and tables with the results of our analyses.

\section{The LTB model and curvature function}\label{sec:theory_intro}
\subsection{The metric}
Traditionally spherically symmetric void models have been studied using LTB metrics which are exact solutions of General Relativity that can be studied both
analytically and numerically. One advantage of using LTB metrics is that the spherical inhomogeneous LTB patches can be pasted onto a homogeneous FLRW
metric consistently~\cite{Khakshournia:2002hc,BN1,Biswas:2007gi}. As far as we know, in all analyses that have been performed so far, the LTB metric is patched to a flat EdS universe either at a finite ``void size'', or asymptotically as we approach the Hubble radius. One important modification we are going to implement is to remove such a restriction and study solutions where the LTB is embedded inside an open, flat or a closed universe. We refer to them as ``Curved Void'' models as opposed to the traditional ``EdS Void'' models.

Technically, the LTB metric is described by three different free functions of the radial coordinate $r$, but the picture can be simplified with a gauge choice and a physical requirement. As explained in~\cite{BN1, Biswas:2007gi} one function (the``mass function'') can be made trivial by a redefinition of coordinates, while another function (the ``bang'' function which sets the time of big bang at each value of $r$) can be set to zero, if we restrict to metrics which become more and more homogenous in the past. In this case, we are left with only one function $k(r)$, the curvature function.
Intuitively  it is like having an independent scale factor
corresponding to each (comoving) radial coordinate, $r$, which is evolving  as an independent FLRW
metric with a given spatial curvature $k(r)$. The function $k(r)$ is arbitrary and also
determines the density profile.

The LTB metric is given in general by
\begin{equation}
ds^2=-dt^2 + S^2(r,t)dr^2 + R^2(r,t)(d\theta^2 + \sin^2 \theta d\varphi^2) , \label{eq:14} \,
\end{equation}
Here we have employed comoving coordinates ($r,\theta,\varphi$) and proper time $t$.
The functions $S(r,t)$ and  $R(r,t)$ are completely specified by the ``curvature'' function $k(r)$. $S(r,t)$ can be determined from $R(r,t)$:
\begin{eqnarray}
S(r,t) &=& {R'(r,t)\over \sqrt{1+2r^2k(r) \bM^2}} , \label{eq:15}
\end{eqnarray}
where the prime denotes partial differentiation with respect to $r$, and $\bM$ is an arbitrary mass scale which  does not appear in any physically observable quantity. $R(r,t)$ in turn is governed by an evolution equation resembling the Hubble equation:
as we mentioned before, intuitively the LTB model can be understood as a continuous collection of FLRW universes, one for each radial coordinate.  We may define an $r$-dependent scale factor and a Hubble parameter as:
\be
a(r,t)\equiv {R(r,t)\over r}\qquad \mx{ and } \quad H(r,t)\equiv {\dot{a}(r,t)\over a(r,t)} \,\, .
\ee
Note that this scale factor is the one associated with the angular part of the metric, while the radial expansion is in general different.

We find from the Einstein equations that this scale factor satisfies an $r$-dependent Hubble equation
\be
\LF{\dot{R}\over R}\RF^2=\LF{\dot{a}\over a}\RF^2=H^2(r,t)= {8\pi\bM^2\over 3}\LT{1\over a^3(r,t)}+{3k(r) \over 4\pi a^2(r,t)}\RT\equiv{8\pi\over 3m_p^2}\LT{M_0^4\over a^3(r,t)}+{3k(r) M_0^4\over 4\pi a^2(r,t)}\RT
\label{r-hubble}
\ee
The above equation is identical to the Hubble equation in a matter dominated  FLRW cosmology with spatial curvature,  the only difference being  the possible $r$-dependence coming from $k(r)$.
Accordingly, the solution is also identical to that of a FLRW universe in the following implicit form:
\bea
a(r,t)={2\pi\over 3k(r)} &[&\cosh u(r,t)-1]\,\, ,\label{scale-factor}\\
t={\sqrt{2} \pi\over 3k(r)^{3/2}\bM}&[&\sinh u(r,t)-u(r,t)] \,\, , \label{time}
\eea
except that the intermediate variable $u$ also depends on the radial coordinate.
Note that $u$ can be  imaginary when $k(r)$ is negative. In this case we have to take the positive root of $-1$ while taking the square root of $k$:
$$k^{3/2}=(\sqrt{k})^{3}=|k|^{3/2}i^3=-i|k|^{3/2}\mx{ for } k<0 \ .$$

Substituting (\ref{scale-factor}) in (\ref{r-hubble}) we find that the $r$-dependent  Hubble parameter is given by
\be
H(r,t)={3k(r)^{3/2}\bM\over \sqrt{2}\pi}{\sinh u(r,t)\over[\cosh u(r,t)-1]^2}\label{H} \,\, .
\ee
When the profile is matched exactly to FLRW we have that $k(r)=k_b$ for $r>L$, where $k_b$ is a constant and $L$ is the radius of matching with FLRW. So, in the outer region, $r>L$, the $r$-dependence vanishes in all the quantities and it is possible to see that $H(r>L,t)$ corresponds to the ``standard'' Hubble parameter that is defined for the homogeneous FLRW background.
In the next section, we will see that the $r$-dependent Hubble parameter at the centre, $H(0,t)$ coincides with the ``observational'' definition of the ``local'' Hubble parameter using the luminosity-distance vs. redshift relation.
We note that our sign convention is such that positive $k_b$ corresponds to a spatially open embedding FLRW universe, and negative $k_b$ corresponds to a spatially closed embedding FLRW universe.
\subsection{Density Profile and Consistency}
We can compute the matter density and abundance for the LTB metric. Using the exact expression for the density function~\cite{BN1,Biswas:2007gi}, we have
\be
 \rho (r,t) = {\bM^2M_{p}^2 r^2 \over R'(r,t) R^2(r,t)}  \label{eq:17} \,\, ,
 \ee
 where $M_p$ is the Planck mass. One can rewrite this as
 \be \rho(r,t) = M_0^4\LT{3k\over2\pi(\cosh u-1)}\RT^3 ={M_0^4\over a^3(r,t)}\, .
\label{central-density}
\ee
as long $k'(r)=0$, which, as we shall soon see, is valid at the centre and in the background for $r>L$. We have defined here the mass scale $M_0^4\equiv \tilde{M}^2 M_p^2$.  From now on we will also use the convention that  $a(L,t_0)=1$, with $t_0$ referring to the present epoch, so that $\tilde{M}$ is fixed by the value of the present energy density $\rho(L,t_0)$.

To have an intuitive understanding of the  kind of density profiles that are modeled by LTB metrics, let us assume $L\ll r_H$ (the Hubble radius) and consider a LTB region embedded in an EdS background.  In this case we have
\be \rho(r,t)\approx \frac{\langle\rho\rangle(t)} {1+(t/t_0)^{2/3}\e(r)} \, ,
\qquad  {\rm where} \, \, \, \langle\rho\rangle(t) \equiv \frac{M_p^2}{6 \pi t^2} \, ,
\label{matterdensity}  \ee
which is approximately valid as long as
\be
\epsilon(r)\equiv 3 k(r)+ r k'(r)\ll 1 \, \mx{ and }t<t_0\equiv {1\over \bM\sqrt{6\pi}} \, .
\ee

We observe that the EdS behaviour for the density is given by the factor $\langle\rho\rangle(t)$, while the
fluctuations are provided by the presence of $\e(r)$ in the denominator. When $\e(r)$ is close to
its maximum value we have a void, while when it is close to  its negative minimum, it signals an
overdensity. Note that at early times the density contrast $\de(r,t)\equiv (\rho(r,t)-\langle \rho
\rangle(t))/ \langle \rho\rangle(t)$, defined in the usual way,  grows as $t^{2/3}$, in agreement
with the prediction of cosmological perturbation theory. On the other hand at late times, when
$(t/t_0)^{2/3}\e(r)\sim \cO(1)$,  the density contrast grows rapidly (and this result is  the same
as found within the Zeldovich approximation ) and diverges. Nevertheless, for our purposes, as long as we do not reach
this singularity time, LTB metrics adequately capture the  effects of non-linear stucture formation on
photon propagation.

 Now,  we are mostly going to be interested in using a $k(r)$ which starts off from a maximum at $r=0$ and falls off to a constant value at $r=L$ such that
\begin{eqnarray}
 k'(0)&=&k'(L)=0 \, , \label{cond1} \\
 k(L)&=&k_b=\frac{4}{3\pi}\frac{\Omega_k}{1-\Omega_k}.
\label{junction}
\end{eqnarray}
The conditions~(\ref{cond1}) on the derivatives \cite{Khakshournia:2002hc,BN1,Biswas:2007gi} guarantee that there is no density cusp at the centre and that there is exact matching to the external FLRW region.
Here $\Om_k$ is the curvature abundance of the FLRW background to which the LTB metric can consistently be
matched, as explained in section~\ref{sec:relations}. %
(Note that in our conventions  $\Om_k>0$ corresponds to an open universe.)
We will also consider, later on, profiles which are not exactly matched to an FLRW.

The above class of functions describes a spherical void region surrounded by a compensating shell-like structure.
The simplest choice of the curvature function that we employ to model the inhomogeneities and fit the supernova data is
given by\footnote{The exponent of $r/L$ has been chosen to be equal to 4, but the reader may note that any exponent $n>1$ satisfies the constraints (\ref{junction}) and approximately describes the same physical situation of having an underdense central core surrounded by a compensating overdense shell. Thus we did not expect $n$ to play an important role in fitting the various observations. In general the larger the value of $n$ the flatter the void and narrower the structure. For the purpose of illustration we therefore first chose to work with (\ref{profile}) which was previously shown to provide  a good fit to the supernova data~\cite{ABNV}. In section~\ref{sec:exotic} we explore the effect of changing the profile in greater details. We also note that we could also play with the overall exponent in (\ref{profile}) which we have set to 2 for similar considerations. }
\be k(r)=\left\{\begin{array}{cr}
k_{\mt{max}}\LT1-\LF{r\over L}\RF^4\RT^2 +k_b& \mx{ for } r\leq L\\
k_b& \mx{ for } r\geq L
\end{array}\right.\label{profile} \, . \ee
so that the value of ``curvature'' at the boundary is just $k_b$, and at the centre
\be
k_0=k_b+k_{\mtx}.
\ee
We can check that
(\ref{profile}) satisfies (\ref{junction}).
There are two important physical parameters, $L$ and
$k_{\mt{max}}$, which correspond to the length-scale and amplitude of fluctuations respectively.

In the
rest of the paper, we will use this profile to gain insight into the basic physics, although some of the analytical
results are general for any $k(r)$.
In section~\ref{sec:exotic}, we will specialize to more elaborate profiles.


\section{Fitting Observations to Void Models}\label{sec:theory_detail}
\subsection{Overview \& Basic Physics}
In the previous sections we have discussed the basic properties of the model and in this section we want to present an overview of how we proceed to test our model against the various cosmological observations. It is well known that the $\La$CDM model provides a reasonably good fit to all the important cosmological data. Thus we will use the $\La$CDM model as our standard of comparison. We will proceed as follows: we will first look at the data coming from SN (Distance vs.~redshift), CMB and BAO, together with the HST measurement.
Next, we will also try to include  the Large Scale Structure data, albeit after making some simplifying assumptions. These assumptions   need to be verified in the future and this would require developing a perturbation theory about the LTB metric, which is still lacking.

The main reason why inhomogeneous models can be interesting for the dark energy problem is because an underdense region tends to become more and more underdense (that is how structure formation works in GR). In other words, the space in an underdense region expands  a little faster than the rest, and therefore if we happen to be living in a local  underdensity this effect can reproduce the faster expansion rate as inferred by comparing nearby supernovae to high-redshift supernovae. Our first aim is to check, with the advent of new supernovae, which inhomogeneous models can still provide a good fit to the SN data. In~\cite{ABNV} a Small Void profile, extending only up to $z\sim 0.1$,  was proposed but the newer supernovae data~\cite{Kessler:2009ys} seem to rule them out. As we will see indeed small EdS void models can no longer fit the supernovae data.  However a Curved Void models still can, which is a  first indication that the background curvature may play an important role in Void cosmology.
It turns out however that the small voids require an open background universe. This unfortunately  is incompatible with CMB which prefers a flat/closed universe.
Thus, we shift our focus mainly to Curved Voids with larger size, and as we will see, they provide much better combined fits.

The second issue we wanted to address is that, although in the literature~\cite{ABNV,GarciaBellido:2008nz,Zibin:2008vk} void models can provide a reasonably good combined fit\footnote{In~\cite{ABNV} the $p$-value (often referred to as goodness of fit) of the combined data sets was found to be around 26\%.} of CMB and SN, these fits are significantly worse as compared to the $\La$CDM fits. We again find  that  allowing for a curved background  improves the fit remarkably by allowing us more freedom on the distance to the last scattering surface: compared to the $\La$CDM + curvature model, for the 3403 combined data points the EdS Void model has a $\De \chi^2\sim 38$, where as for the Curved Void model this difference goes down to $\De \chi^2\sim 7$. This should not come as a total surprise as it is known that the closed universe model, albeit with a very low Hubble parameter,  provides an excellent fit to the WMAP data~\cite{Bennett:2003bz}.
The CMB also receives a non-negligible correction to the effective monopole temperature when the  void becomes large (this is in contrast to the small void scenario studied in~\cite{ABNV}), which further changes  the distance to the last scattering surface. We incorporated this effect in our numerical code to obtain an accurate fit to the data.

Next, there is the issue of the  position of the Baryon Acoustic Oscillations peak in the void models. It is well known that the FLRW models without Dark Energy cannot be made consistent with the BAO data~\cite{Blanchard:2005ev,ABNV}. However, when the void is large the BAO data fall inside the Void, and the distance rulers which are inside the void are very different from the usual FLRW ones. We  provide a detailed derivation of the relevant distance measurements for a general LTB patch. Using these accurate physical quantities, in conjunction with having the freedom to choose the background curvature, remarkably, allows us to find a good fit to the BAO peak. In fact, for certain inhomogeneous profiles the fit can be made even better than the $\La$CDM model, see for instance profile C in table \ref{tab:chi2_exotic}.

In section~\ref{risult}, after discussing the fits to supernovae, CMB and BAO, we look at the local Hubble measurement coming from HST. We find that the simplest inhomogeneous  profiles give us a relatively low Hubble parameter which is difficult to reconcile with HST measurements; however we construct other inhomogeneous profiles which provide a higher value of $h$. Depending on which value is taken for HST (different values have been published ranging from $h\sim 62$ to $h\sim 74$), it is possible to provide a consistent combined fit of SN, WMAP, BAO and HST. For instance, the combined fit of profile D is only slightly worse than the $\La$CDM fit, $\De \chi^2\sim 2$ for a total of 3406 data points, when using the HST value of  $62.3 \pm 6.3$ km s$^{-1}$ Mpc$^{-1}$ from Ref.~\cite{Sandage:2006cv}.

Finally, we discuss compatibility with  Large scale structure data coming from the SDSS main sample and from the LRG data.  The situation here is involved, because  in void models the  matter density is no longer homogeneous and we cannot directly  apply the known FLRW   results on growth of structures. Although progress have been made \cite{Zibin:2008vj,Clarkson:2009sc,Clarkson:2010uz} in understanding these issues in LTB models, it is still going to be extremely involved to analyze the large scale structure data (LRG and SDSS) in LTB models both theoretically as well as numerically. In this paper we took a much easier approximate route to deal with this problem: we
treat the underdense region in the LTB patch as an effective open FLRW model for the purpose of computing the growth of structures relevant for SDSS.
We discuss our findings and possible caveats in section \ref{results:LSS}.

In the following subsections we discuss some of the physical quantities that are relevant for fitting our model to the observational data, and explain how they can be calculated in the framework of LTB metrics.
\subsection{Distance vs. Redshift relation for Supernovae}
The  Void models became interesting in the first place because  they could fit the SN data without Dark Energy. Each supernova is a data-point which measures a luminosity distance $D_L$ at a given redshift $z$.
We have included the $D_L-z$ relationship in our MCMC code. This firstly involves numerically solving for a radial photon trajectory:
\be
{dt(r)\over dr}=-{R'(r,t(r))\over \sqrt{1+2k(r)\bM^2r^2}} \, .
\label{t-radial}
\ee
Next, the redshift as a function of the radial coordinate is obtained by solving~\cite{Celerier:1999hp,BN1}
\be
{dz\over dr}={(1+z)\dot{R}'\over \sqrt{1+2k(r)\bM^2r^2}} \, , \label{eq:z-radial}
\ee
Finally, the angular distances of the supernovae can be calculated quite easily   as they are simply given by~\cite{Celerier:1999hp,BN1}
\be
D_A(r)=R(r,t(r))\mx{ and }D_L=(1+z)^2D_A\ .
\ee
Thus,  once $t(r),z(r)$ is known, so is $D_L(r)$. Combining $D_L(r)$ and $z(r)$, we obtain the usual $D_L(z)$ plot.

For the supernova data the key physical parameters in our model are $k_0, L, k_b$ and $t_0$, or equivalently the central density contrast of the void, $\de_0$, the boundary redshift, $z_b$, the background curvature abundance, $\Om_k$, and the background Hubble parameter, $H_\mto$. (We will discuss in more technical detail the relation between the LTB metric parameters and the more physical parameters in section~\ref{sec:relations}.) For small voids ($z_b\lesssim 0.05$) and/or small inhomogeneities ($\delta_0<0.1$), very good analytical approximations for the redshift as a function of the radial coordinate exist, see Appendix~\ref{app:ana_approx}. We have compared our numerical results with the analytical results in Appendix~\ref{app:ana_approx} in figure~\ref{fig:app_dt_t_2}. As we can see, both for small redshifts and small inhomogeneities, the numerical results agree very well with the analytical approximations.
\subsection{Local Measurement of Hubble Parameter}
For local Void models, the essential point is that at the centre of the void the Hubble parameter is slightly larger in comparison to the value it has in the external region at the same LTB time-slice. In this context we note that the HST measurements go only up to a redshift of around 0.1, so that they lie well within the core underdense LTB region. Another important remark is that, although we provided an intuitive radially dependent expression for the transverse Hubble parameter, observationally the local Hubble parameter, $H_0$, is defined as
\be H_0^{-1}\equiv \lim_{z\ra 0}{D_A(z)\over
z}\ = {3000\ {\rm Mpc}\over h}. \label{Hubble} \,
\ee
In section~\ref{subsec:localhub} we show that the intuitive definition of the Hubble parameter coincides with the formal definition (\ref{Hubble}), \ie $H_0=H(0,t)$. We also explain how the local Hubble parameter is  related to $H_{\mt out}$ via $\delta_0$ and $\Om_k$ in a general way which is independent of the details of the profile. These are therefore the only quantities needed to fit the HST measurements.

The fact that the local Hubble parameter is larger than the background Hubble parameter in LTB models is rather useful because, as we will see, the value in the external region needed for the CMB fit is  very low, and incompatible with HST measurements. It turns out that for the simplest LTB profiles the enhancement of the Hubble parameter is still not enough to be consistent with HST, but we have found other LTB profiles which can yield  a much higher local value of the Hubble parameter making it compatible with HST, at least using the HST value of  $62.3 \pm 6.3$ km s$^{-1}$ Mpc$^{-1}$ from Ref.~\cite{Sandage:2006cv}..  This will be discussed in section~\ref{sec:exotic}.
\subsection{The Cosmic Microwave Background Radiation}
\label{sec:CMB}
Performing an analysis of the CMB data might seem a challenging task in the light of the presence of large scale inhomogeneities such as the one we are considering. For the purpose of this paper however we are going to assume that we are located very close to the centre, so that by symmetry there are no additional anistropic  effects on the photon redshift.\footnote{This symmetry is broken when considering perturbations, and in fact an effect that we should consider is the Integrated Sachs-Wolfe (ISW) effect, due to the secondary effect of density fluctuations at late time. Such an effect is absent in EdS, but it is expected to be generically present in a Void model, because the growth of perturbations is different, in general. Since we do not have a treatment of the growth of perturbation in LTB, we keep the FLRW treatment in the code. The reader should remember, however, that the low-$\ell$ part of the CMB spectrum cannot be fully trusted because of this approximation.} So, in fact, the CMB can be analyzed in terms of an effective FLRW observer. Let us explain this in more details.

The formation of the CMB spectrum depends primarily on three factors, the  primordial spectrum of metric fluctuations, the epoch of matter-radiation equality, and the speed  of the sound waves in the  baryon-photon plasma at the epoch of recombination which provide us with the characteristic physical length  scale $L_s$, the sound horizon of acoustic oscillations at the time of recombination. Throughout this paper we consider an adiabatic primordial power spectrum characterized by an amplitude $A$, a spectral index $n_{\rm s}$ and a running $\alpha$. Both the epoch of matter-radiation equality and the sound velocity at recombination can be determined if we know the composition of the universe at the recombination time. Now, since the last scattering surface is located in the FLRW background, all  we  need to know are the energy densities of the various components of the universe ({\it i.e.} that of curvature, dark matter, baryons  and radiation) today in the background region. Then we can just use the standard FLRW codes to ``recreate'' the recombination epoch from the background universe today.

There is however one subtlety: While $H_\mto,\Om_{k, {\rm out}}$ and $\Om_{b,\mt{out}}$ specify the first three of the energy density components, for radiation we need to compute the photon temperature in the background, $T_\mto$,
which is not given by the usual value $T_0=2.726 K$. The temperature seen by the background FLRW observer is in general different, which means that effectively the last scattering surface is located at a different redshift. This effect was neglected in~\cite{ABNV}, since the Void considered there was much smaller and the correction to the temperature was tiny, since it goes as $(L/r_{hor})^2$. In this paper we calculate the ratio $T_\mto/T_0$ which boils down to calculating the extra redshift an LTB observer sees compared to the FLRW observer. We use  numerical results for our MCMC code, but we ensured that our numerics agreed with  analytical approximations, as discussed in Sections~\ref{sec:technicalresults}~and~\ref{sec:num_analysis}.

Finally, we need to evolve the physical length scales at the CMB epoch to angular scales that we observe in the sky today. This requires the knowledge of the angular distance, $D_A$, of the last scattering surface. Fortunately, in the LTB metric this is precisely given by $R(r,t(r))$ which reduces to its FLRW form\footnote{There is actually one non-trivial point about this calculation. Although $R(r,t(r))$ reduces to its FLRW form at last scattering, the function $t(r)$ is different in the LTB patch compared to the FLRW background. To account for this mismatch in time, we have to readjust the time coordinate in the effective FLRW model. This is explained in more details in section \ref{sec:technicalresults}.} with $k(r)=k_b$ at last scattering. Thus the CMB spectrum that we see in the LTB model is identical to the background  effective FLRW model, and we modified our MCMC code accordingly. To find the parameters for this effective FLRW model we need to relate the outer parameters to the inner parameters, as a function of the Void profile, density contrast and radius, as we will explain in Section~\ref{sec:technicalresults}.

Finally, in order to perform fits to the CMB data we also included the other usual parameters, which do not depend on the presence of the void. Thus the complete list of parameters in our model relevant for CMB is given by $\{A_{\rm s}, n_{\rm s},\alpha_{\rm s},H_\mto,\Om_{b,\mt{out}},\Om_k,\de_0,z_b,\tau, A_{\rm SZ}\}$, where $\tau$ is the optical depth at re-ionization epoch, and $A_{\rm SZ}$  is the normalization of the Sunyaev-Zeldovich template from~\cite{Komatsu:2002wc}, which is treated as a nuisance parameter since from the WMAP observations alone it is not possible to distinguish between primordial anisotropies and those induced by the thermal SZ-effect~\cite{Hinshaw:2006ia}.

\subsection{Large Scale Structure}
\label{sec:LSS}
The correct way to treat the Large Scale Structure (LSS) data would be to compute the growth of perturbations in an LTB model.
This is a challenging task, since the perturbations around LTB have not been fully studied, and since they involve a non-trivial interplay between radial and transverse modes. Although this has been approached by some authors\cite{Zibin:2008vj,Clarkson:2009sc,Clarkson:2010uz} a full treatment is still missing and we do not attempt it here.
We take the more modest approach of treating perturbations in an effective FLRW model to get a sense of  how easy or difficult it is going to be to fit the LSS data in the LTB models.
As mentioned before, in fact, the LTB metric can be qualitatively understood as a continuous collection of FLRW universes, one for each $r$. The transverse Hubble constant $H(r,t)$ is given by equation~(\ref{r-hubble}). Dividing all terms by $H(r,t_0)^2$, we can define  effective abundances $\Omega_m(r)$ and $\Omega_k(r)$ via
\begin{align}
1&= {8\pi\bM^2\over H^2(r,t_0)}\LT{a^3(L,t_0)\over a^3(r,t_0)}+{3k(r) a^2(L,t_0) \over 4\pi a^2(r,t_0)}\RT,\nonumber\\
\Omega_{m}(r) &=  {8\pi\bM^2\over H^2(r,t_0)}{1\over a^3(r,t_0)}\ , \mx{ and}\label{eq:om_mat_r}\\
\Omega_{k}(r) &=  {8\pi\bM^2\over H^2(r,t_0)}{{3 k_0   \over 4\pi a^2(r,t_0)}},\label{eq:om_k_r}
\end{align}
where we are continuing to use the normalization $a(L,t_0)=1$. In the case of an uncompensated void that only asymptotically goes to FLRW, we replace  $a(L,t_0)$ with  $a(r_*,t_0)$, as described in Appendix~\ref{app:t_ast_CMB}.

When the void is very large such that $z_B\gg z_{max}$, where $z_{max}$ is the highest redshift  for a given  LSS survey, and  the void profile $k(r)$ is sufficiently flat near the origin, then it seems a reasonable approximation to calculate the matter power spectrum inside the void treating it as an effective open FLRW universe. Thus in our MCMC simulations to estimate the growth of structures we indeed use the effective open universe model with  parameters $\Omega_{m, {\rm in}}\equiv \Omega_{m}(r=0)$, $\Omega_{k, {\rm in}}\equiv \Omega_{k}(r=0)$ and $H_{\rm{obs}} = H_{\rm in} = H(0,t_0)$. We also assume that  $\Omega_b / \Omega_{cdm}$ remains a constant throughout the universe. Clearly such an approach does not take into account the different behavior between radial and transverse modes and this should be taken into account in a more refined analysis.

To complete the story, apart from the parameters discussed above we also need the two other usual ``nuisance'' parameters to capture the non-linear structure formation:  $b^2$, the square of the bias (between luminous and dark matter) in the region outside the void, and $Q_{nl}$, which  parameterizes small scale non-linearities on the matter power spectrum.
\subsection{Baryon Acoustic Peak}
\label{sec:BAO}
What is measured by several collaborations \cite{Eisenstein:2005su,Percival:2009xn} is a feature in the correlation function of the galaxy distribution, corresponding to the sound horizon $L_S$ at CMB time.
This can be quoted  as a measurement of the  combination $(\Delta \theta)^2 \Delta z$ at some redshift (so far at 0.2 and 0.35), where $\Delta\theta$ is the angle in the sky and $\Delta z$ is the interval in redshift space corresponding to the comoving sound horizon $L_S$.

Thus, we have to construct what is the value of the quantity $(\Delta \theta)^2 \Delta z$ in our LTB model.
In order to do this, we first have to compute the value of the sound horizon $L_S$ corresponding to our cosmological parameters in the relevant location. For this purpose we again use the effective abundances~(\ref{eq:om_k_r}) evaluated at the appropriate  $r$  corresponding to the location of the BAO observations.  Next we compute $L_S$ using a Boltzmann code (CAMB)~\cite{Lewis:1999bs} under the approximation of an effective FLRW model. Finally, we  calculate $(\Delta \theta)^2 \Delta z$  using the LTB metric, taking into account properly that the expansion in the radial and transverse directions are different. This is explained in detail in the next section.

\subsection{List of the parameters}
\label{sec:parameters}
To summarize, let us  enumerate all the parameters of our model that are needed to fit the five observations (SN, CMB, HST, LSS, BAO) we discussed: (i) $H_{0,\mt{out}}$, this is the Hubble parameter outside the void in the FLRW background at time $t_0$, which is the time for the LTB observer located at $r=0$, (ii) $\Om_{m,{\rm out}}$, this is the matter (dark matter+baryons) abundance in the FLRW background today. (iii) $\Om_{b,{\rm out}}$, this is the baryonic abundance in the FLRW background today.
(iv) $\de_0$, the density contrast at the centre compared to the matter density in the background, evaluated today. (v) $z_b$, the redshift corresponding to  the boundary of the void. (vi)  $n_{\rm s}$, (vii) $\al_s$, and (viii) $A_{\rm s}$, are the spectral tilt, the running,  and the amplitude of the primordial power-spectrum respectively. (ix) $\tau$, this is the optical depth at re-ionization epoch. For the matter power spectrum we always marginalize over (x) $b^2$, the square of the bias (between luminous and dark matter) in the region outside the void, and over (xi) $Q_{\rm nl} b^2$, where $Q_{\rm nl}$ parameterizes small scale non-linearities on the matter power spectrum. Finally, for the CMB we always marginalize over the Sunyaev-Zeldovich template normalization, (xii) $A_{\rm SZ}$.


\section{Technical results: Relating LTB models to observations}\label{sec:technicalresults}
In this section we address some important technical results and approximations which are relevant for our analysis.
However, due to their slightly technical nature. The reader who is interested mainly in the basic picture and results may decide to skip them.
\subsection{The Effective FLRW model for CMB}\label{sec:effective}
As explained before, the CMB spectrum  obtained in the LTB model is identical to the spectrum  obtained in an  effective FLRW model.

This is because our observed CMB spectrum primarily depends on four physical ingredients: (A) The primordial spectrum, characterized by amplitude $A_{\rm s}$, spectral index $n_{\rm s}$ and possible running of the spectral index $\alpha_{\rm s}$ ; (B) The epoch of matter-radiation equality, which has an important effect on the growth of fluctuations as sub-Hubble density fluctuations start to grow as soon as  matter starts to dominate. (C) The sound speed at the epoch of recombination, which gives us the characteristic length scale for the acoustic oscillations, and (D) the angular distance of the last scattering surface which relates the angular spectrum that we observe in the sky with the physical length scale of the acoustic oscillations at CMB. Now (A)  is independent of the ``late-time'' background cosmology, and therefore it makes no difference whether we replace the LTB model with FLRW or not. For (B) and (C),  since the last scattering surface lies in an FLRW region, all we need to know is the energy budget of the FLRW background (the different energy components).  We can express this as quantities at a time $t_0 \equiv 2/(3 H_{\mt out})$ and then back-track and re-create the recombination epoch. Finally, in order for the FLRW model to reproduce the LTB spectrum the angular distance of the last scattering surface (D) must be identical to the LTB model. Since the angular distance in an LTB model is just given by the metric function $R(r,t(r))$, and the function $R(r,t)$ reduces to its FLRW form in the background, it seems that we should indeed be able to replace the LTB model with the background FLRW model as far as the CMB spectrum is concerned.

So, we need to find out which FLRW parameters describe an effective observer which sees the same CMB sky.
Now, $R(r,t)$ has the same functional form as an FLRW model once we substitute $k(r)=k_b$ in the outer region, so if we want the same angular distances in the outer region, we have to find an effective FLRW model such that  the trajectory function $t(r)$ that appears in the argument of $R(r,t)$ is also identical to the inhomogeneous scenario.
In general if $t(r)$ coincides in the outer region it will be different in the inner region.
So the FLRW observer at $r=0$ will have a time coordinate $t(0)\equiv t_{\mt{eff}}$, different from the LTB observer's time $t_0$.
For  similar reasons,  the background radiation temperature, which is one of the quantities that we need in order to recreate the last scattering surface, is going to be different from the observed CMB temperature $T_0$, because a photon which passes through the inhomogeneous LTB patch suffers an extra redshift compared to its passage in an FLRW region.

The prescription to find $t_{\mt{eff}}$ and the redshift is as follows. Given an LTB observer located at radial coordinate $r=0$ and time $t=t_0$, we can go backward in time along a geodesic $t(r)$ from the centre outwards. Once we cross the Void radius, we reach  the background FLRW region. We can easily compute the time coordinate, the redshift and the angular distance at this boundary, let us denote them by $t_b,z_b$ and $D_{Ab}$ respectively. We can now define the effective FLRW model as follows:
Let us integrate forward in time in the background FLRW model along a geodesic till we reach the center at $r=0$. The final value of time $t(r=0)=t_{\rm eff}$ defines our effective fictitious observer. For any object which lies outside the void, this fictitious FLRW observer measures the same angular distance  as the LTB observer. The time $t_{\rm eff}$ will be in general different from $t_0$.

Having defined our effective model and observer, let us return to the question of the appropriate background temperature. This is given by the following
\be
T_{\rm eff} = T_{\rm{obs}} \frac{1 + z_{b}}{1 + z_{Fb}}\ ,
\label{eff-obs}
\ee
where  $T_{\rm{obs}} = T_0 = 2.726$ K, and $z_{Fb}$ is the redshift of the boundary, $r=r_b$, for the fictitious FLRW observer.
In subsequent subsections we show analytically (for small $z_b$ and/or small $\de_0$) how to compute $ t_{\rm eff},T_{\rm eff}$ from $t_0,T_0$.

To summarize, the effective FLRW model that we need to use is the same as the background FLRW model, but the ``present'' epoch is no longer given by $t_0$, but is shifted by a small amount to $t_{\rm eff}$. Accordingly, in our CAMB code we need to rescale the Hubble parameter, and the abundances so as to reflect their values at $t_{\rm eff}$, and not at $t_0$:
\begin{align}
 H_{0, {\rm eff}} &= H(r_b,t_{\rm eff}), \\
  \Omega_{m, {\rm eff}} &=  \Omega_{m, {\rm out}}\left( \frac{H_{0,\mt{out}}}{H_{0, {\rm eff}}}\right)^2\left( \frac{  R(r_b,t_0) }{  R(r_b,t_{\rm eff}) }\right)^3,
\end{align}
where at  $r_b$ we can substitute $k(r)=k_b$. The radiation temperature is given by \ref{eff-obs} and we  keep the ratio $\Omega_b / \Omega_{\rm dm}$ constant throughout the universe.

As it turns out, in practice $\{ \Omega_{m, {\rm eff}},  H_{0, {\rm eff}} \} \simeq \{ \Omega_{m, {\rm out}},  H_{0, {\rm out}} \}$ for most of the parameter space.

\subsection{Relation between the parameters  \texorpdfstring{$\{H_{0,\mt{out}},\Om_{m,\mto},\de_0,z_b\}$}{\{Ho, Omega-m, delta0, zB\}} and the LTB parameters  \texorpdfstring{$\{k_b, k_0, t_{0},L\}$}{\{kb, k0, t0, L\}}}\label{sec:relations}
In this section we relate the parameters in our void model to the parameters that will be used in the Monte-Carlo runs to obtain fits for the different observational data.

Although the first set of quantities,$\{H_{0,\mt{out}},\Om_{m,\mto},\de_0,z_b\}$, are physically more transparent, technically it is easier to work with parameters characterizing the LTB metric. Therefore, first we will explain how the LTB parameters,  $\{k_b, k_0, t_{0},L\}$, can be obtained from the above ``input'' parameters.
Next, we will show how  the different  observational quantities can be computed from the LTB parameters.

As we have mentioned, we choose $a(L,t_0)=1$, \ie set the value of the scale factor today in the background to be one\footnote{This simply corresponds to making a specific choice for the arbitrary mass parameter $\bM$:
$$
\bM^2={3H^2_{0,\mt{out}}\over 8\pi(1+3k_b/4\pi)}
$$}. Setting $a=1$ in (\ref{central-density}) and (\ref{r-hubble}) it is now easy to calculate the matter and curvature abundance in the background FLRW cosmology:
\bea
\rho_{m,\mto}=M_0^4\Ra \Om_{m,\mto}&=&{1\over 1+{3k_b\over4\pi}}\label{omega-m}\,\, ,\\
\mx{ and }\Om_{k,\mto}=1-\Om_{m,\mto}&=&{{3k_b\over4\pi}\over 1+{3k_b\over4\pi}}\label{omega-k}\,\, .
\eea
Thus knowing $\Om_{m,\mto}$ or $\Om_{k,\mto}$, we obtain
\be
k_b=\frac{4}{3\pi}\frac{\Omega_k}{1-\Omega_k}.
\ee

Further, (\ref{scale-factor}) determines $u_b$ in terms of $k_b$:
\be
u_b=\cosh^{-1} \LF1+{3k_b\over 2\pi}\RF\,\, .
\label{ub-kb}
\ee
From (\ref{time}) we can now obtain $t_0$ from the background Hubble parameter:
\be
t_0={\sqrt{2}\over H_{0,\mt{out}}}\sqrt{1+{3k_b\over 4\pi}}\LT{(\sinh u_b-u_b)\over (\cosh u_b-1)^{3/2}}\RT
\label{H-out}\,\, .
\ee
It is reassuring to check that as $u_b,k_b\ra0$, we recover the EdS limit for the relation between the Hubble parameter and proper time.

We can also obtain $k_0$ from the central density contrast. Dividing  the central density,(\ref{central-density}), by the background density $M_0^4$, we obtain $\de_0$:
\be
\de_0\equiv {\rho_0-\rho_{\mto}\over \rho_{\mto}}=\LT{3k_0\over2\pi(\cosh u_0-1)}\RT^3-1=\LT{3k_b\over2\pi(\cosh u_0-1)}\RT^3\LF{\sinh u_0-u_0\over \sinh u_b-u_b}\RF^2-1
\label{f-de}
\, . \ee
where $u_0\equiv u(0,t_0)$ and where we have used (\ref{time}). Since we already know $k_b,u_b$, this implicitly determines $u_0$ in terms of $\de_0$, which in turn determines  $k_0$ via (\ref{time}):
\be
{\sinh u_0-u_0\over k_0^{3/2}}={\sinh u_b-u_b\over k_b^{3/2}}
\label{u0-k0}\,\, .
\ee
We have just described an algorithm to obtain $\{k_b, k_0,t_{0}\}$ from   $\{H_{0,\mt{out}},\Om_{m,\mto},\de_0\}$ . We can also check that using (\ref{ub-kb}), (\ref{H-out}), (\ref{omega-m}),  (\ref{f-de}) and (\ref{u0-k0}) we can solve the inverse problem as well.

Finally, we obtain $z(L) = z_B$ using a numerical iterative procedure: Since $z(r)$ is only obtained after explicit integration of the geodesic equations in a metric for a certain $L$, we employ the following procedure. We have to set some $L$, and define a new $\tilde L(L)$ as the radius $r$ for which $z(r)=z_B$. Blueshift may occur in the shell near the edge of the LTB metric, such that $r(z_B)$ is not unique. But since blueshift will not occur in the embedding FLRW metric, it is unambiguous to define $\tilde L(L)$ as the largest of all possible $r(z_B)$. Numerical inversion is then applied to find the right $L$ for which $\tilde L(L) = L$. In practice this means that the code performs about twenty integrations before the right $L$ is found as a function of $z_B$, and $L$ is usually accurate up to one part in $10^8$.  Since the inversion of $\tilde{L}(L)$ already involves solving the geodesic equations, we have the full solution $\{ t(r), z(r) \}$ as soon as $L(z_B)$ is obtained. This process determines the duration of the calculation, which is of the order of $10^{-2}$ s and has negligible impact on the duration of the MCMC runs.
\subsection{The implicit function  \texorpdfstring{$f$}{f} and its derivatives}
Although for fitting the various observations with the void models we resort to numerical computations, analytical approximations which are valid in different regimes serve as an extremely useful check on the accuracy of the numerical codes. In the LTB model (or in general in any inhomogeneous model) there are two important parameters, $L/r_{\mt{hor}}$ (where $r_{\mt{hor}}$ is the horizon radius)  determining the size of the inhomogeneity and $k(r)$, which is related to the spatial curvature/density contrast. In principle, we can perform a perturbative power-series expansion in either or both the variables. The leading order correction in  $L/r_{\mt hor}$ is the Newtonian approximation which is a good approximation for small voids. On the other hand, usual cosmological perturbation theory in small potentials is closely linked to an expansion in $k(r)$~\cite{Biswas:2007gi}. It turns out that, to obtain these perturbative expansions, it is convenient to introduce a function $f(k\ga^2\tau^2)$, where $\tau$ stands for the EdS ``conformal time'':
\be
\tau\equiv (\bM t)^{1\over 3}\mx{ and } \ga\equiv \LF{9\sqrt{2}\over \pi}\RF^{1/3} \,\, .
\ee
This function and its derivatives appear on numerous occasions  while solving for the photon trajectory as well as obtaining relations between different physical quantities. Let us therefore try to see why this function is useful  in a little more quantitative detail.

$f(x)$ is implicitly defined as
\be
1+f(x)\equiv {2(\cosh u-1)\over x} \qquad \mx{ where } \, \, x^{3/2}=6(\sinh u-u)\,\, .
\label{defnf}
\ee
The function is expandable in a power series of the form:
\be
f(x)=\sum_{n=0}^{\infty}R_{n}x^n, \qquad \mx{ with } R_2={1\over 20}\ , R_4={-3\over 2800} \ ,R_6={23\over 504000} \ ,R_8={-947\over 388080000} \ ,
\ee
and so on. What this suggests is that approximately unless $x>20$, it will be a very good approximation to replace $f$ with just the leading order terms. For future purposes we also  define its derivatives as
\be
f_n(x)\equiv {d^nf(x)\over dx^n}\,\, .
\ee

In terms of this function we can re-express the scale factor  as
\be
a(r,t)={\pi\over 3}\ga^2\tau^2\LT1+f(k(r)\tau^2\ga^2)\RT \qquad \,\, ,
\ee
To see how useful the function $f$ or rather its Taylor series truncations can be, let us compute $x_0=k_0\tau_0^2\ga^2$, which is the maximum value of $x$ that appears in the evolution. (Both $k(r)$ and $\tau$ are largest at the centre.) Let us consider first an EdS background with a central underdensity contrast $\de_0=-0.5$. In terms of $f$ we have a rather simple expression for the density contrast:
\be
\de_0=\LF{1+f_{\mto}\over 1+f_0}\RF^3-1\im  1+f_0=(1+f_{\mto})(1+\de_0)^{-1/3} \,\, ,
\ee
where
\be
f_0\equiv f(k_0\ga^2\tau_0^2)\mx{ and }f_{\mto}\equiv f(k_b\ga^2\tau_0^2) \,\, .
\ee
Now for EdS, $f_{\mto}=0$
$$\Ra f_0\approx 0.26\mx{ for }\de_0=-0.5\,\, .
$$
By inverting the function $f(x)$, we find that this corresponds to $x\approx 6$ and it is clear that a truncation of $f$ up to a couple of terms will give rise to an excellent approximation. We have verified this numerically.

Next let us consider the case when we have a curved background, $\Om_k=0.3$ open universe. Using (\ref{omega-k}) and (\ref{ub-kb}) we then have
\be
u_b=\cosh^{-1}\LF1+{2\Om_k\over 1-\Om_k}\RF\approx 1.23\mx{ for }\Om_k=0.3 \,\, .
\ee
From (\ref{defnf}) we can numerically determine that this corresponds to $x_b\approx 1.5$ and $1+f_{\mto}\approx 1.08$. This gives us $f_0\approx 0.2\Ra x_0\approx 9$ which is still sufficiently small. Similarly for $\Om_k=-0.3$ closed universe we also find $f_0\approx 0.2\Ra x_0\approx 9$. Numerically we find  that at least up till $|\Om_k|<0.3$, both for open and closed universes the truncation of $f$ up to 4th order terms is a very good approximation.
\subsection{Local Hubble parameter\label{subsec:localhub}}
The locally observed Hubble parameter is defined via
\be
H_0^{-1} \equiv \lim_{z \rightarrow 0} {d \over dz}d_L(z)= \lim_{z \rightarrow 0} {d \over dz}d_A(z)\ .
\ee
where the last equality holds because as $z\ra 0$ all the different distance measures coincide. From the geodesic equations~(\ref{t-radial},\ref{eq:z-radial}), we then have
\begin{align}
H_{\rm obs} &=  \LF{d\,d_A\over dz}\RF^{-1}_{z=0} =\left.\frac{dr}{d\,d_A}\frac{dz}{dr}\right|_{z=0} =\left. \frac{1}{R'(r(z),t(z))}\frac{(1+z)\dot R'(r(z),t(z))}{\sqrt{1 + 2 E(r)}}   \right|_{z=0}\nonumber\\
&=\left.\frac{\dot R'}{R'}  \right|_{z=0}=\left.\frac{\dot R}{R}  \right|_{z=0}\equiv \left.H(r,t)  \right|_{z=0},
\end{align}
where we used that for $z\rightarrow 0$ we have $k'(r)=0$ by definition, which locally corresponds exactly to an FLRW universe, for which indeed $\dot R' / R' = \dot R / R$.

For fitting the inhomogeneous LTB models we need to be able to compute the observed local Hubble parameter, $H_0$, in terms of the input parameters of the model. This can be done in the following way: First, as discussed in section \ref{sec:relations}, we can obtain the LTB parameters $k_0, k_b$ and $t_0$ from $\de_0,H_\mto$ and $\Om_k$. Using (\ref{f-de}) we can then obtain $u_0$.
Now, using (\ref{time}) and (\ref{H}) we find an expression for $H_0$ in terms of $u_0,t_0$:
\be
H_0={1\over t_0}{\sinh u_0(\sinh u_0-u_0)\over(\cosh u_0-1)^2}
\label{Hlocal}\,\, .
\ee
This gives us a prescription  to calculate the local Hubble parameter in our model.

Alternatively, we can rewrite the local Hubble parameter in terms of $f$'s:
\be
 H_{\mt{0}}={2\over 3t_0}\LF{1+f_{0}+\tau_0^2\ga^2k_0f_{1,0}\over 1+f_{0}}\RF={2\over 3t_0}\sqrt{4+\tau_0^2\ga^2k_0[1+f_0]\over 4[1+f_0]^3}\ ,
\ee
where the subscript $0$, simply means that $f$ and its derivatives are evaluated at $r=0$. This expression turns out to be more useful for numerical evaluations.
\subsection{Redshift \& The ``Background'' Temperature}
We have explained before why the background temperature that we have to use in the MCMC code to compute the CMB map which is fit to WMAP is different from the observed CMB temperature. We also provided a numerical prescription on how to compute the effective background temperature, but for a better understanding of the relation between $T_0$ and $T_\eff$, here we provide an analytical treatment of the same in the perturbative regimes. Also, since the difference between the observed and the effective temperatures is very small for most of the parameter space,  the analytical results served as a crucial check on our numerics.

Since we are able to treat the problem analytically in the regime of small $k(r)$ by perturbatively expanding the functions involving $f(k\ga^2\tau^2)$, we were able to compute the correction to the monopole temperature in this regime.  For technical simplicity, here we will focus on void models embedded in EdS backgrounds, please see  appendix \ref{app:t_ast_CMB} for more details. First, let us compare the difference in redshift between an LTB and FLRW observer that are both placed at the centre of the void, $r=0$,  and have the same observational epoch, $t=t_0$.

The leading order term in $\rb\equiv \bM r\sim r/r_H$ vanishes at the boundary~\cite{Biswas:2007gi}, and we have to look into $\cO(\rb^2)$ corrections.
The result for an outer EdS metric, derived in Appendix~\ref{approximations},  is given by
\be
1+z=(1+z_E)\exp\LT -{2\tau_1\over \tau_E}+ 2\al R_2\ga^2\tau_E\rb k(\rb)+2\al^2 R_2\ga^2k_1(\rb)\RT  \, ,
\ee
where $z_E$ is the redshift for the EdS metric \ie, with $k=0$. We have also defined the following quantities:
\bea
\tau_E&=&\tau_0-\al\rb\ \, ,\\
k_n(r)&\equiv& \int_0^{\rb} d\rb \rb^n k(r)\ ,\mx{ and} \, , \\
\tau_1&=&-\al\LT R_2 \ga^2(\tau_F^2\rb k+2\al\tau_0 k_1)-6k_2/5 \RT
\, .
\label{tau1}
\eea
Note that $\tau_E$ is simply the conformal time in the EdS metric and that $\tau_1$ is the time difference between the LTB metric and the FLRW metric, with the same observer's time $t_0$.

The other limit when we can reliably compute the redshift correction is for small $\rb$, and it is sufficient to keep only the $\cO(\rb^2)$ term. This corresponds to Next-to-Newtonian approximation. Again, the results are derived in Appendix \ref{app:ana_approx} and here we quote the result:
\bea
1+z&=&(1+z_E)\exp\left\{ {2\al^2\over \tau_0^2}\int_0d\rb\ \rb f(\tau_0^2\ga^2k)+{2\al\over \tau}[f(\tau^2\ga^2k)+\tau^2\ga^2kf_1(\tau^2\ga^2k)]\right.\nonumber\\
&+&{2\al^2\over\tau_0^{2}}\int_0d\rb\ \rb [3\tau_0^2\ga^2kf_1(\tau_0^2\ga^2k) +2(\tau_0^2\ga^2k)^2f_2(\tau_0^2\ga^2k)-f(\tau_0^2\ga^2k)]\nonumber\\
&\times&\left.[1+f(\tau_0^2\ga^2k)+\rb k' \ga^2\tau_0^2f_1(\tau_0^2\ga^2k)]\right\} \, .
\eea

Now, as explained in section \ref{sec:effective}, the effective FLRW observer must have a different  observer time, $t_\eff$, compared to the LTB observer time $t_0$ to get the angular scales to match.  In conformal time, the difference  is precisely given by
\be
\tau_\eff=\tau_0-\tau_1\,\, ,
\ee
since $\tau_1$ was the difference between the LTB and the FLRW patches in conformal time that a photon takes to reach $r=0$ from $r=r_b$.  This is given by (\ref{tau1}) in the small $k(r)$ approximation, while in the small $\bar{r}$ approximation we have (Eq.~(\ref{eq:tau1_appendix}) in Appendix~\ref{approximations})
\begin{align}
\tau_1=-\al\LF\rb f(\tau^2\ga^2k)
+2\al\ga^2\tau_0\int d\rb\ \rb k f_1(\tau_0^2\ga^2k)\LT1+  f(\tau_0^2\ga^2k)+\rb k'\ga^2\tau_0^2f_1(\tau_0^2\ga^2k)\RT\RF \,\, .
\end{align}

Now, in EdS we have $1+z\sim \tau^{-2/3}$.  Therefore,
\be
1+z_E=(1+z_\eff)\LF{\tau_\eff\over \tau_0}\RF^{2\over 3}\ ,
\ee
so that finally we can obtain $T_\eff$ from $T_0$ using (\ref{eff-obs}).

For small voids, $z\lesssim 0.2$, in EdS background the Next-to-Newtonian approximation is reliable, but for larger voids it is no longer valid. On the other hand, for curved FLRW backgrounds  and density contrasts that we will ultimately be interested in, linear perturbation theory cannot be trusted. Thus we have to resort to numerical computations in the general cases. However, the two approximate limits indeed served as useful checks to the numerical results. In Appendix~\ref{approximations} and figure \ref{fig:app_dt_t_2} we show the agreement of the numerical results with our approximations.


\subsection{The BAO scale}
\subsubsection{FLRW case}
Following \cite{Percival:2009xn} the BAO data consists in fitting two numbers, the ratio
\be
\te\equiv {L_S\over D_V(z)} \, ,
\label{theta}
\ee
for two values of $z=0.2$ and $z=0.35$. Here $L_S$ is the comoving sound horizon scale at recombination and $D_V$ is a combination of angular and radial distance defined as follows:
\be
D_V=[(1+z)^2D_A^2D_z]^{1/3} \, .
\ee
In FLRW the ``radial distance'' is  simply given by
\be
D_z\equiv {z\over H(z)} \, ,
\ee
where $H(z)$ is the Hubble rate expressed as a function of $z$:
\be
H(z)=H_0\sqrt{\Om_m(1+z)^3+\Om_k(1+z)^2+\Om_{\La}}\equiv H_0 h(z) \, .
\ee
To define the angular diameter distance, $D_A$, let us first define the ``comoving distance'' as
\be
r(z)=\int{dz\over H(z)}={1\over H_0}\int{dz\over h(z)} \, .
\ee
The angular diameter distance for an open FLRW universe is then given by
\be
D_A= {\sinh(H_0\sqrt{\Om_k}r(z))\over H_0\sqrt{\Om_k}(1+z)}={\sinh\LF\sqrt{\Om_k}\int{dz\over h(z)}\RF\over H_0\sqrt{\Om_k}(1+z)} \, .
\ee

The expression for $D_V$ then simplifies to
\be
D_V={1\over H_0}\LT{z\sinh^2\LF\sqrt{\Om_k}\int{dz\over h(z)}\RF\over \Om_kh(z)}\RT^{1/3} \, .
\ee

In particular a rather useful ratio to consider is
\be
{\cal R}\equiv {D_V(0.35)\over D_V(0.2)} \, .
\label{BAOratio}
\ee
This quantity only depends on $\{\Om_m,\Om_k,\Om_{\La}\}$ and therefore provides a rather useful bound. We can check that we cannot reproduce the measured ratio without $\La$. In fact even the conventional $\La$CDM does not fit the number very well. The measured value~\cite{Percival:2009xn} is about $1.812 \pm 0.060$ , while the $\La$CDM value is $1.67$ (with $\Omega_\La=0.75$). An open empty universe gives about 1.5.

Having defined how to compute the ``BAO observable'' in FLRW models, let us now describe how to proceed in the more general case of the  LTB metric, especially when the BAO data is inside the Void.

\subsubsection{Computing the model-independent observable,  \texorpdfstring{$(\Delta\theta^2\Delta z)^{1/3}$}{delta theta2 delta z}}

Most papers on BAO observations  quote  numbers for $D_V$, and this is what we defined above for FLRW models. However, what is actually measured by different collaborations (\cite{Eisenstein:2005su}, \cite{Percival:2009xn}), is a model-independent physical observable, the product $(\Delta \theta)^2 \Delta z$ at some redshift (0.2 or 0.35).  $\Delta\theta$ is an angle in the sky, and $\Delta z$ is an interval in redshift corresponding to the comoving $L_S$ evaluated at the redshift relevant for the BAO measurements. Only \cite{Gaztanaga:2008de} gives a value of $\Delta z$ alone and \cite{GarciaBellido:2008yq} points out that in the future $\Delta z$ (the radial BAO) may be more constraining for void scenario, but we do not discuss that here.
We are first going to present an algorithm to obtain the quantity $(\Delta \theta)^2 \Delta z$ in a general cosmological model.

Let us start with the  acoustic scale $L_S$. This is a scale for physical processes which happen at $z\sim 1100$. It is usually given as a comoving length and its value is about $100 \, {\rm Mpc}/h$.

Given $L_S$ we can reconstruct the value for $(\Delta \theta)^2 \Delta z$ at some redshift $z_{\rm BAO}$. Let us do it first in an FLRW  model.
This goes as follows: The physical scale at the epoch of $z=z_{\rm BAO}$ is  $L_S^{\rm phys}=1/(1+z_{\rm BAO}) L_S$ so that the FLRW angular distance is
\be
D_A^{\rm FLRW}=\frac{L_S^{\rm phys}}{\Delta \theta} = \frac{L_S}{\Delta \theta (1+z_{\rm BAO})} \, ,
\ee
and in terms of the expansion rate $H^{\rm FLRW}$ we have
\be
L_S=\frac{\Delta z}{H^{\rm FLRW}(z_{\rm BAO})}  \label{deltazFLRW} \, .
\ee
Therefore
\be
(\Delta\theta^2 \Delta z)^{1/3}=  r_{S} \left[\frac{H^{\rm FLRW} (z_{\rm BAO}) }{(1+z_{\rm BAO})^2 D_A^{{\rm FLRW} \, 2}(z_{\rm BAO})}\right]^{1/3} \, , \label{thetazetaFLRW}
\ee
which agrees with~\cite{SarkarHunt}.

One can now check that this quantity reduces to
\be
(\Delta\theta^2 \Delta z)^{1/3}= \frac{z^{1/3} L_S}{D^{\rm FLRW}_V(z)} \, .
\ee
for FLRW universe.

So, now if we consider~\cite{Percival:2009xn}, the measured numbers are:
$L_S/D_V(0.2)=0.1905\pm0.0061$, $L_S/D_V(0.35)=0.1097\pm0.0036$, which is all we need to have the two datapoints $(\Delta\theta^2 \Delta z)^{1/3}(z=0.2)$ and $(\Delta\theta^2 \Delta z)^{1/3} (z=0.35)$.  We actually use a $2\times 2$ covariance matrix, as in~\cite{Percival:2009xn} in the MCMC code.%
\footnote{Another way of expressing this, for instance at $z=0.35$, is to consider the measurement  $R(0.35)=\frac{D_V(0.35)}{D_M(1089)}=0.0979\pm0.0036$ \cite{Eisenstein:2005su}, where $D_M$ is the comoving angular diameter distance.
From WMAP we know that the first acoustic peak is detected at an angle of $0.59^\circ$, which is $0.01$ radians.
This means that:
$L_S/D_M(1089)\simeq 0.01 \, ,$
in a model-independent way. Therefore:
$\frac{L_S}{D_V(z=0.35)}\simeq\frac{1}{100 R}$ 
and so we can compute $(\Delta\theta^2 \Delta z)^{1/3}(z=0.35)$.}

Having explained how we can obtain the model independent $(\Delta\theta^2 \Delta z)^{1/3}$ values from $\te$'s, our next task is to determine a way to compute this quantity for the LTB metric.

\subsubsection{\texorpdfstring{$(\Delta\theta^2\Delta z)^{1/3}$}{delta theta2 delta z} in LTB}

We want to find what our model {\it predicts} for $(\Delta \theta)^2 \Delta z$ at some redshift $z_{\rm BAO}$.
So, first of all, given a set of cosmological parameters for the outer FLRW region, we can compute, using  the usual fitting formulae from Ref.~\cite{Eisenstein:1997ik}, the sound horizon in our model $L^{\rm LTB, rec}_S$ at recombination time, and the redshift to recombination, $z_{\rm{rec}}$.
What we really want is the sound horizon at the radial coordinate $r_{\rm BAO}$ which corresponds to the BAO observations ($z=0.35$ and $z=0.2$).

We assume here that $z_{\rm{rec}}(r)$ and $L_S^{\rm LTB, rec}(r)$ are not uniform inside the void, but at each radius they are approximately determined by the usual fitting formulae from Ref.~\cite{Eisenstein:1997ik}, except that we have to use $\Omega_m(r)$, $\Omega_k(r)$ and $\Omega_b(r)$ determined by the equations~(\ref{eq:om_mat_r},  \ref{eq:om_k_r}). This additional $r$-dependence is a crucial difference between the LTB and the FLRW models, and it is also the main reason why the LTB models can give a better fit to the BAO as compared to their FLRW cousins.  Moving on, the radius $r$ and time $t$ are determined as a function of $z_{\rm BAO}$ by the geodesic equation describing the geodesic between the observer at $(r=0, t_0)$ and the BAO scale at $r(z_{\rm BAO}), t(z_{\rm BAO})$.

Then, we just need to see how the scale $L^{\rm LTB, rec}_S$ is stretched at the epoch $z_{\rm BAO}$ and position $r_{\rm BAO}$, and how this translates to a prediction for $\Delta \theta^2 \Delta z$.
In the transverse direction the physical size of the scale at the BAO time is:
\be
L_S^{\rm transverse}= \frac{R_{\rm BAO}}{R(r(z_{\rm BAO}),t(z_{\rm{rec}}))} L^{\rm LTB, rec}_S \, ,
\ee
while in the radial direction it is
\be
L_S^{\rm radial}= \frac{R'_{\rm BAO}}{R'(r(z_{\rm BAO}),t(z_{\rm{rec}}))} L^{\rm LTB, rec}_S \, ,
\ee
where we used the short notation $R(r(z_{\rm BAO}),t(z_{\rm BAO}))\equiv R_{\rm BAO}$ and  $R'(r(z_{\rm BAO}),t(z_{\rm BAO}))\equiv R'_{\rm BAO}$.

Finally from these two numbers we have to reconstruct the predictions for $\Delta \theta$ and $\Delta z$.
This is as follows:
\be
\Delta \theta = \frac{L_S^{\rm transverse}}{D_A^{\rm LTB}} =  \frac{1}{R(r(z_{\rm BAO}),t(z_{\rm{rec}}))} L^{\rm{LTB, rec}}_S \, ,
\ee
where we have used the fact that:  $D_A^{\rm LTB}=R_{\rm BAO}$. For the radial direction (in the approximation $E(r)\ll 1$) using~(\ref{eq:z-radial}) leads to:
\be
\Delta z = (1+z_{\rm BAO}) \dot{R}_{\rm BAO}' \Delta r_{\rm radial} \, ,
\ee
where $\Delta r_{\rm radial}$ is the coordinate distance which corresponds to the physical length $L_S^{\rm radial}$ at $(r=r_{\rm BAO}, t=t_{\rm BAO})$.
The two are related by:
\be
 \Delta r_{\rm radial}= \frac{L_S^{\rm radial}(r=r_{\rm BAO}, t=t_{\rm BAO})}{ R'_{\rm BAO}}= \frac{1}{R'(r(z_{\rm BAO}),t(z_{\rm{rec}}))} L^{\rm{LTB, rec}}_S \, .
\ee
Putting everything together we get
\be
(\Delta \theta^2 \Delta z)^{1/3} = \left[ (1+z_{\rm BAO}) \dot{R}_{\rm BAO}' \frac{1}{R'(r(z_{\rm BAO}),t(z_{\rm{rec}})) R^2(r(z_{\rm BAO}),t(z_{\rm{rec}}))}\right]^{1/3}  \frac{L^{\rm LTB}_S}{(1+z_{\rm{rec}})}\label{eq:z_bao_3d} \, .
\ee

In the above equations we need to specify what $t_{\rm{rec}}$ is, {\it i.e.} at what coordinate time recombination happens at a given position $r$.
We define, for equation (\ref{eq:z_bao_3d}), $t_{\rm{rec}}$ as the time for which the volume element in the position $r$ is diluted by a factor $(1+z_{\rm{rec}})/(1+z_{\rm BAO})$:
\be
1+ z_{\rm{rec}} = (1 + z_{\rm BAO}) \left(\frac{R'_{\rm BAO} R_{\rm BAO}^2 }{ R'(t_{\rm{rec}},r_{\rm BAO}) R^2(t_{\rm{rec}},r_{\rm BAO})}\right) ^{\frac 1 3} \, ,
\ee

These are the equations we use in our code. More precisely, we define the following quantity in the MCMC code:
\be
Q\equiv(\Delta\theta^2 \Delta z/z)^{1/3} \, ,
\ee
and we compare with the covariance matrix in \cite{Percival:2009xn}.

Finally, let us make a few remarks about the location of the BAO observations. Let us consider a ``small'' Void ({\it i.e.} if it extends only up to $z=0.15$ at most):
in this case the BAO data are all in the FLRW region, therefore we can simply use the FLRW  limit of the above expressions.

Using the fact that $D_A=r a$ we get in this case the same expression as ~(\ref{thetazetaFLRW}).

In the more general case in which the Void is larger we take the value of $Q(z)$ in the full LTB model at the average redshift ($z=0.2$ or $z=0.35$).
This should be a good approximation in  the limit of very large voids, that is when the profile does not change too rapidly in the range  $0.2\leq z \leq 0.4$.
A more refined treatment could be done in principle for more general situations, performing some kind of averaging of this quantity in this redshift range, weighted for instance by the number density of observed objects:
\be
Q_{\mt average}\equiv \frac{\int Q(z) n(z) dz}{\int n(z) dz} \, ,
\ee
where $n(z)$ is a function which describes the number density of observed objects at a given redshift.

\section{Numerical Analysis}\label{sec:num_analysis}
\subsection{Description of our numerical code}
In order to prevent the mathematical limitations of analytical approximations from biasing our results, we have written a code that performs a full numerical integration of the geodesics in LTB space-times. Our program computes the redshift to any distance.  The geodesic equations and the analytical expressions for all the necessary derivatives of the background functions $R(r,t)$ and $S(r,t)$ have been given in section \ref{sec:theory_intro}.
Analytical approximations have been used to test the results of our code in all possible limits, as briefly discussed in the end of  appendix~\ref{app:ana_approx}. In particular, we have analytical control in two regimes: the approximation for small $k(r)$ (which corresponds to first order in the gravitational potential, as in the usual perturbation theory around an FLRW metric~\cite{Biswas:2007gi}) and the approximation for  $L\ll r_{\mt{hor}}$, which corresponds to a Newtonian expansion.
Our code reproduces all analytical results up to high accuracy, and calculates all possible quantities to $\mathcal{O}\left(10^{-2}\right)$ accuracy.

As input parameters that specify the cosmological model, we use the set\\ $\left\{\Omega_{m, {\rm out}}, \Omega_{b, {\rm out}}, H_{0, {\rm out}}, z_b, \delta_0,n_{\rm s},\al_s,A_{\rm s},\tau,b^2,Q_{nl},A_{\rm SZ}\right\}$, as defined in section \ref{sec:parameters}. For comparison the $\Lambda$CDM model does not have the $z_b$ and $\delta_0$ parameters, but it has an additional parameter $\Omega_{\Lambda}$, so finally it has the set\\ $\left\{\Omega_{m}, \Omega_{b}, \Omega_{\Lambda, }, H_{0, {\rm out}}, n_{\rm s},\al_s,A_{\rm s},\tau,b^2,Q_{nl},A_{\rm SZ}\right\}$, with one parameter less than the Void.
We will also consider in section \ref{sec:exotic} more complicated Void profiles with one or two additional parameters, which will lead to significantly better fits.

In solving the geodesic equations, we can choose different options as the integration variable, such as the affine parameter $\lambda$,  the time $t$, or the coordinate $r$. We chose $r$ as our integration variable since the metric is analytically defined by $k(r)$, and  that there is no numerical ambiguity in determining when the photon enters the LTB patch.
Note also that one should not choose $z$ as a time parameter along the geodesic because in a significant portion of parameter space the shell of the void is such that the photon experiences a blueshift;
the over-dense shell contracts faster than the Hubble expansion. This also implies an additional complication: for redshifts around the radius of the void, the relation between distance and redshift is not unique and typically three different distances correspond to the same redshift. The Supernovae observations give us a luminosity distance as a function of redshift. Hence, the theory may sometimes predict three different luminosity distances corresponding to the same redshift. In Appendix~\ref{app:snchi2} we construct the likelihood for such a situation, and explain why such a likelihood forces us to numerically marginalize over $H_0$, as opposed to doing it analytically, for each fit to the supernovae. Such a marginalization is customary, since the Hubble parameter acts as an unknown normalization for the luminosity distance - redshift curve from supernovae.

\subsection{Interfacing with CAMB and COSMOMC}
By integrating  the geodesic equations we can straightforwardly determine $\left\{\Omega_{m, {\rm in}}, \Omega_{b, {\rm in}}, H_{0, {\rm in}}\right\}$ which are the effective parameters inside the Void, while the parameters  $\left\{\Omega_{m, {\rm{eff}}}, \Omega_{b, {\rm{eff}}}, H_{0, {\rm{eff}}}, T_{{\rm{eff}}}\right\}$, are used to fit the CMB, as explained in the previous sections. Finally, we numerically compute  $\left\{\Omega_b(r_{\rm BAO}), \Omega_m(r_{\rm BAO}), H(r_{\rm BAO})\right\}$ necessary for fitting the BAO observations. We implemented our numerical code as a module in {\sc cosmomc}, which we will publicly release at { {\href{http://web.physik.rwth-aachen.de/download/valkenburg/}{http://web.physik.rwth-aachen.de/download/valkenburg/}}}. For any set of parameters, {\sc cosmomc} first calls our module, which performs the integration and returns  $\left\{\Omega_{m, {\rm{eff}}}, \Omega_{b, {\rm{eff}}}, H_{0, {\rm{eff}}}, T_{{\rm CMB, eff}}\right\}$. Then {\sc camb} is run with these effective parameters. If desired, also $\left\{\Omega_{m, {\rm in}}, \Omega_{b, {\rm in}}, H_{0, {\rm in}}\right\}$ are returned by the module, such that {\sc camb} can be called a second time, this time for a calculation of the Large Scale Structure power spectrum inside the void. This way we perform a full Monte Carlo Markov Chain (MCMC) analysis in order to obtain Bayesian estimates of parameter ranges, and simultaneously the best fit $\chi^2$ that can be achieved for a certain model.

\subsection{Datasets}
For the CMB we use the $TT$ and $TE$ correlation spectra from the WMAP 7-year data release~\cite{Komatsu:2010fb}. For the supernovae we use the SDSS-II 1st-year SN compilation, where we choose the {\sc mlcs2k2}  lightcurve fitter~\cite{Kessler:2009ys}. For comparison we also include a few runs with the {\sc salt-ii} lightcurve fitter. The former  is supposed to be more cosmology independent and more conservative, with larger error bars. We use  galaxy power spectrum from the SDSS main sample~\cite{Tegmark:2003uf} as large scale structure data, although we also include some runs fitting to the DR4 LRG power spectrum from SDSS~\cite{Tegmark:2006az}. We use the BAO data as presented in Ref.~\cite{Percival:2009xn}. For the local value of the Hubble constant we use the value   $62.3 \pm 6.3$ km s$^{-1}$ Mpc$^{-1}$ coming from HST measurements as quoted in Ref.~\cite{Sandage:2006cv} (labeled HST$_{62\pm6}$), although later  we also perform runs with  higher values of $H_0$ viz., $72 \pm 8$ km s$^{-1}$ Mpc$^{-1}$~\cite{Freedman:2000cf} (labeled HST$_{72\pm8}$) and $74.2 \pm 3.6$ km s$^{-1}$ Mpc$^{-1}$~\cite{Riess:2009pu}\footnote{In this case the quoted value is an effective value, since what is actually fit here is the value of $1/d_A(z=0.04)$, with $d_A$ the angular diameter distance.} (labeled HST$_{0,74\pm3}$).

\subsection{Priors}\label{subsec:priors}
We use flat priors on the parameters displayed in Table~\ref{tab:priors}. Some of the parameters are only relevant when fitting to the CMB or LSS, namely $\{A_{\rm s}, n_{\rm s}, \alpha_{\rm s}, \tau \}$, hence these parameters are fixed to a central value when these datasets are not included in the fits. We still allow $H_{0,\mt{out}}$ to vary even when fitting to SN only, in order to explicitly test that our numerical marginalization over $H_{{\rm out}}$ works as expected. The parameters $H_{\rm{out}}$, $\Omega_{b, \rm{out}} h^2$ and $\Omega_{dm, \rm{out}} h^2$ are defined in the embedding FLRW universe at time $t_0$ as explained in section~\ref{sec:theory_detail}. In the void scenario, $\Omega_{k, {\rm out}}\equiv 1 - \Omega_{\rm m, out}$, and in the $\Lambda$CDM scenario $\Omega_{\Lambda}\equiv 1 - \Omega_{\rm m} - \Omega_{k}$.

\subsubsection{Coordinate divergences}
\label{divergences}
For certain extreme combinations of $\delta_0$ and $z_B$, it is very well possible that the metric functions $S(r,t)$ and $R(r,t)$ go to zero for coordinates in the massive shell surrounding the void. This happens when shells at different radii cross, such that the choice of coordinates comoving with the matter is no longer valid. Since the region in parameter space for which such situations occur is sufficiently far from the region in which we find good fits, we simply reject models for which shell crossing occurs. This implies however, that our prior on $z_B$ and $\delta_0$ is not exactly as stated in Table~\ref{tab:priors}, but is bounded by some a priori unknown and non-trivial boundary, depending on all cosmological parameters. Therefore we still do have a flat prior on $z_B$ and $\delta_0$, but we do not know exactly on which volume we take the flat prior. As a consequence it becomes rather complicated to calculate the Bayesian Evidence with a code such as {\sc multinest}~\cite{Feroz:2008xx}, which is not compatible with points that are rejected by a prior during an assessment of the posterior likelihood. In other words, a Bayesian evidence calculation with an a priori unknown prior volume needs further investigation, which we leave to future work.

\TABLE{
\begin{tabular}{ll|r|r}
\phantom{{\bf All scenarios:}}&Parameter & Lower bound & Upper bound\\
\hline
\hline
\multicolumn{4}{l}{All scenarios:}\\
\hline
&$\Omega_{b, \rm{out}} h^2$ & 0.01 & 0.03\\
&$\Omega_{\rm dm, out} h^2$ & 0.01 & 0.2\\
&$H_{0, \rm{out}}$ [km s$^{-1}$ Mpc$^{-1}$] & 20 & 100\\
&$\tau$ & 0 & 0.3\\
&$\log \left[ 10^{10} A_{\rm s} \right]$ & 2.5 & 3.5\\
&$n_{\rm s}$ & 0.75 & 1.0\\
&$\alpha_{\rm s}$ & -0.2 & 0.1\\
\hline
\multicolumn{4}{l}{Void:}\\
\hline
&$z_{\rm B}$ & 0 & 3\\
&$\delta_0$ & -1 & 0\\
\hline
\multicolumn{4}{l}{$\Lambda$CDM:}\\
\hline
&$\Omega_k$ & -0.1 & 0.1
\end{tabular}
\caption{Free parameters on which we take a flat prior, and their priors: $h\equiv H_{0, \rm{out}}/100$, the optical depth to the last scattering surface $\tau$, the amplitude of the primordial spectrum of scalar perturbations $A_{\rm s}$, its tilt $n_{\rm s}$ and the running of its tilt $\alpha_{\rm s}$. The parameters $H_{0, \rm{out}}$, $\Omega_{b, \rm{out}} h^2$ and $\Omega_{dm, \rm{out}} h^2$ are defined in the embedding FLRW universe at time $t_0$ as explained in section \ref{sec:theory_detail}. In the void scenario, $\Omega_{k, {\rm out}}\equiv 1 - \Omega_{\rm m, out}$, and in the $\Lambda$CDM scenario $\Omega_{\Lambda}\equiv 1 - \Omega_{\rm m} - \Omega_{k}$. We chose our pivot scale for the primordial spectrum at $k_{\rm pivot}=0.05$ Mpc/h}\label{tab:priors}
}

\section{Results with simplest Profiles} \label{risult}
In this section we show results for the simplest Void profile given by Eq.~\eqref{profile}, which has an underdensity and a compensating shell and then is matched exactly to FLRW.
In the next section we discuss more elaborated density profiles, which will allow to find better fits.

We will fit the data with  voids    embedded  either  (1) in an EdS background (EdS Void), or (2)  in  a universe containing dust and curvature (Curved Void). In Table~\ref{tab:chi2} we show, for the best fit parameters of each model, a list of $\chi^2_{\rm eff}\equiv -2 \ln \mathcal{L}$, with $\mathcal{L}$ denoting the likelihood, and differences $\Delta \chi_{\rm eff}^2$ with respect to the best-fit  $\Lambda$CDM model. Our $\Lambda$CDM model is actually a little different from the ``Standard'' $\Lambda$CDM model: since we want to compare our models with $\Lambda$CDM, to make the comparisons meaningful we wanted the models to resemble each other  as closely  as possible. Hence we added two additional parameters, the running of the tilt, $\al_S$, and curvature, $\Om_k$, to the Standard  $\Lambda$CDM model.

We chose to perform a full MCMC analysis for each of the displayed combination of datasets, in order to show in a most explicit and exact way what the effect of including each dataset is on the performance of a given model. Let us go through the table from top to bottom, addressing the effects of all datasets.

\TABLE{
\begin{tabular}{l|r|r|r|r|}
Datasets				&\#dof&$\Lambda$CDM + $\Omega_k$&EdS Void& Curved Void \\
\hline
\phantom{CMB+BAO+}SN                                                                		& 288 	& (236.5)		& +2.0  	& -2.4	 \\
{\phantom {CMB+BAO+}}SN\phantom{+SDSS}+HST$_{62\pm6}$	& 289 	& (236.5)		& +2.4 	& -2.4	\\
\phantom{CMB+}BAO+SN                                                         		& 290 	& (239.0)	 	& +3.7 	& -2.7	  \\
{\phantom {CMB+}}BAO+SN\phantom{+SDSS}+HST$_{62\pm6}$    & 291 	& (239.1)		& +5.3 	& -2.8	  \\
CMB                                                               						& 3115 	& (3371.2)	& +35.0	& +4.0 	 \\
CMB+BAO                                               						& 3117 	& (3372.5)	& +39.1	& +7.1 	  \\
CMB+BAO\phantom{+SN+SDSS}+HST$_{62\pm6}$			& 3118 	& (3380.3)    	& +32.2	& +3.9	  \\
CMB{\phantom {+BAO}}+SN                     						& 3403 	& (3608.9)    	& +37.8	& +6.8 	\\
CMB{\phantom {+BAO}}+SN\phantom{+SDSS}+HST$_{62\pm6}$	& 3404 	& (3608.9)    	& +38.6	& +12.2	 \\
CMB+BAO+SN                                            						& 3405 	& (3613.6)    	& +40.2	& +6.2 	\\
CMB+BAO+SN\phantom{+SDSS}+HST$_{62\pm6}$			& 3406 	& (3614.1)    	& +40.5	& +10.2	  \\
CMB{\phantom {+BAO}}+SN+SDSS           					& 3423 	& (3627.9)    	& +49.0	& +19.2	  \\
CMB{\phantom {+BAO}}+SN+SDSS+HST$_{62\pm6}$			& 3424 	& (3628.4)    	& +50.4 	& +28.5	  \\
\hline
CMB{\phantom {+BAO+SN}}+LRG+HST$_{62\pm6}$                       	& 3425 	& (3638.5)     	& +30.0	& +7.2	 \\
CMB{\phantom {+BAO}}+SN+LRG+HST$_{62\pm6}$                       	& 3426 	& (3639.1)     	& +30.3	& +12.7	
\end{tabular}
\caption{$\Delta \chi^2$ for different models against standard $\Lambda$CDM allowing for a nonzero curvature. The values in parentheses are the actual $-2 \ln \mathcal{L} \equiv \chi^2_{\rm eff}$, in the correct normalization such that $\lim_{\chi^2_{\rm eff} \rightarrow 0} \mathcal{L} = 1$. For complicated likelihood estimations such as for the CMB, the quantity $-2 \ln \mathcal{L}$ is not $\chi^2$-distributed, but we still denote it by $ \chi^2_{\rm eff}$ for an intuitive interpretation.}\label{tab:chi2}
}

\subsection{Fitting the Supernovae}
One of the difficulties inhomogeneous models face is that, according to the standard analysis of the growth of perturbations starting from the primordial inflationary spectrum, the existence of large scale inhomogeneous structures with $\cO(1)$ density contrasts is exponentially suppressed (see for instance~\cite{SarkarHunt}). This suppression increases as the size of the void increases. In~\cite{ABNV} a Minimal Void (MV) model was proposed which could fit the SN and CMB data available at that time. The  Minimal Void, however, failed to fit the newer datasets~\cite{Kessler:2009ys}. In our present study we find that if we include spatial curvature, the best fit value for the size of the void is still rather small, with $ z_B = 0.221 $ with a corresponding density contrast of $ \delta_0 = -0.322 $ and a global spatial curvature $ \Omega_{k, {\rm out}} = 0.693 $. The Curved Void models give a consistent fit to the SN data, having  $\De\chi^2\sim -2$ compared to  $\La$CDM. In comparison, the EdS based void model gives a worse fit than the $\La$CDM, $\De\chi^2\sim +2$. These values indicate that the SN at higher redshifts can be well fit by a very open FLRW universe. The void itself plays a role in fitting the low redshift part of the luminosity distance diagram.

On the other hand, the marginalized posterior likelihoods for these parameters are $ z_B = 1.48^{+1.45}_{-1.40} $, $\delta_0 = -0.516^{+0.448}_{-0.325} $ and $ \Omega_{k, {\rm out}} = 0.340^{+0.558}_{-1.180} $. The discrepancy between the best fit values and the central marginalized likelihood values, tells us that $z_B$ is bound from above by our prior, not by the data. If the void is very large, say $z_B \sim 3$, which is well beyond the highest redshift of the SNe, the SN are practically fit by the effective open FLRW universe inside the void.

We also verified that by including only HST along with SN, the goodness of fit does not change at all, indicating that the SN alone do not constrain $H_0$, and the numerical marginalization is implemented correctly.

\subsection{Adding the BAO}
In~\cite{ABNV} it was noted that inhomogeneous models based on small Voids cannot be consistent with the  BAO peaks observed at $z\sim 0.2$ and at $z\sim 0.35$. We find here that EdS based void models, see table \ref{tab:chi2},  have a rather poor fit of the BAO peak position, and this effect gets even more pronounced when the HST$_{62\pm6}$ data are included: for a combined SN+BAO+HST$_{62\pm6}$ best-fit the EdS Void model gives a $\De\chi^2\sim +5$. However, this situation changes quite dramatically once we allow the background geometry to be curved. In fact, the best-fit Curved Void model has $\De\chi^2\sim -3$ with respect to $\La$CDM.

While this is encouraging news,
we also have to be cautious before drawing conclusions: the improvement in the fit is mainly due to the fact that for a Void in EdS a large $z_b$ is needed to fit the SN, while in our model the freedom in $\Omega_k$ allows a good fit of the supernova even with a lower value of $z_b$ in conjunction with having an open outer FLRW universe (in fact we saw above that any value for $z_b$ is allowed). This allows 
the MCMC code to find a good fit to the BAO with $z_b\sim 0.3$. Since typically around the overdensity the distance scales are very different from the ones in the background or in the Void, the existence of an overdense shell at $z\sim 0.3$ provides enough freedom to find a good fit to BAO. This is consistent with our general understanding of the problem of fitting BAO without dark energy in FLRW models. Unless, the FLRW universes around $z\sim 0.2$ and $z\sim 0.35$ are significantly different from each other we cannot hope to get a good fit of the ratio of the $D_V$'s (\ref{BAOratio}) at these two redshifts. Fortunately, LTB models has enough latitude to reconcile the two measurements.

However,  it is clear that the good fit that we obtained is a somewhat  fine-tuned result: first, because it is quite dependent on the shape of the profile of the overdense shell at the BAO location, second because if the BAO scale changes rapidly it is no longer a good approximation to take the distance scale at one single redshift point, which is what we do. More importantly, as we will discuss below, when we add the CMB data,  the best-fit model will no longer correspond to  $z_b\sim 0.3$, essentially  because the open outer FLRW model does not provide  a good fit to the CMB. Thus we will not reproduce this very good fit of the BAO. As a general lesson however,  we see that we can  fit the BAO scale well, and that this fit becomes especially good if we allow the LTB metric to have some kind of feature at the redshift of the BAO. We also saw that $\Om_k$ can be a crucial parameter in void models, and this will become even more apparent in the next subsection.

\subsection{Adding the CMB}
Apart from the $\La$CDM model, also an FLRW closed universe (without dark energy) does fit the CMB data well, albeit with a very small Hubble parameter. LTB models offer a unique opportunity to vary curvature of the universe in such a way that the last scattering surface be situated in a closed FLRW background, while the local universe mimics an open universe. This also increases the value of the local expansion rate and helps us fit the supernova data in a manner similar to the EdS based inhomogeneous models.
Note also that a Minimal Void does not lead to a good fit, because an open curvature in the outer region would be needed to fit the Supernova which lie outside the Void, while a closed curvature is required to fit the CMB.
We find that $\Om_k$ plays a decisive role while fitting the combined data sets of SN, CMB and BAO. As we can see from Table~\ref{tab:chi2}, the $\Om_k=0$  best fit model has a $\Delta\chi^2\sim 40$ as compared with $\La$CDM, while our Curved Void model gives  $\Delta\chi^2\sim 6$. The combined data set contains 3405 data points, and thus if we consider the $\chi^2$ per degree of freedom, the difference between $\La$CDM and the Curved Void model is not large.

We should also point out that while fitting the CMB, the Void plays an important role by affecting the monopole of the CMB temperature: in fact, as explained in section \ref{sec:BAO}, we describe the CMB via an effective FLRW model with a correction on the monopole temperature.
This effect was neglected in~\cite{ABNV}, because the Void considered there was much smaller and the correction to the monopole was tiny, since it goes as $(L/r_{hor})^2$, but when we consider larger Voids this effect is not negligible anymore (see \cite{Zibin:2008vk}). This effect is such that the photons in the effective FLRW have a slightly higher temperature than what is observed in the Void, as shown in Table~\ref{tab:CBHS}. This means that effectively the surface of last scattering is closer  than in a $\Lambda$CDM universe with the same parameters otherwise. A same size of the sound horizon at decoupling would hence be observed at a larger angle for an observer in the Void. Note that this correction to the monopole temperature goes in the direction of making the CMB fit better ({\it i.e.} makes the distance to last scattering larger) but note also that this happens only if the Void boundary is roughly $z\lesssim 2$. For larger Voids, in fact,  the correction changes sign, and therefore too large Voids ar disfavoured in the MCMC analysis, as can be seen from Figure~\ref{fig:marge_prim_CBHS}. For a discussion of the sign of the correction see subsection \ref{asymp}.

To summarize, the freedom in $\Omega_k$, $\delta_0$, $z_{\rm B}$ and $H_{0, {\rm out}}$ allows the large Curved Void model to fit the CMB almost as well as $\Lambda$CDM, while the EdS Voids cannot be made consistent with SN, CMB, and BAO all at once. This is one of the main positive results of our study. The overall trend, when following the Table further down, is that $\Lambda$CDM does a better job everywhere when the CMB is included, but the Curved Void follows closely.

\subsection{Adding HST, Parameter constraints from CMB+BAO+HST+SN}
Let us now turn our attention to the HST  measurement, and in particular see whether our model can be consistent with the combined data sets of  CMB+BAO+HST+SN. In this subsection we discuss the HST value  $62.3 \pm 6.3$ km s$^{-1}$ Mpc$^{-1}$ from Ref.~\cite{Sandage:2006cv}. The higher values  of $72 \pm 8$ km s$^{-1}$ Mpc$^{-1}$ from \cite{Freedman:2000cf} and $74.2 \pm 3.6$ km s$^{-1}$ Mpc$^{-1}$ from~\cite{Riess:2009pu} will be discussed later in section \ref{sec:exotic} and will be clearly more difficult to fit.  This is actually the largest choice of datasets where we can trust the various assumptions and approximations we made in our analysis. For instance, we have correctly computed the distance to the last scattering surface and the energy content of the universe at last scattering, which makes our fit to the CMB reliable. The distance measure to the SN is calculated as good as exactly. The value of the Hubble parameter is measured at such  low redshifts in HST,  that the model prediction at $r=0$ is a good estimate. And finally the BAO is a relatively model independent measurement of a combination of the expansion of space  in angular and radial (redshift) directions. The only assumption that went into our calculation and prediction of the BAO inside the void, is that the local universe at the drag epoch is correctly modeled by an FLRW-metric. Thus we can be quite confident about our analysis of the combined data set consisting of CMB, BAO, SN and HST$_{62\pm6}$.  We will discuss in the next subsection why LSS data, possibly the only other important observational set,  cannot accurately be analyzed within our theoretical framework of the LTB universe, although we will try to perform an estimate.

\newlength{\totfigwidth}
\setlength{\totfigwidth}{1.9\textwidth}
\TABLE{ 
\resizebox{\textwidth}{!}{
{\huge{
\begin{tabular*}{\totfigwidth}{llr}
\begin{tabular}{@{\extracolsep{\fill}}l|rrr|r}
Model&CMB&BAO&SN&total $\chi^2$\\
\hline
$\Lambda$CDM		& 3371.5 & 2.3 & 239.7 &     3613.5\\
Curved Void  	& 3377.1 & 3.6 & 238.9 &     3619.6\\
EdS Void	& 3408.8 & 4.3 & 240.7 &     3653.8\\
\end{tabular}
&{\phantom{a}}&
\begin{tabular}{@{\extracolsep{\fill}}l|rrrr|r}
Model&CMB&BAO&SN&HST$_{62\pm6}$&total $\chi^2$\\
\hline
$\Lambda$CDM & 3371.1 & 3.1 & 239.5 & 0.4 &     3614.1\\
Curved Void  	& 3377.4 & 4.0 & 238.9 & 4.1 &     3624.4\\
EdS Void	& 3408.7 & 4.2 & 240.9 & 0.8 &     3654.6
\end{tabular}
\vspace{.5cm}
\\
\begin{tabular}{@{\extracolsep{\fill}}l|rrr|r}
Model&CMB&SN&SDSS&total $\chi^2$\\
\hline
$\Lambda$CDM		& 3371.2 & 237.1 & 19.6 &     3627.9 \\
Curved Void  	& 3377.0 & 255.8 & 14.3 &     3647.1 \\
EdS Void	& 3411.4 & 252.8 & 12.7 &     3676.9
\end{tabular}
&{\phantom{a}}&
\begin{tabular}{@{\extracolsep{\fill}}l|rrrr|r}
Model&CMB&SN&SDSS&HST$_{62\pm6}$&total $\chi^2$\\
\hline
$\Lambda$CDM		& 3371.3 & 236.7 & 19.6 & 0.8 &     3628.4\\
Curved Void  	& 3379.3 & 257.8 & 12.7 & 7.1 &     3656.9\\
EdS Void	& 3411.2 & 252.4 & 12.9 & 2.3 &     3678.8
\end{tabular}
\end{tabular*}
}}
}
\caption{A breakdown of the various contributions to the total best fit $\chi^2$ for the different models, when simultaneously fit to CMB+BAO+SN (top left), CMB+BAO+SN+HST$_{62\pm6}$ (top right), CMB+SN+SDSS (bottom left) and CMB+SN+SDSS+HST$_{62\pm6}$ (bottom right).
When we compare from left to right, we see that the inclusion of HST$_{62\pm6}$ hardly affects the goodness of fit to SN, BAO or SDSS, but it does weigh in on the goodness of fit to the CMB in the Curved Void case. Comparing from top to bottom, we see that exchanging the BAO for the SDSS has repercussions on the goodness of fit to the SN as well. That is, the (oversimplified) assumption of taking the cosmological parameters at $r=0$ to fit an FLRW cosmology to the observed SDSS, has a strong influence on the allowed range of $\delta_0$, handicapping the luminosity distance curve.}\label{tab:chi2_breakdown}
} 

In Figure~\ref{fig:marge_prim_CBHS} we show the marginalized one-dimensional posterior probabilities of all free parameters on which we had a flat prior except for $\Omega_{k, {\rm out}}$, which is only a free parameter in $\Lambda$CDM and a derived parameter in the Void scenarios, and $\Omega_{\rm dm, out} h^2$, which is a derived parameter for the EdS-Void model. A comparison between $\Lambda$CDM, the Eds-Void model and the Curved Void model is provided. In Figure~\ref{fig:marge_sec_CBHS} we show the same for all derived parameters. In Table~\ref{tab:chi2_breakdown} we give a breakdown of the total $\chi^2$ in terms of the $\chi^2$ against the different datasets. In Table~\ref{tab:CBHS} we show the best fit parameters and marginalized parameter ranges for all three models.

It is clear that our Curved Void model now provides a significantly worse fit compared to the $\La$CDM model: the effect of including HST$_{62\pm6}$ is large, even though HST$_{62\pm6}$ only includes one datapoint. In comparison the effect of adding the BAO and SN to the CMB on the $\Delta \chi^2$ of the Curved Void  model versus $\Lambda$CDM is not so large; only $\sim+3.1$ and $\sim+2.8$ respectively for 2 and 288 extra data points respectively, please see Table~\ref{tab:chi2}.

Allowing $\Omega_k$ to vary, significantly opens up parameter space in the sense that this model can reconcile the two different constraints on the distance to the last-scattering surface and the constraint on $\Omega_{\rm m, eff} h^2$. However, since the best fit point happens to be a slightly closed universe,  the much better fit to the CMB is achieved  at the cost of a lower $h$ and therefore a worse fit to HST$_{62\pm6}$, as we see in Table~\ref{tab:chi2_breakdown}. This is exactly the same problem that the closed FLRW model encounters in reconciling CMB with other observations. The profile of the curvature inside the void as well as the density along the past light cone are shown in Figure~\ref{fig:kofr}. The inclusion of $\Omega_k$ clearly allows for much larger density contrasts, favoured by the SN and CMB. Even after boosting the Hubble expansion rate with the inhomogeneous void, the Hubble parameter remains inconsistent with HST$_{62\pm6}$. However, as we will see in the next section, once we allow the void profiles to have some additional features, or deal with very large inhomogeneous  models, the problem of a low Hubble value ameliorates considerably. Figure~\ref{fig:marge_prim_CBHS} shows that the redshift of the boundary of the void is not really constrained from above, except from the effect on the CMB monopole temperature as discussed in section \ref{sec:CMB}, reflected by a thick tail to the right in the posterior distribution of $z_B$.

In Figure~\ref{fig:marge_sec_CBHS}, one might be intrigued to see that in EdS, the void actually seems to do a better job at fitting HST$_{62\pm6}$ ($H_{0, {\rm in}}$) than the Curved Void. What is going on is a struggle between two well-constrained characteristics of the CMB: the position of the peaks and the matter to radiation density ratio. The former is most relevant for constraining $\Omega_{\rm m, eff} h^2$, the latter for constraining a different combination of $H_{\rm 0, {\rm eff}}$ and $\Omega_{m, {\rm eff}}$. Since in this model $\Omega_{\rm k, eff}=0$ and $\Omega_{\rm m, eff} =1$, $\Omega_{\rm m, eff} h^2$ and $H_0$ are strictly related to each other. The matter to radiation density ratio favours a value of  about $\Omega_{\rm m, eff} h^2\sim 0.14$, while the position of the peaks favours  $h\sim 0.45$. These two requirements are in contradiction with each other and eventually, the larger value of $H_{0, {\rm out}}$ and $H_{0, {\rm in}}$ is only a fortunate consequence of an otherwise bad fit to the CMB.

\subsection{Large Scale Structure}
\label{results:LSS}
For the large scale structure data (LSS) we use both the SDSS main sample, which have a mean redshift of $z\sim0.1$ and the LRG data which go much further away, with an average around $z\sim0.4$. For the SDSS data in all the relevant cases, the size of the void, $z_{\rm B}$, is so large that all the datapoints are contained inside the void. For the LRG this is  only partially true, but we proceed in the same way just to get an estimate.
This means that we can try to fit the LSS with the approximate effective FLRW universe that is built up from the parameters $\Omega_{i, \rm in}$ and $H_{0, {\rm in}}$ at the centre of the void, as explained in sect. \ref{sec:LSS}. Let us emphasize the caveats to this approach. To begin with, the large scale structure data does not lie at the centre of the void but instead at a slightly  higher mean redshift, hence our prescription is not so accurate, especially for the LRG.
More importantly, the growth of perturbations inside an LTB metric is not yet understood. This means that the $\chi^2$ values when fitting the void to SDSS are to be taken only as a rough indication. In Table~\ref{tab:chi2} we show the $\chi^2$  for the different SDSS and LRG runs. The void does badly when fitting against the LSS, be it with $\Omega_{k, {\rm out}}=0$ or with $\Omega_{k, {\rm out}}\neq0$. In fact, both matter power spectra favour a value for the combination $\Omega_{\rm dm, in} h$ of about $0.2$, while the Curved Void  from the fit with CMB+BAO+SN+HST$_{62\pm6}$
has a best fit value of about $0.09$. Note, however, that the fit is significantly better for LRG than for SDSS.  In fact,  the void does predict too much power on large scales
and the LRG data prefer this, compared to the SDSS data.

In Figures~\ref{fig:marge_prim_CHSS},~\ref{fig:marge_der_CHSS} and~\ref{fig:CHSS_all} and Table~\ref{tab:CHSS} we show the parameter constraints resulting from a fit to CMB+SN+SDSS+HST$_{62\pm6}$. The value of $\Omega_{dm, in} h$ is now twice as much, with respect to the fit without the SDSS. This is due to the fact that the LSS data prefer a higher value of $\Omega_{dm, in} h$, which means a lower density contrast. This  poses difficulties for the SN as well as for the CMB. However, as long as no perturbation theory in the LTB-metric has been developed, there is not much to say about the goodness of this fit.  The strongest conclusion we can draw from this, is that it gives a hint that the void may not be in agreement with observed LSS perturbations, but a correct treatment of the perturbations is really needed before we can say anything definitevely. The main message of this paragraph hence is that such perturbation theory will be of crucial importance for testing the Void versus $\Lambda$CDM.

The effect of the large scale structure on the redshift of the boundary of the void, $z_B$, is that smaller radii no longer give a better fit than larger radii, hence pushing $z_B$ up towards the regions in which it is not constrained by the data but only by the prior (and not even by the monopole effect, since this is smaller due to a smaller density contrast). This is apparent if we compare Figure~\ref{fig:marge_prim_CHSS} with Figure~\ref{fig:marge_prim_CBHS}.

\subsection{Baryon density and BBN}
We note that the baryon density in the outer FLRW $\Omega_{b, {\mt eff}} h^2$ is constrained by the CMB to be in the right range for the Big Bang Nucleosynthesis constraints, and from Table~\ref{tab:CBHS} we indeed find this number to be very close to the $\Lambda$CDM result.
This number tells us what the baryon-to-photon number density ratio, $\eta=n_b/n_{\gamma}$, was outside the Void.
Therefore we are constraining $\eta$ in the outer region ($\eta_{\mto}$) and our value of $\Omega_{b, {\mt eff}} h^2 \sim 0.02$ is in good agreement with BBN constraint.
In this context we note that  measurements of the abundances of the light elements relevant for BBN  are taken at high redshift ($z\gg 1$) \cite{Steigman:2007xt,Amsler:2008zzb,Pettini:2008mq}, and therefore $\eta_{\mto}$ is the appropriate quantity to consider. The only measurements which are closer to us are the ones relevant to $^7Li$, which are known to be in disagreement~\cite{Cyburt:2008kw} with the BBN, so we shall not consider them.
Actually~\cite{Regis:2010iq} we could even take the point of view that those measurements are in disagreement with the rest because they are taken inside the local Void and we could try and explain the $^7Li$ problem as a consequence of the hypothesis that we are living in a Void.

\subsection{Significance of  \texorpdfstring{$\Delta \chi_{\rm eff}^2$}{effective delta chi2}}
So far we have only quoted the effective $\Delta \chi_{\rm eff}^2\equiv -2 \ln \mathcal{L}$. It is still difficult to translate these numbers into a fair model comparison. We could not do a Bayesian Evidence calculation, as explained earlier. In order to get a taste of the meaning of the quoted $\chi^2$ values, let us quote (approximate) $p$-values, that is the probability that the goodness of fit can be worse than the one obtained, given that the hypothesis is true. This has to be taken with caution here: we already stressed that we are quoting an effective $\chi^2$, because the likelihood of the CMB is not a true $\chi^2$ distribution. Nevertheless, we calculate the $p$-value as if everything were exactly $\chi^2$, for simplicity. The numbers serve to give the reader a feeling for the difference between the models. For true $p$-values one would have to simulate all data and explicitly probe how often a worse fit is obtained, when these models are the true models.

The $p$-values against the observations
 are quoted in Table~\ref{tab:pval1}. We take only the best fit parameters obtained from the fit to CMB + BAO + SN + HST$_{62\pm6}$, and calculate the p-values of these models when fitting to some datasets alone and some combinations of datasets, as an illustration. As done in the WMAP paper~\cite{Larson:2010gs}, we consider the CMB as three independent tests of the model, namely the high $l$ TT $C_l$ spectrum, the high $l$ TE $C_l$ spectrum, and the combined low $l$ pixel based likelihood analysis (CMB all low $l$). As could be expected from the $\chi^2$-values in Table~\ref{tab:chi2}, $\Lambda$CDM has the best performing $p$-values. It is still interesting to see  that when we demand a rejection at 99\% C.L., the Curved Void model along with $\Lambda$CDM is not rejected, as opposed to the EdS Void model which is rejected by the larger combination of datasets.

\TABLE{ 
\resizebox{\textwidth}{!}{
\begin{tabular}{l|r|r|r|r}
Model   & CMB TT high $l$ & SN & CMB TT high $l$  & CMB TT high $l$ \\
(Bestfit to CMB+   &  & &  + SN &  + CMB all low $l$  \\
  BAO+SN+HST$_{62\pm6}$) &  & &  & + SN + BAO +HST$_{62\pm6}$  \\
   \hline
$\Lambda$CDM & $ 5.4 \times 10^{-2} $ & $ 0.96 $ & $ 0.29 $ & $ 4.2 \times 10^{-2} $\\
Curved Void  & $ 5.1 \times 10^{-2} $ & $ 0.96 $ & $ 0.28 $ & $ 2.5 \times 10^{-2} $\\
EdS Void  & $ 2.2 \times 10^{-2} $ & $ 0.95 $ & $ 0.17 $ & $ 1.3 \times 10^{-2} $
\end{tabular}}
\caption{The roughly approximated $p$-values of the $\chi^2$ of the best fit models obtained when fitting to CMB+BAO+SN+HST$_{62\pm6}$, against the different datasets, considering $\chi^2$-values of the datasets alone and in combination. As done in the WMAP paper~\cite{Larson:2010gs}, we consider the CMB as three independent tests of the model, namely the high $l$ TT $C_l$ spectrum, the high $l$ TE $C_l$ spectrum, and the combined low $l$ pixel based likelihood analysis (CMB all low $l$). The $p$-value we quote here for the high $l$ TT spectrum is a factor three lower than the value that the WMAP team obtained, probably due to our flawed assumption that the likelihood is actually $\chi^2$. The last column considers all datasets except for the CMB TE spectrum.
If we demand a $p$-value of at least 5\%, all models are rejected at 95\% C.L. by the observations (last column). The high $p$-values against the SN, indicate that the error bars in the MLCS2k2 pipeline from the SDSS SN survey are most likely overestimated. These values serve as an illustration of the meaning of the $\chi^2$ values in Table~\ref{tab:chi2}, but are by no means to be considered correct.}\label{tab:pval1}
} 

\setlength{\totfigwidth}{0.8\textwidth}
\FIGURE{
\begin{tabular}{lr}
	\includegraphics[width=.5\totfigwidth]{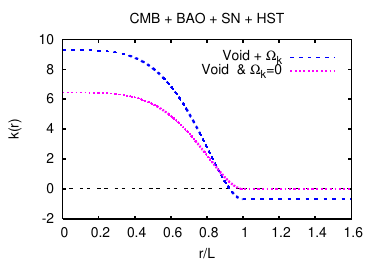} &	\includegraphics[width=.5\totfigwidth]{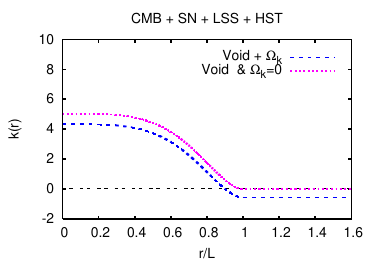} \\
	\includegraphics[width=.5\totfigwidth]{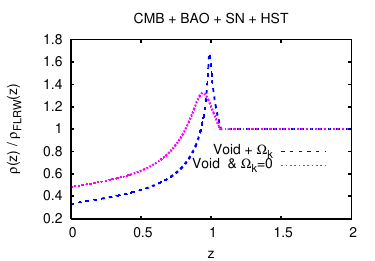} & \includegraphics[width=.5\totfigwidth]{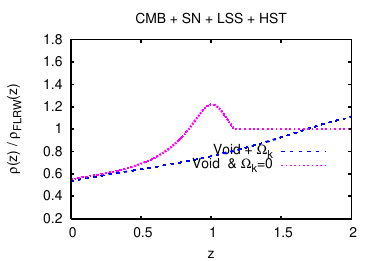} \\	
\end{tabular}
\caption{Curvature profiles as a function of coordinate radius $r$ (top) and density profile as a function of redshift $z$ along a geodesic (bottom), comparing the best-fit CMB+BAO+SN+HST$_{62\pm6}$ to CMB+SN+SDSS+HST$_{62\pm6}$. The density is normalized to the FLRW density that would be measured at that redshift by an observer in a pure FLRW at the same time of observation as the observer in the centre of the void. Comparing the figures from left to right, it is clear what the restrictive power of the large scale structure inside the void is. }\label{fig:kofr}
}

\section{Results with Profiles with higher \texorpdfstring{$H_0$}{H0}}\label{sec:exotic}
It is important to realize that with infinite freedom in the function $k(r)$, we can construct almost any expansion history on the past light cone\footnote{With the exception that one always gets deceleration locally at the origin $r=0$~\cite{Vanderveld:2007cq,Vanderveld:2009eu}.}. Therefore it may be possible to fix the problem that we have found in the previous sections, namely to find profiles which can fit HST as well as the other datasets. There are two ways this can be achieved. The first way is to modify the profile at very low redshift, say $z\lesssim 0.03$, since the observed value of $h$ is measured at these low redshifts, and such a change should not affect any other measurements, such as SN and BAO.
The other way is to try and get a much larger correction to the monopole of the CMB.  Since it is the CMB which constrains $H_{\rm out}$ to be so small, by altering the distance to the last scattering surface significantly, we can hope to obtain a different value of $H_{\rm out}$. This can be done if we use  non-compensated profiles which asymptotes very slowly to FLRW,  and only at high redshifts, say $z\gg 3$. This ensures that the  compensating overdense shell, which in general cancels the monopole shift from the underdense void,  is absent and we in turn can get a much larger monopole correction. Further, this modification does not affect other measurements such as SN and BAO.
We explore in detail the two options in the following subsections.  It may be possible to combine them together, but we leave that as a future excercise.\footnote{Eventually we could perform a principal component analysis, describing the void profile by nodes. This way we would find which features in a void profile are most important and best constrained, as is similarly done for the equation of state of Dark Energy~\cite{Huterer:2002hy}.}

\subsection{Modified profiles}
In this section we will consider curvature profiles that exhibit a larger curvature very close to the centre of the void, in other words with a more peaked density contrast. Effectively this scenario looks like a `Void inside a Void'.

We have tested the profiles given below in Equations~(\ref{eq:profA}--\ref{eq:profD}), where profile A is the profile considered in the previous sections, given by Eq.~\eqref{profile}. Profiles B, C, D and E add two more degrees of freedom. For B, C and D  these are  the curvature $k_{\rm max, 2}$ in the inner void and its radius $L_2 = f\,L$. Instead of $L_2$ there is a parameter $\alpha$ in profile E, which determines the rate at which the profile converges to FLRW at high redshift.

The profiles C, D and E only reach FLRW asymptotically. In the case of profile C and D this happens at such a high rate, that there is no significant effect, compared to the exactly compensated profiles. Profile E on the other hand, only reaches the asymptotic metric at a very slow slow rate, and it does so with a very large `overdense' region, which is discussed in the next subsection.
\renewcommand\arraystretch{2}%
\begin{align}
&\mbox{Profile A:{} } &k(r)&=\left\{\begin{array}{cr}k_{\mt{max}}\LT1-\LF{r\over L}\RF^4\RT^2 +k_b& \mx{ for } r\leq L\\
k_b & \mx{ for } r > L  \end{array}\right.\label{eq:profA} \, , \\
&\mbox{Profile B:{} } &k(r)&=\left\{\begin{array}{cr} k_{\mt{max, 2}}\LT1-\LF{r\over L_2}\RF^4\RT^2 +k_{\mt{max}}\LT1-\LF{r\over L}\RF^4\RT^2 +k_b & \mx{ for } r\leq L_2\\
	k_{\mt{max}}\LT1-\LF{r\over L}\RF^\alpha\RT^\beta +k_b& \mx{ for } L_2 < r \leq L  \\
k_b & \mx{ for } r > L\end{array}\right.\label{eq:profB} \, , \\
&\mbox{Profile C:{} } &k(r)&=\left\{\begin{array}{cr}k_{\mt{max}}\LT1-\LF{r\over L}\RF^4\RT^2 +  k_{\mt{max,2}} \exp \left[ \left(\frac{r}{L_2}\right)^2   \right]+k_b& \mx{ for } r\leq L \, , \\
 k_{\mt{max,2}} \exp \left[ \left(\frac{r}{L_2}\right)^2   \right]+k_b & \mx{ for } r > L  \end{array}\right.\label{eq:profC}\, , \\
&\mbox{Profile D:{} } &k(r)&=\left\{\begin{array}{cr}k_{\mt{max}}\LT1-\LF{r\over L}\RF^4\RT^2 +  k_{\mt{max,2}} {1\over 1+(\frac{r}{L_2})^2}  +k_b& \mx{ for } r\leq L\\
k_{\mt{max,2}}\frac{1}{1+(\frac{r}{L_2})^2}  +k_b & \mx{ for } r > L  \end{array}\right.\label{eq:profD}  \, , \\
&\mbox{Profile E:{} } &k(r)&=k_{\mt{max}}{1\over 1 +  \left(\frac{r}{L}\right)^2  } +  k_{\mt{max,2}} {1\over \LF 1+ {r\over L}\RF^\alpha } \LF \tanh\LT \tfrac{10}{3} {r-L\over L} \RT -\tanh\LT  -\tfrac{10}{3} \RT \RF  +k_b& \forall r  \label{eq:profE} \, .
\end{align}
\renewcommand\arraystretch{1}%

The profiles for these configurations are illustrated in Figure~\ref{fig:Exoticprofs}, using their best fit parameters. The best fit parameters and the marginalized likelihoods are presented in Table~\ref{tab:Exotic}. We find that the `Void inside a Void' does indeed improve the goodness of fit to the HST observation, but not only that. It also improves the goodness of fit to the BAO by a large factor. In fact, the improvement for the BAO is such that it actually makes the Void truly competitive with $\Lambda$CDM, as we see in Table~\ref{tab:chi2_exotic} and Figure~\ref{fig:Exotic_BAO_SN}. $\Lambda$CDM has a best fit $\chi^2$ value of 3614.1 when fit to CMB+BAO+SN+HST, where profile D is capable of getting a $\chi^2$ of 3616.1.
This difference is due to the fact that the profile changes rapidly around the redshift of the BAO scale.
However, as we said in Section~\ref{risult}, we stress that we have to be cautious in this case: when the BAO scale changes rapidly it is not anymore a good approximation to take the distance scale at one single redshift point.

Another interesting feature in these profiles is apparent in Figure~\ref{fig:Exotic_BAO_SN}. The inner void clearly has its radius, and hence its shell, somewhere in the middle of the observed supernovae.  Voids are preferred to be large since the inclusion of the SDSS Supernovae~\cite{Kessler:2009ys} which filled the gap at intermediate redshifts. But the lesson to learn here, is that the preference for large Voids is not due to the shell no longer being allowed inside the observer SNe, but only due to the overall shape of the distance-z relationship,  $\mu(z)$, which favours a very open universe up to high redshifts. A feature in $\mu(z)$ is not ruled out at all.

\setlength{\totfigwidth}{\textwidth}
\FIGURE{ 
       \includegraphics[width=\totfigwidth]{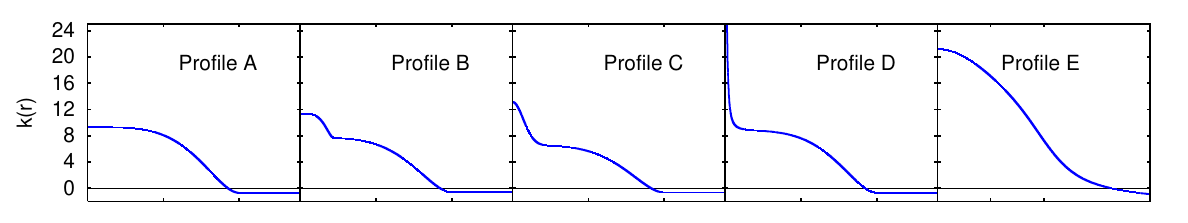}\vspace{-0.03\totfigwidth}\\
       \includegraphics[width=\totfigwidth]{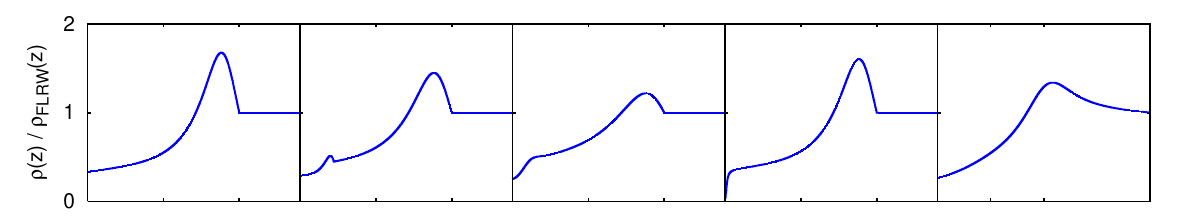}\vspace{-0.03\totfigwidth}\\
       \includegraphics[width=\totfigwidth]{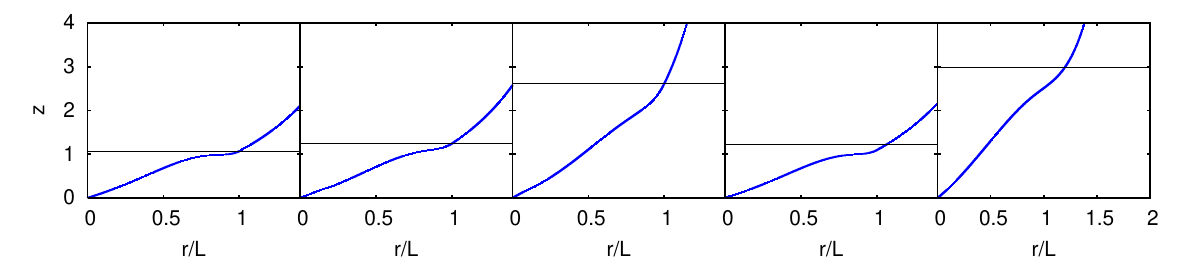}\\
       \includegraphics[width=\totfigwidth]{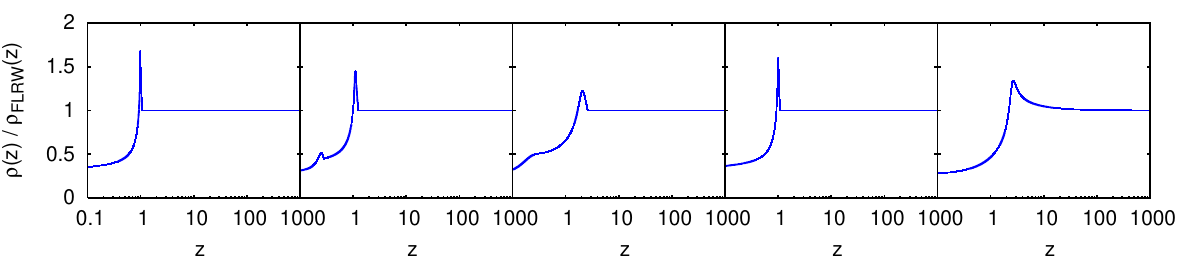}
       \caption{The best-fit curvature profiles (top), density profiles
(second from top) and redshift histories (third from top) as a
function of coordinate radius $r/L$, and the density as a function of
redshift (bottom). The horizontal black lines in the third panel
indicate the redshift radius $z_B$ of the void. In the second panel it
seems that for profile E the asymptotic curvature is not reached
before the surface of last scattering, when we look along coordinate
radius $r$. In the bottom panel, however, as a function of the
physically relevant redshift, it is clear that also for this profile
the asymptote is as good as reached at the surface of last scattering.
}\label{fig:Exoticprofs}
}

\TABLE{ 
\begin{tabular*}{\textwidth}{@{\extracolsep{\fill}}  l|rrrr|r}
Model&CMB&BAO&SN&HST$_{62\pm6}$&total $\chi^2$\\
\hline
$\Lambda$CDM & 3371.1 & 3.1 & 239.5 & 0.4 &    {\bf 3614.1}\\
Profile A (Curved Void) & 3377.4 & 4.0 & 238.9 & 4.1 &     3624.4\\
Profile B & 3377.0 & 0.2 & 237.9 & 2.2 &     3617.3\\
Profile C & 3376.9 & 0.7 & 237.7 & 1.9 &     3617.2\\
Profile D & 3377.5 & 3.6 & 233.7 & 1.3 &     {\bf 3616.1}\\
Profile E & 3380.2 & 3.3 & 241.4 & 0.8 &     3625.7
\end{tabular*}
\caption{A breakdown of the total $\chi^2$ for each dataset, for fitting simultaneously to CMB + BAO + SN + HST$_{62\pm6}$. The main improvement achieved with the `Void inside a Void' is on the $\chi^2$ of the BAO, more than on that of HST$_{62\pm6}$.}\label{tab:chi2_exotic}
} 

\FIGURE{ 
\raisebox{0.01\textwidth}{	\includegraphics[width=0.23\textwidth]{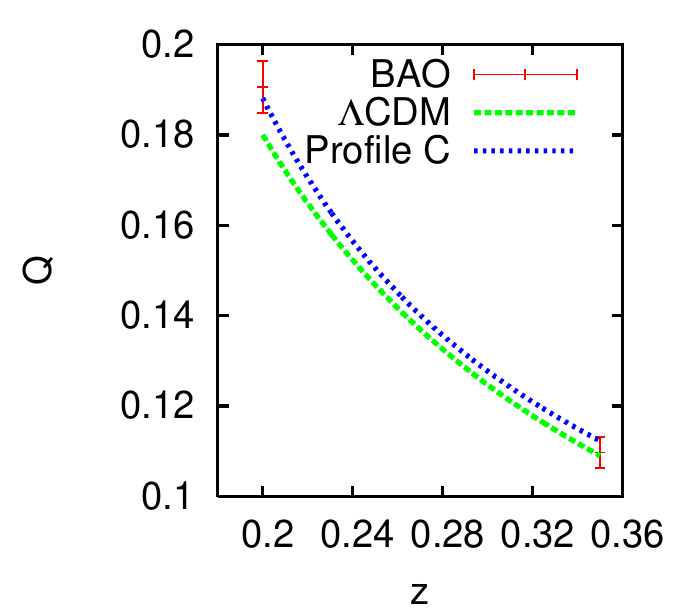}  }
	\includegraphics[width=0.64\textwidth]{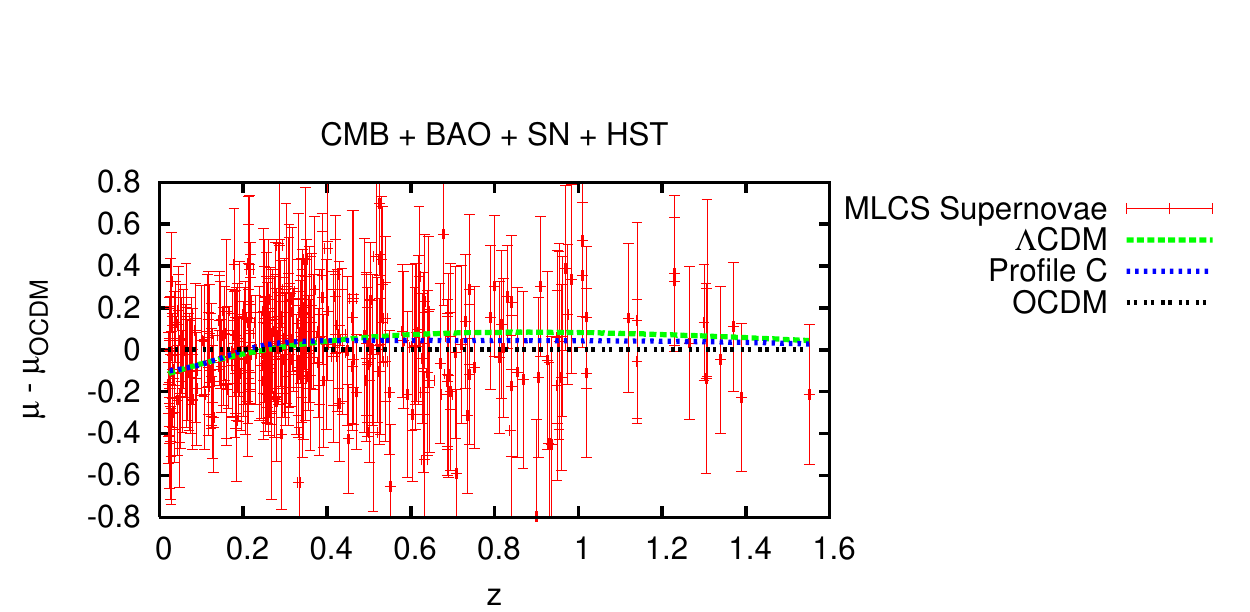}
\caption{The best-fit theoretical prediction of a Void with profile C compared to $\Lambda$CDM, for the BAO (left) and SN (right). Clearly, this Void profile has less difficulty fitting the BAO than $\Lambda$CDM has. An interesting feature in the right figure, is that the inner void has its radius in the middle of the SN, which is no problem for getting a very good $\chi^2$. This implies that the reason that single Voids need to be large, is not the disagreement with a feature in the middle of the redshift range, but the overall shape of the SN curve that favours a very open universe.}\label{fig:Exotic_BAO_SN}
}

\label{asymp}
\FIGURE{ 
	\includegraphics[width=0.6\textwidth]{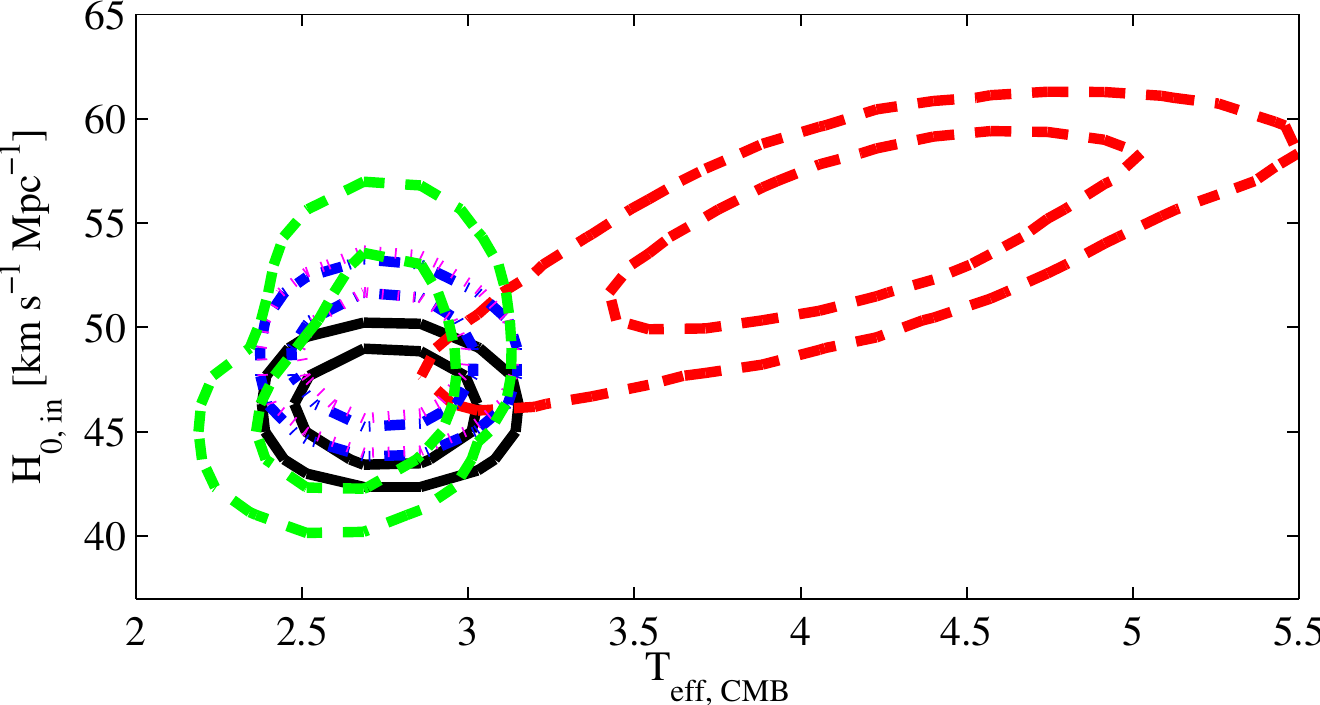}
	\caption{The 2D marginalized posterior likelihoods of the locally observed Hubble constant and the temperature of the CMB for the effective observer, for all profiles: A (solid black), B (dashed blue), C (dotted magenta), D (dashed green) and E (dashed red). Only profile E predicts significantly different values, notably a higher observed value for $H_0$.}\label{fig:exotic_asymp_TCMB}
}

We also analyze a class of profiles $k(r)$ that does not match exactly to FLRW, but which goes only asymptotically to FLRW at very large $r$, \ie  $k(r)\rightarrow k_b$ at $r\rightarrow \infty$.
It has been shown~\cite{Zibin:2008vk} that within this class of profiles we can have a large effect on the monopole temperature, due to the fact that such profiles do not have a perfectly compensating shell, which in the usual case almost exactly cancels the effect of the Void. Instead, such models have an overcompensating shell, such that the total mass contained in the LTB patch is actually larger than what would be contained in an equally large background FLRW patch.

However, if the profile goes to a constant  very quickly after some distance $L$, we get an effect very similar to the previously analyzed case. Therefore, to get something non-trivial we need a profile which goes slowly to a constant,  as exemplified by profile E.
It is possible to show, already from the analytical approximations given in Appendix~\ref{approximations}, that the dominant contribution to the monopole effect (in the flat case $k_b=0$) for such a Void is given by the following integral:
\begin{equation}
k_2\equiv \int k(r) r^2 dr \, .
\label{kdue}
\end{equation}
It turns out that this integral enters with a negative sign in the final expression of the monopole, so if $k_2$ is positive this makes the effective CMB temperature lower, rather than higher.
This is precisely what happens in the compensated case as well; since this term starts dominating when the Void has a very large radius, the fit becomes worse (see sect.\ref{results:LSS}).
In the case of an asymptotic profile the way to get a positive correction to the monopole in (\ref{kdue}) is to have a $k(r)$ which goes negative and then  asymptotes  to zero from below.
In the curved case, the profile has to become more negative than $k_b$ and then approach  $k_b$ asymptotically from below.

We find that in such cases the correction to the monopole is much larger than the previous ${\cal O}(1\%)$, going up to $30\%-40\%$. This drastically changes the other parameters, most notably $H_0$. In figure~\ref{fig:exotic_asymp_TCMB}, we see that profile E changes the temperature of the CMB so much, that the effective outer universe allows for a higher value of $H_{\mt out}$ (and accordingly also $H_{0 }$) compared to the other profiles.

\subsection{Other data sets}
\subsubsection{A higher value for  \texorpdfstring{$H_0$}{H0}}
For the sake of completeness, we performed a brief analysis including more CMB data, and other values for  $H_0$. In the previous sections, we used the value $H_0 = 62.3 \pm 6.3$ km s$^{-1}$ Mpc$^{-1}$ from Ref.~\cite{Sandage:2006cv}.  The results that we  obtained using $H_0 = 72 \pm 8$ km s$^{-1}$ Mpc$^{-1}$ and $H_0 = 74.2 \pm 3.6$ km s$^{-1}$ Mpc$^{-1}$, as  quoted in Ref.~\cite{Freedman:2000cf} and Ref.~\cite{Riess:2009pu} respectively, are shown in table~\ref{tab:chi2_hstl} and  table~\ref{tab:chi2_hstl10}.

Profile E manages to fit the higher HST value $H_0 = 72 \pm 8$ km s$^{-1}$ Mpc$^{-1}$ at $1.7 \sigma$, at the cost of a worse fit to the BAO and SN than profile C does. The change in goodness of fit to $H_0$, is mainly due to the shape of the profile at very large distances. Due to the large effect on the monopole of the CMB, profile E allows for a higher overall value of $H_0$, hence also a higher value inside the void. Since the distances at which this monopole-effect comes to play are different from the distances at which the profiles fit the BAO and SN, it may be possible to combine the good features of profile C and E into one new profile, thereby fitting all datasets better at the same time.

When fitting $H_0 = 74.2 \pm 3.6$ km s$^{-1}$ Mpc$^{-1}$ from Ref.~\cite{Riess:2009pu}, one does not actually fit $H_0$, but rather the value of $z/d_A(z)$ at $z=0.04$, with $d_A$ the angular diameter distance.
Fitting this quantity is more model independent, since $H_0 \equiv \lim_{z \rightarrow 0} z/d_A$, and the actual observation is at nonzero redshifts. So the value of $h=74$ makes sense only when assuming $\Lambda$CDM in deriving $H_0$.
That this is in fact the case, is reflected in the resulting $\chi^2$. For Profile E for example, the value of $H_{0, {\rm in}}$ for both fits (Tables~\ref{tab:chi2_hstl}~and~\ref{tab:chi2_hstl10}) is $H_{0, {\rm in}}=58$ km s$^{-1}$ Mpc$^{-1}$. But fitting against the effective
$H_0 = 74.2 \pm 3.6$ km s$^{-1}$ Mpc$^{-1}$, its $\chi^2$ is slightly less then $(74.2-58)^2/(3.6)^2 = 20.3$, namely $\chi^2=18.5$.

Putting this minor effect aside, the higher value for $H_0$ with smaller error bars leads to no surprises: the Void Profiles assessed here perform in a similar manner overall, but are punished for the lack of a high value for $H_{0, in}$.

\TABLE{ 
\begin{tabular*}{0.7\textwidth}{@{\extracolsep{\fill}}  l|rrrr|r}
Model&CMB&BAO&SN&HST$_{72\pm8}$&total $\chi^2$\\
\hline
$\Lambda$CDM				& 3371.8 & 2.3 & 239.5 & 0.3 &     3613.9\\
Profile C						& 3377.1 & 4.0 & 235.9 & 3.6 &     3620.6\\
Profile E						& 3382.0 & 5.9 & 242.2 & 3.0 &     3633.1
\end{tabular*}
\caption{A breakdown of the total $\chi^2$ for each dataset, for fitting simultaneously to CMB + BAO + SN + HST$_{72\pm8}$, where HST$_{72\pm8}$ indicates that we used $H_0 = 72 \pm 8$  km s$^{-1}$ Mpc$^{-1}$ from Ref.~\cite{Freedman:2000cf}, keeping all other data sets equal to the previous sections. Profile E fits $H_0$ at $\sqrt{3.0} \sigma = 1.7 \sigma$, which is good compared to the other profiles, due to its high-density asymptotic profile as discussed above. Profile C gives a much better fit to the BAO and SN, such that its overall fit is still slightly better than profile E. }\label{tab:chi2_hstl}
} 

\TABLE{ 
\begin{tabular*}{0.7\textwidth}{@{\extracolsep{\fill}}  l|rrrr|r}
Model&CMB&BAO&SN&HST$_{74\pm4}$ &total $\chi^2$\\
\hline
$\Lambda$CDM 		& 3372.7 & 1.8 & 239.7 & 2.1 &     3616.3\\
Profile C 				& 3380.7 & 1.9 & 238.4 & 20.7 &     3641.7\\
Profile E 				& 3380.2 & 3.2 & 242.0 & 18.5 &     3643.9
\end{tabular*}
\caption{A breakdown of the total $\chi^2$ for each dataset, for fitting simultaneously to CMB + BAO + SN + HST$_{74\pm4}$, where HST$_{74\pm4}$ indicates that we used effectively $74.2 \pm 3.6$ km s$^{-1}$ Mpc$^{-1}$ from Ref.~\cite{Riess:2009pu}, keeping all other data sets equal to the previous sections. The quantity that is acutally fit in stead of $H_0$ is $z/d_A(z)$ with $z=0.04$, which is more model independent. }\label{tab:chi2_hstl10}
} 

\TABLE{ 
\begin{tabular*}{0.7\textwidth}{@{\extracolsep{\fill}}  l|rrrr|r}
Model&CMBe&BAO&SN&HST$_{72\pm8}$&total $\chi^2$\\
\hline
$\Lambda$CDM				&  3531.4 & 2.3 & 239.7 & 0.3 &     3773.7\\
Profile C						& 3537.7 & 4.2 & 236.1 & 3.3 &     3781.3 \\
Profile E						& 3537.9 & 2.4 & 242.0 & 6.7 &     3789.0
\end{tabular*}
\caption{A breakdown of the total $\chi^2$ for each dataset, for fitting simultaneously to CMBe + BAO + SN + HST$_{72\pm8}$, where HST$_{72\pm8}$ again indicates that we used $H_0 = 72 \pm 8$  km s$^{-1}$ Mpc$^{-1}$, and CMBe indicates that we use the extra data sets as discussed in the text.}\label{tab:chi2_cmbe}
} 

Again, let us quote the approximate $p$-values we obtain by assuming that the goodness of fit is actually $\chi^2$-distributed, and that our $\chi_{\rm eff}^2$ values  are the actual $\chi^2$'s. The values are listed in Table~\ref{tab:pval2}. In this case, when using $74.2 \pm 3.6$ km s$^{-1}$ Mpc$^{-1}$ from Ref.~\cite{Riess:2009pu}, we see that the extra freedom in Profiles C and E allows these profiles to obtain similar $p$-values as profile A in Table~\ref{tab:pval1}, in spite of the much worse fit to the higher $H_0$-value. In the last column we provide a comparison with the HST$_{62\pm6}$-runs for these models, where the $\chi^2$ is very close to that of $\Lambda$CDM. The higher number of free parameters in the void profiles, compared to $\Lambda$CDM, still pushes the $p$-value down relatively to $\Lambda$CDM.

\TABLE{ 
\resizebox{\textwidth}{!}{
\begin{tabular}{l|r|r|r|r||r}
Model   & CMB TT high $l$ & SN & CMB TT high $l$  & CMB TT high $l$ & CMB TT high $l$ \\
(Bestfit to CMB+   &  & &  + SN &  + CMB all low $l$ &  + CMB all low $l$  \\
  BAO+SN+ &  & &  & + SN + BAO& + SN + BAO   \\
  HST$_{74\pm4}$) &  & &  &  +HST$_{74\pm4}$&  +HST$_{62\pm6}$  \\   \hline
$\Lambda$CDM & $ 5.2 \times 10^{-2} $ & $ 0.96 $ & $ 0.28 $ & $ 4.2 \times 10^{-2} $ & $ 4.2 \times 10^{-2} $\\
Profile C & $ 4.2 \times 10^{-2} $ & $ 0.95 $ & $ 0.26 $ & $ 1.3 \times 10^{-2} $& $ 3.4 \times 10^{-2} $\\
Profile E & $ 4.1 \times 10^{-2} $ & $ 0.93 $ & $ 0.23 $ & $ 9.7 \times 10^{-3} $ & $ 2.2 \times 10^{-2} $
\end{tabular}}
\caption{The roughly approximated $p$-values of the $\chi^2$ of the best fit models, against the different datasets, as in Table~\ref{tab:pval1}. The CMB, BAO and SN give no surprises with respect to the previous section. In the second column from the right we see that $\Lambda$CDM has no difficulty fitting the higher $H_0$-value, and that the extra freedom in profiles C and E is necessary to keep them at the same level with respect to $\Lambda$CDM, when compared to profile A in Table~\ref{tab:pval1}. In the last column we show for comparison the $p$-values obtained in the runs with HST$_{62\pm6}$, using the best fit parameters from those runs. There the difference with $\Lambda$CDM is much smaller, although the void profiles are still punished for their extra parameters, since the number of degrees of freedom used for the $p$-value calculation is $N_{\rm dof} = N_{\rm data\,points} - N_{\rm free\,parameters}$. Note that these $p$-values are very rough estimates, which serve as an illustration only.}\label{tab:pval2}
} 

\subsubsection{Small scale CMB data and additional polarization data}
On top of the higher value for $H_0$, we also added CMB data from BICEP~\cite{Chiang:2009xsa}, ACBAR~\cite{Reichardt:2008ay} and CBIPol~\cite{Sievers:2005gj} along with the already used WMAP data. The resulting best fit $\chi^2$ values are quoted in Table~\ref{tab:chi2_cmbe}. Comparing to Table~\ref{tab:chi2_hstl}, we immediately see that the inclusion of other CMB data has no role in the difference in $\chi^2$ between $\Lambda$CDM and the Void-models. An improvement of the goodness of fit must be searched in the prediction for the observed value of $H_0$, as well as on the large scales of the CMB, which is exactly where the late-time ISW-effect is important. However, given the ignorance about perturbation theory in the LTB metric, at this point we cannot say precisely what is the impact of the ISW effect, and how the void profile should change in order to improve the goodness of fit to the CMB.

\section{CMB Dipole and the ``Dark Flow''}\label{sec:otherpreds}
When considering a non-homogeneous large scale model for our Universe, it is unlikely that we should occupy a special position, the centre. However the more off-centre we place the observer, the more we spoil isotropy. Therefore the observer's position is constrained by observations.
The most constraining observation is given by a velocity that a comoving observer acquires with respect to the CMB frame. Barring special cancellations (the observer may have a small-scale peculiar velocity which cancels the large scale LTB velocity) this puts a strong constraint~\cite{Alnes:2005rw}.
So, let us consider our observer to be located slightly off-centre, at $r=r_O$.
In this case the non-zero radial velocity of the observer will lead to a dipole moment in the CMB~\cite{Alnes:2005rw, ABNV} of the order of:
\be
{\de T\over T}\sim v_O=\dot{d_O} - \dot{d_F}
\, ,
\ee
where the proper radial distance, $d_O$, of the observer~\footnote{For an exact calculation of the dipole moment see~[?] for instance. For the purpose of estimation however it is sufficient to calculate the dipole moment from radial velocities.}  is given by
$$
d_O=\int_0^{r_O}dr\ {R'\over\sqrt{1+2(\bM r)^2k(r)}} \, ,
$$
and $d_F$ is the proper radial distance of an observer at the same position but at rest in the FLRW background, such that $d_O =d_F$ but  $\dot d_O \neq \dot d_F$.

Now, in our profile $k(r)$ remains almost a constant for almost the entire underdense region. Assuming we are living in this ``constant'' underdense region, we have
\be
d_O=a_0(t)\int_0^{r_O}dr\ {1\over\sqrt{1+2(\bM r)^2k(r)}} =a_0(t){\tanh^{-1}(\bM\sqrt{2k_{0}}r_O)\over \bM\sqrt{2k_{0}}} \approx a_0(t)r_O\, .
\ee
where $a_0(t)$ refers to the scale factor at $r=0$, and the simplification occurs because $u$ and hence $R'$ becomes only a function of time, and  $\bM r_O$ is expected to be very small.
Taking the time derivative  we  simply find
\be
\dot{d_O}= \dot a_0(t_0) r_O=H_0d_O
\label{ddot}
\, .
\ee
In the FLRW background, we similarly have
\be
\dot d_F = \dot a_F(t_0) r_F = H_{{\rm out}}\, d_O
\ee

Thus at the present epoch the estimated dipole moment comes out to be
\be
{\de T\over T}\sim (H_0-H_{{\rm out}})d_O\approx {2\pi(\cosh u_0-1)(H_0-H_{{\rm out}})r_O\over 3k_{0}} \label{eq:dipole_ana} \, .
\ee
We note that $u_0$ can be obtained by inverting (\ref{H}). In Table~\ref{tab:dipole} we compare approxation~\eqref{eq:dipole_ana} to numerical results for the maximum distance of the observer to the centre of the void, for which the dipole is at most equal to the dipole observed in the CMB, 3.355~mK~\cite{Hinshaw:2008kr}. For the profiles A and B, which behave similarly at the centre and have close to constant $k(r)$ for small $r$, approximation~\eqref{eq:dipole_ana} is good up to a few percent.  The observer is thus constrained to be within $\mathcal{O}(10$~Mpc$)$ of the centre of the void.

\TABLE{ 
\begin{tabular*}{0.8\textwidth}{@{\extracolsep{\fill}}l|rrrrr}
Profile & $z_B$ & $L$ [Mpc] & $r_{3.355 {\rm mK}}$ [Mpc] & $r/L$& Eq.~\eqref{eq:dipole_ana} [mK]\\
\hline
A  &     1.024  &  4540.964  &    17.104  &   0.00377  &   3.02155 \\
B  &     2.776  &  7400.249  &    13.814  &   0.00187  &   3.13626 \\
C &     2.912  &  7586.902  &    12.663  &   0.00167  &   3.22672 \\
D &     1.012  &  4536.971  &     5.476  &   0.00121  &   3.27521 \\
E &     2.129  &  5077.986  &    13.394  &   0.00264  &   3.29353
\end{tabular*}
\caption{Dipole constraint on the position of the observer for the best fit parameters of the different curvature profiles. $r_{3.355 {\rm mK}}$ indicates the numerically obtained radius at which the dipole due to the off-centre position is (up to numerical precision) exactly equal to the observed dipole in the CMB, 3.355~mK~\cite{Hinshaw:2008kr}. The last column gives the dipole as estimated in Eq.~~\eqref{eq:dipole_ana}, which assumes a constant $k(r)$ for small $r$. }\label{tab:dipole}
} 

While the CMB dipole moment constrains our model, the fact that we are located slightly off the centre of the void can also account for a rather surprising observation, the ``dark flow''. In~\cite{Kashlinsky:2008us} the authors observed a systematic difference in the kSZ effect from light sources as one varied the angle keeping the distances fixed. The authors had already subtracted the contribution coming from the motion of the observer (this gives the CMB dipole moment discussed above) and therefore this additional Doppler shift was interpreted by the authors to indicate a ``dark flow'' of all the light sources towards a particular direction. This can be understood as follows: If all the objects have a common peculiar velocity, say $v_S$, along a particular direction, then the expected Doppler shift will be given by
\be
\De z_{\mt{kSZ}}\sim v_S\cos \te\,\, ,
\ee
where $\te$ is the angle between the position and the velocity vector of a given light source. In fact, in~\cite{Kashlinsky:2008us} a cosine variation of the kSZ effect was extracted on a scale of at least $300$ Mpc/$h$, and the authors estimated the bulk velocity to be $600-1000$ km/s from their observations, which is way too large to be consistent with $\Lambda$CDM.

In the LTB model however, the above effect would naturally arise: if we are located slightly off the centre of the void, then the light sources which are approximately the same distance away from us, will not be at the same distances from the centre. Now, in the void region all the objects move away from the centre towards the spherical structure at the edge of the void, and more importantly, their outward velocity is proportional to their distances from the centre. Thus we, the off-centre observers,  will start seeing a cosine modulation in their velocities (Doppler shift). Let us try to provide a simple estimate of the effect. Approximately using simple Eucleadian geometry we find
\be
r_S\approx r'_S\LF 1+{{r_O\over r'_S} \cos \te}\RF\,\, ,
\ee
where $r_S,r'_S$ are the radial distances of a given light source from the centre and the off-centre observer, $\te$ is the angle of the sources as measured from the observer. Now, if we assume that both the observer and the light sources are located in the core of the void region where $k(r)\approx k_0$, then we can use the same Doppler shift formula (\ref{eq:dipole_ana}) that we have derived above:
\be
v_S=\dot{d_S}\approx  {2\pi(H_0-H_{{\rm out}})(\cosh u_0-1)r_S\over 3k_{0}}\approx  {2\pi(H_0-H_{{\rm out}})(\cosh u_0-1)r'_S\over 3k_{0}}\LF 1+{{r_O\over r'_S}{ \cos \te}}\RF
\label{dsdot}
\, .
\ee
Please note that since $r_O\ll r_S$, the radial directions subtended by the light sources to the centre and the observer are approximately the same.

The term multiplying one is simply a monopole term which is filtered out of the kSZ dipole observations, but the second term, containing $\cos \te$, provides an additional contribution to the kSZ dipole. Thus in the void model we will have
\be
\De z_{\mt{kSZ}}\sim \cos\te {2\pi(H_0-H_{{\rm out}})(\cosh u_0-1)r_O\over 3k_{0}}=\cos \te\,\LF{\de T\over T}\RF_{\mt{cmb-dip}}\,\, .
\ee
Thus we actually see a connection between the dark flow velocity and the dipole measured in CMB in an LTB model, which gives precisely a flow of order $10^{-3}$ times the speed of light on very large scales, roughly consistent with the measurement of~\cite{Kashlinsky:2008us}. In fact~\cite{Kashlinsky:2008us} mentions that the direction of the bulk velocity also seems to be roughly aligned with the dipole velocity, again what we would expect in our model. A prediction of our model is that we should continue to observe the same ``dark flow'' velocity all the way out to Gpc scales till we reach the end of the underdense region.

\section{Conclusions\label{sec:conclusions}}

In the present paper we have investigated the possibility of fitting most of the available cosmological data without Dark Energy, with a non-homogenous density distribution in which we are located at the centre of a Gpc scale Void that can mimic cosmic acceleration.
The new features of this analysis are a combined MCMC fit of many datasets (WMAP, Supernova, BAO data, HST and an estimate of the Large Scale Structure data)  for which we wrote a publicly released module in {\sc cosmomc}.
We also enlarged the parameter space of Void models by allowing for the presence of a background curvature,  and we have included the additional redshift that the photons experience when traveling in a Gpc scale Void which  modifies the distance to the last-scattering surface.
We have shown that the SN, BAO and CMB can be fit well already with the simplest Void profile, which includes a Void and an outer overdense shell with two parameters, matched exactly to FLRW. In this case the best-fit Void extends upto $z\sim 1$, it has a density contrast of $\delta\sim -0.65$ and the external background FLRW is slightly closed,  with $\Omega_k\sim -0.2$. The primordial spectra in the best fit are consistent with the usual almost scale-invariant ones.
Finally, the Baryon abundance computed in the outer region is consistent with BBN.

However, for the simplest profile, the measured value of the local Hubble constant of about $h\simeq 0.6\sim 0.7$ is difficult to fit, since the CMB forces the Hubble constant to be very low ($h\simeq 0.45$), in order to fit the angular distance to the last-scattering surface.
We have then shown that modifying the Void profile by adding more parameters we could accommodate for a larger value of $H_0$, upto $h\simeq0.57$, in the process obtaining  an overall better fit to all the cosmological data that becomes comparable to the $\Lambda$CDM fit. This can be achieved in two ways. Either one can modify the profile at very small redshift, thereby increasing the local density contrast and  obtaining a higher local Hubble constant. Incidentally, this leads also to a better fit of the BAO data.
Or, one can modify the profile at very high $z$ introducing a large overdense region which goes to FLRW only asymptotically at large distance. Such profiles can change the distance to the last-scattering surface significantly, thereby allowing us to fit the CMB with a higher $H_0$.

Finally we have shown that the model is consistent with the CMB dipole, if the position of the observer is at most about ${\cal O}(10)$ Mpc/$h$ away from the centre.
On the other hand this would account for the recently measured large scale "bulk flow" which seems inconsistent with $\Lambda$CDM.

We have therefore demonstrated that there exist Void models based on LTB metrics, which may be a viable alternative to $\Lambda$CDM (although still with a rather low value of the local Hubble parameter), and which ought to be tested and constrained by future observations.

\section*{Note Added} After the submission of the first version of the this manuscript two papers, \cite{Clarkson:2010ej} and \cite{Moss:2010jx}, appeared in close succession which address similar issues. While a detailed comparison of the similarities and differences between the approaches and results of these three papers requires more time, we  make here a few remarks which are already apparent. While we agree with~\cite{Moss:2010jx} that SN+CMB+$H_0$ rule out EdS based void models with simple profiles, we find that Curved void models can be consistent, although the value of $H_0$ is still somewhat low compared to observations. In contrast \cite{Moss:2010jx} concludes that void models can be ruled out based on these three measurements. In our opinion there are four possible reasons for this disagreement: (i) We have an enlarged parameter space where we include background curvature. \cite{Moss:2010jx} has only looked at EdS voids. (ii) We consider ``deep'' voids where the central matter abundance  can be less than 0.1. We did not find any convincing reason to exclude deep voids, which have also been considered previously in literature~\cite{Clifton:2009kx}. (iii) We considered a range of radial profiles. While not all profiles can produce a sufficiently large local Hubble parameter, some apparently can. (iv) Finally, we implemented an integrated MCMC approach. That is, we fit all the relevant data sets simultaneously with MCMC simulations. In contrast in \cite{Moss:2010jx} only SN and CMB were fit using MCMC and other constraints were later applied by means of importance sampling, or compared {\em a posteriori} with the best fit parameters obtained from SN+CMB. When fitting only SN+CMB, the MCMC chain has no interest in going into areas in parameter space that embody a higher value for $H_{0, {\rm in}}$ or give a better fit to the BAO. In our work we find models that fit the CMB worse than $\Lambda$CDM, but fit the BAO better, leading to a net $\chi^2_{\rm eff}$ that is competitive with that of $\Lambda$CDM. When only considering CMB, these parameters are rejected. In other words, the authors of~\cite{Moss:2010jx} have taken a prior on parameter space that reflects the favoured parameters by SN+CMB, and then fit this biased parameter space to $H_0$, BAO and other observables. This does not necessarily lead to the same result as taking a completely flat prior on parameters and then simultaneously fitting CMB+BAO+SN+HST. %
This is probably the reason for having different values for $H_{0,{\rm in}}$ in EdS voids quoted in this work and in~\cite{Moss:2010jx}.
Finally, we agree with~\cite{Moss:2010jx}, that void models may be in conflict with LSS measurements, although an analysis which considers perturbations of LTB would be needed to make firm conclusions, as we have stressed.

The authors of~\cite{Clarkson:2010ej} argue that if the void extends to the surface of last scattering, radiation can no longer be treated as a test field, and its effect on the metric should be included in the calculations. We acknowledge that our analysis and other analyses to date may be oversimplified in that respect.

\section*{Acknowledgments}

We would like to thank Julien Lesgourgues, Suman Bhattacharya, Paul Hunt, Jan Hamann, Yvonne Wong, Luca Amendola, Juan Garcia-Bellido, Morgan Le Delliou and Cinzia di Porto for useful conversations. We also would like to thank Valerio Marra, Adam Moss, Marco Regis, Enea Romano, Douglas Scott and Jim Zibin for useful comments and correspondence after the first version of this paper. Part of the numerical simulations were performed on the MUST cluster at LAPP (CNRS \& Universit\'e de Savoie).

\appendix

\section{Analytical Approximations for Conformal times and Redshifts}
\label{approximations}
LTB metrics form a class of exact solutions of General Relativity, which contains only dust, {\ i.e.} zero pressure source. They contain three arbitrary functions of the radial coordinate $r$: however one of them describes an inhomogeneous big bang time and we are not going to use it here. Another function can be eliminated redefining a new radial coordinate. So, we are left with one physically meaningful arbitrary function, $k(r)$.
The metric functions are known exactly, but only implicitly, and therefore one cannot find closed form analytic expressions for photon trajectories, $t(r),z(r)$, in the LTB metric. Approximate analytical expressions exist in two perturbative regimes, either when (1)  the inhomogeneities extend only a small distance compared to the Hubble radius, or when (2) the LTB space-time is only a small deviation from the EdS space-time. (1) corresponds to a Newtonian expansion which in our model is given by $L\ll r_{\rm hor}$, while (2) is just the traditional cosmological perturbative expansion in small potentials and corresponds to having a small curvature function, $k(r)$.

The perturbative expansions have previously been used to analytically fit the supernovae data for small voids~\cite{ABNV}. For our purpose, these analytical expressions provide important checks on our numerical simulations. Also, they provide physical insight into the various corrections to the redshift and distances in the presence of inhomogeneities.

In the following, we use the definitions $\rb\equiv \tilde M r$ and $ \bar L \equiv \tilde M L$.

\subsection{Next to leading Order Newtonian Approximation:  \texorpdfstring{$\cO(\Lb^2)$}{O(L2)}  \& all orders in  \texorpdfstring{$k$}{k}}\label{app:ana_approx}
In~\cite{ABNV} Newtonian expansions were obtained for voids with  small curvatures everywhere. Here we generalize this to profiles with general $k(r)$ and also compute the corrections up to $\cO(\Lb^2)$ which is essential to computing the monopole correction due the presence of the LTB void.

The general expressions for $R$ and $R'$ can be written as a power series in  $k(r)$:
$$
R={\pi\over 3}r\ga^2\tau^2\LF1+\sum_1^{\infty}R_{2n} \ga^{2n}\tau^{2n} k^n\RF \, ,
$$
\be
R'={\pi\over 3}\tau^2\LT1+\sum_1^{\infty}R_{2n} \ga^{2n}\tau^{2n} (rk^n)'\RT \, ,
\label{Rp-large}
\ee
and
\be
\dot{R}'=2\al{\bM\over\tau}\LT1+\sum_1^{\infty}(n+1)R_{2n} \ga^{2n}\tau^{2n} (rk^n)'\RT \, .
\label{Rpd-large}
\ee

The evolution equation for $\tau$ is then given by (\ref{t-radial}) and (\ref{Rpd-large})
\be
{d\tau\over d\rb}=-{\al\LT1+\sum_1^{\infty}R_{2n} \ga^{2n}\tau^{2n} (\rb k^n)'\RT\over \sqrt{1+2k\rb^2}}\approx -\al\LT1+\sum_1^{\infty}R_{2n} \ga^{2n}\tau^{2n} (rk^n)'\RT+\cO(\rb^2)
\label{tau-series} \, .
 \ee

From (\ref{tau-series}) we find the  expression
\be
\tau=\tau_E+\tau_1=\tau_E-\al\sum_1^{\infty} R_{2n}\ga^{2n}\int d\rb\ \tau^{2n}(\rb k^n)'
\, . \ee

The above integral can be evaluated  as follows:
\begin{eqnarray}
I_n &\equiv&\int d\rb\ \tau^{2n}(\rb k^n)'= \tau^{2n}\rb k^n-2n\int d\rb\ \tau^{2n-1}\rb k^n{d\tau\over d\rb}  \nonumber \, , \\
& \approx & \tau^{2n}\rb k^n+2n\al\int d\rb\ \tau^{2n-1}\rb k^n\LT1+\sum_1^{\infty}R_{2m} \ga^{2m}\tau^{2m} (rk^m)'\RT+\cO(\rb^4) \nonumber \, , \\
& \approx & \tau^{2n}\rb k^n+2n\al\tau_0^{2n-1}\int d\rb\ \rb k^n\LT1+\sum_1^{\infty}R_{2m} \ga^{2m}\tau_0^{2m} (rk^m)'\RT+\cO(\rb^3) \nonumber \, , \\
&=& \tau^{2n}\rb k^n+2n\al\tau_0^{2n-1}\LT\int d\rb\ \rb k^n+\sum_1^{\infty}R_{2m} \ga^{2m}\tau_0^{2m} \int d\rb\ \rb k^n(rk^m)'\RT \, .
\end{eqnarray}

Putting everything together we have
\begin{eqnarray}
\tau_1=-\al\sum_1^{\infty} R_{2n}\ga^{2n}\left(\tau^{2n}\rb k^n+2n\al\tau_0^{2n-1}\LT\int d\rb\ \rb k^n+\sum_1^{\infty}R_{2m} \ga^{2m}\tau_0^{2m} \int d\rb\ \rb k^n(rk^m)'\RT\right) \, . \nonumber \\
\end{eqnarray}
Using the definitions of $f_n$'s  we can condense the above equation to
\bea
\tau_1=-\al\LF\rb f(\tau^2\ga^2k)
+2\al\ga^2\tau_0\int d\rb\ \rb k f_1(\tau_0^2\ga^2k)\LT1+  f(\tau_0^2\ga^2k)+\rb k'\ga^2\tau_0^2f_1(\tau_0^2\ga^2k)\RT\RF \, . \label{eq:tau1_appendix}
\eea
We  cannot simplify the expression any further, but  we should be able to compute it because both the functions $f,f_1$ are known, and so is the profile. Also note that at the boundary where $k=0$, the first term do not contribute.

We are now ready to look at the corrections to the redshift. From (\ref{eq:z-radial}) we have
\be
{dz\over 1+z}={2\al\over\tau}\LT1+\sum_1^{\infty}(n+1)R_{2n} \ga^{2n}\tau^{2n} (rk^n)'\RT d\rb \, ,
\ee
so that
\be
\ln(1+z)=2\al\LT\int_0d\rb\ {1\over \tau}+\int_0d\rb\ \sum_1^{\infty}(n+1)R_{2n} \ga^{2n}\tau^{2n-1} (rk^n)'\RT \, .
\ee
We can evaluate these integrals using similar techniques as for $\tau$:
$$I_0\equiv \int_0d\rb\ {1\over \tau}=\int_0d\rb\ {1\over\tau_E}\LF 1+{\de \tau \over \tau_E}\RF^{-1}=\int_0 {d\rb\ \over\tau_E}+\al\int_0d\rb\ {\rb f(\tau_0^2\ga^2k)\over \tau_0^2}+\cO(\rb^3) \, . $$
Or,
\be
J_0={1\over 2\al}\ln(1+z_E)+{\al\over \tau_0^2}\int_0d\rb\ \rb f(\tau_0^2\ga^2k) \, .
\ee

For the second integral we have
$$J_n\equiv \int_0d\rb\ \tau^{2n-1} (rk^n)'=\tau^{2n-1} \rb k^n+(2n-1)\int_0d\rb\ \tau_0^{2n-2}\rb k^n\al\LT1+\sum_1^{\infty}R_{2m} \ga^{2m}\tau_0^{2m} (rk^m)'\RT+\cO(\rb^3)\, .$$
Putting everything together we have
\bea
&\ln&(1+z)=\ln(1+z_E)+{2\al^2\over \tau_0^2}\int_0d\rb\ \rb f(\tau_0^2\ga^2k)+2\al\sum_1^{\infty}(n+1)R_{2n} \ga^{2n}\tau^{2n-1} \rb k^n\nonumber  \\
&+&2\al^2\tau_0^{2n-2}\sum_1^{\infty}(n+1)(2n-1)R_{2n} \ga^{2n}\int_0d\rb\ \rb k^n\LT1+\sum_1^{\infty}R_{2m} \ga^{2m}\tau_0^{2m} (k^m+m\rb k^{m-1}k')\RT\nonumber \, .
\eea
We can again rewrite the above expression in a more concise form:
\bea
1&+&z=(1+z_E)\exp\left\{ {2\al^2\over \tau_0^2}\int_0d\rb\ \rb f(\tau_0^2\ga^2k)+{2\al\over \tau}[f(\tau^2\ga^2k)+\tau^2\ga^2kf_1(\tau^2\ga^2k)]\right.\nonumber  \\
&+&{2\al^2\over\tau_0^{2}}\int_0d\rb\ \rb [3\tau_0^2\ga^2kf_1(\tau_0^2\ga^2k) +2(\tau_0^2\ga^2k)^2f_2(\tau_0^2\ga^2k)-f(\tau_0^2\ga^2k)]\nonumber \\\\
&\times& [1+f(\tau_0^2\ga^2k)+\rb k' \ga^2\tau_0^2f_1(\tau_0^2\ga^2k)]\Big\}  \label{eq:app_dt_t_1} \, .
\eea
Note that this is a useful
\subsection{ \texorpdfstring{$\cO(k)$}{O(k)} \& all orders in  \texorpdfstring{$\rb$}{rbar}}
In this subsection we will look at the opposite regime, when $L$ can be arbitrarily large but the curvature is always small. Thus, we are going to keep only the linear order terms in $k$.  The metric functions are thus approximated as
$$
R={\pi\over 3}\ga^2r\tau^2(1+R_2 u_0^2)={\pi\over 3}\ga^2r\tau^2(1+R_2 \ga^2\tau^2 k) \, ,
$$
$$
R'={\pi\over 3}\ga^2\tau^2[1+R_2 \ga^2\tau^2 (rk)'] \, ,
$$
and
$$
\dot{R}'={2\pi\over 9}\ga^2{\bM\over\tau}\left[1+2R_2 \ga^2\tau^2 (rk)'\right] \, .
$$

The evolution equation for $\tau$ is then given by (\ref{t-radial})
\be
{d\tau\over dr}=-{{\pi\over 9}\ga^2\bM[1+R_2 \ga^2\tau^2 (rk)']\over \sqrt{1+2k(\bM r)^2}}\approx -\al \bM[1+R_2 \ga^2 \tau^2 (rk)'-k(\bM r)^2]+\cO(k^2)
\, , \ee
 The above equation  leads us to the iterative expression
\be
 \tau=\tau_0-\al\rb-\al\LT R_2 \ga^2\int_0^{\rb}  d\rb\ \tau^2(\rb k)'-\int_0^{\rb} d\rb\ k\rb^2\RT\equiv \tau_E+\tau_1
\, , \ee
where $\tau_1$ now denotes the $\cO(k)$ corrections to the FLRW trajectory coming from the integrals. Let us look at the first integral:
$$
I\equiv\int_0^{\rb}  d\rb\ \tau^2(\rb k)'=\tau^2\rb k-\int_0^{\rb}  d\rb\ 2\tau\rb k{d\tau\over dr}\approx \tau_E^2\rb k+\al\int_0^{\rb}  d\rb\ 2\tau_E\rb k$$
$$=\tau_E^2\rb k+\al\int_0^{\rb}  d\rb\ 2\tau_0\rb k-\al^2\int_0^{\rb}  d\rb\ 2\rb^2 k \, .
$$
If we define the following functions
\be
k_n(r)\equiv \int_0^{\rb} d\rb \rb^nk(r)\ ,
\ee
then we have
\be
I=\tau_E^2\rb k+2\al\tau_0 k_1-2\al^2 k_2 \, .
\ee

and therefore the approximate result
\be
\tau_1=-\al\LT R_2 \ga^2(\tau_E^2\rb k+2\al\tau_0 k_1-2\al^2 k_2)-k_2\RT=-\al\LT R_2 \ga^2(\tau_E^2\rb k+2\al\tau_0 k_1)-6k_2/5 \RT
\, . \ee
We emphasize that the above result is exact in $\rb$.

We can now look at the redshift equation
\be
{dz\over dr}={(1+z)\dot{R}'\over \sqrt{1+2E}}\Ra {dz\over 1+z}=2\al\LT
\tau^{-1}+2R_2\ga^2 \tau (kr)'-k\rb^2\tau^{-1}\RT d\rb \nonumber \\
\, .
\ee
Thus we have
\be
\ln(1+z)=2\al\LT
\int_0^{\rb} d\rb\ \tau^{-1}+2R_2\ga^2 \int_0^{\rb} d\rb\ \tau (kr)'-\int_0^{\rb} d\rb\ k\rb^2\tau^{-1}\RT \, .
\ee
Since we are only interested in $\cO(k)$ corrections, in the second and third integral we can replace $\tau\ra \tau_E$, so that
\be
\ln(1+z)=2\al\LT
\int_0^{\rb} {d\rb\over \tau}+2R_2\ga^2 \int_0^{\rb} d\rb\ \tau_E (kr)'-\int_0^{\rb} d\rb\ {k\rb^2\over\tau_E}\RT\equiv 2\al[J_1+J_2-J_3] \, .
\label{z-integral}
\ee

The first integral simplifies as
\begin{eqnarray}
J_1&\equiv&\int_0^{\rb} {d\rb\over \tau}=\int_0^{\rb} {d\rb\over \tau_E}\LF1-{\tau_1\over\tau_E}\RF+\cO(k^2)=\int_0^{\rb} {d\rb\over \tau_E}-\int_0^{\rb} d\rb{\tau_1\over \tau_E^2}\equiv {1\over 2\al}\ln(1+z_E)-\int_0^{\rb} d\rb{\tau_1\over \tau_E^2}
\nonumber \\
&=&{1\over 2\al}\ln(1+z_E)-{\tau_1\over \al\tau_E}+\int_0^{\rb} {d\rb\over \al\tau_E}{d\tau_1\over d\rb}={1\over 2\al}\ln(1+z_E)-{\tau_1\over \al\tau_E}-\int_0^{\rb} {d\rb\over \tau_E}[R_2\ga^2\tau_E^2(\rb k)'-k\rb^2]
\nonumber \\
&=&{1\over 2\al}\ln(1+z_E)-{\tau_1\over \al\tau_E}-\LT\int_0^{\rb}  d\rb R_2\ga^2\tau_E(\rb k)'-\int_0^{\rb} {d\rb\over \tau_E}k\rb^2\RT
\nonumber \\
&=&{1\over 2\al}\ln(1+z_E)-{\tau_1\over \al\tau_E}- {J_2\over 2}+J_3 \, .
\nonumber
\end{eqnarray}

To summarize, we have
\be
\ln(1+z)=\ln(1+z_E)-2{\tau_1\over \tau_E}+\al J_2 \, .
\ee

Now, $J_2$ can be simplified as follows:
$$
J_2\equiv 2R_2\ga^2 \int_0^{\rb} d\rb\ \tau_E (kr)'=2R_2\ga^2 [ \tau_E k\rb+\al\int_0^{\rb} d\rb\ k\rb]=2R_2\ga^2 [ \tau_E k\rb+\al k_1(\rb)] \, .
$$
Putting everything together we have the final formula for the redshift:
\be
1+z=(1+z_E)\exp\LT -{2\tau_1\over \tau_E}+ 2\al R_2\ga^2\tau_E\rb k(\rb)+2\al^2 R_2\ga^2k_1(\rb)\RT \label{eq:app_dt_t_2} \, .
\ee

\subsection{Comparison between analytical approximations and numerics}
\FIGURE{
\begin{tabular}{lr}
\includegraphics[width=0.5\textwidth]{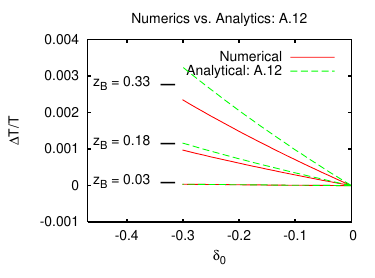}&
\includegraphics[width=0.5\textwidth]{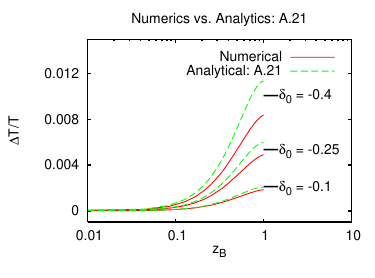}
\end{tabular}
\caption{A comparison of numerical results with analytical approximations. {\em Right:} A comparison with Eq.~\eqref{eq:app_dt_t_1}, keeping $z_B$ constant and varying $\delta_0$. For small $z_B$ there is perfect agreement. {\em Left:} A comparison with Eq.~\eqref{eq:app_dt_t_2}, keeping $\delta_0$ constant and varying $z_B$. Again, this time for small $\delta_0$, there is perfect agreement.}\label{fig:app_dt_t_2}
}

Since at all times the metric functions and their derivatives are defined analytically by Eqs.~(\ref{r-hubble},\ref{scale-factor},\ref{time}), for any pair $\{r,t\}$ we can calculate these functions up to very high accuracy by numerically inverting Eq.~\eqref{time}. The only less controllable calculation is the integration of the geodesic equations, Eqs.~(\ref{t-radial},\ref{eq:z-radial}), backwards in time starting from an observer at $r=0,t=t_0$.

We have checked that the numerical code confirms the analytical approximations in the above-mentioned regimes. In figure~\ref{fig:app_dt_t_2} we see $\Delta T / T \equiv 1 - T_{\rm CMB, in} / T_{\rm CMB, eff}$, where $T_{\rm CMB, in}=2.726$ K and $T_{\rm CMB, eff}$ is the value used for the calculation of the CMB power spectrum. On the left, a comparison with Eq.~\eqref{eq:app_dt_t_1} is made. For three values of $z_B$, we show the difference as a function of increasing $\delta_0$. For tiny $z_B$ the difference is invisible, and for each of the three cases the ratio between the analytical and numerical result is a constant with respect to varying $\delta_0$, indicating that the discrepancy is dependent only on $z_B$. This was to be expected, as  Eq.~\eqref{eq:app_dt_t_1} is at all orders in $k$.

On the right in Figure~\ref{fig:app_dt_t_2} we compare the numerical results with those from Eq.~\eqref{eq:app_dt_t_2}. In this case we calculated $\Delta T / T$ for three fixed values of $\delta_0$, as a function of a varying $z_B$. Again, for small values of the expansion parameter, in this case small $k_{max}$ (hence small $\delta_0$), the numerics and the analytics perfectly agree, and the relative (dis)agreement does not change with changing $z_B$, as Eq.~\eqref{eq:app_dt_t_2} is at all orders in $r/L$.

\section{Determining the CMB temperature  \texorpdfstring{$T_{\rm{eff}}$}{Teff} for an uncompensated void\label{app:t_ast_CMB}}

The exact relation between the observed CMB temperature and the CMB temperature to feed in {\sc camb} is defined in equation~(\ref{eff-obs}), which we repeat here slightly rewritten,
\be	
T_{dec} = T_{\rm{eff}} (1 + z^{\rm{dec}}_{\rm{eff}}) = T_{\rm{obs}} (1 + z^{\rm{dec}}_{\rm{obs}}).
\ee
Let $_{\rm{eff}}$ denote all quantities in the metric of the effective FLRW-observer, and $_{\rm{obs}}$ denote all quantities in the metric of the real observer at the centre of the void.
Then,
\begin{align}
 T_{\rm{eff}} = &T_{\rm{obs}} \frac{1 + z^{\rm{dec}}_{\rm{obs}}}{ 1 + z^{\rm{dec}}_{\rm{eff}}}\nonumber\\
 = & T_{\rm{obs}} \frac{1 + z^{\rm{dec}}_{\rm{obs}}}{ 1 + z^{\rm{dec}}_{\rm{eff}}}\frac{1 + z^{*}_{\rm{obs}}}{ 1 + z^{*}_{\rm{eff}}}\frac{1 + z^{*}_{\rm{eff}}}{ 1 + z^{*}_{\rm{obs}}} \,\, .
\end{align}
Here $z^{*}$ denotes redshift $z$ evaluated on the photon geodesic at time $t_{*}$. The case where the metric outside of the void is exactly FLRW, we had $ \frac{1 + z^{\rm{dec}}_{\rm{obs}}}{ 1 + z^{\rm{dec}}_{\rm{eff}}}\frac{1 + z^{*}_{\rm{obs}}}{ 1 + z^{*}_{\rm{eff}}} = \frac{ a_{dec}/a_{*}}{a_{dec}/a_{*}}=1$. For the void that only asymptotically goes to FLRW, we have
\begin{align}
 \frac{1 + z^{\rm{dec}}_{\rm{obs}}}{ 1 + z^{\rm{dec}}_{\rm{eff}}}\frac{1 + z^{*}_{\rm{eff}}}{ 1 + z^{*}_{\rm{obs}}} = 1+\epsilon,
\end{align}
and
\begin{align}
T_{\rm{eff}}  = & T_{\rm{obs}} \frac{1 + z^{*}_{\rm{obs}}}{ 1 + z^{*}_{\rm{eff}}} (1 + \epsilon),
\end{align}
with $\lim_{t_* \rightarrow t_{dec}} \epsilon = 0$. In the LTB metric, the geodesic equation describing the redshift can be written as
\be
\frac{dz}{dt} = \frac{\dot S(r,t)}{S(r,t)}(1+z),
\ee
such that
\begin{align}
	\frac{1}{1+z_{\rm{(eff,obs)}}(t)} =&1- \int_{t_0}^{t}\frac{dt'}{ 1+z_{\rm{(eff,obs)}}(t') } \frac{\dot S_{\rm{(eff,obs)}}(r_{\rm{(eff,obs)}}(t'),t')}{S_{\rm{(eff,obs)}}(r_{\rm{(eff,obs)}}(t'),t')},\\
	\frac{1 + z^{*}_{\rm{obs}}}{ 1 + z^{\rm{dec}}_{\rm{obs}}}=& (1 + z^{*}_{\rm{obs}}) \left[  \frac{1}{1+z^{*}_{\rm{obs}}} -   \int_{t_*}^{t_{dec}}\frac{dt'}{ 1+z_{\rm{obs}}(t') } \frac{\dot S_{\rm{obs}}(r_{\rm{obs}}(t'),t')}{S_{\rm{obs}}(r_{\rm{obs}}(t'),t')}  \right]\nonumber\\
	=& 1  -   \int_{t_*}^{t_{dec}}dt' \left[ \frac{1 + z^{*}_{\rm{obs}}}{ 1+z_{\rm{obs}}(t') } \frac{\dot S_{\rm{obs}}(r_{\rm{obs}}(t'),t')}{S_{\rm{obs}}(r_{\rm{obs}}(t'),t')}  - \frac{1 + z^{*}_{\rm{eff}}}{ 1+z_{\rm{eff}}(t') } \frac{\dot S_{\rm{eff}}(t')}{S_{\rm{eff}}(t')}  + \frac{1 + z^{*}_{\rm{eff}}}{ 1+z_{\rm{eff}}(t') } \frac{\dot S_{\rm{eff}}(t')}{S_{\rm{eff}}(t')}  \right]\nonumber\\
	=& \frac{1 + z^{*}_{\rm{eff}}}{ 1 + z^{\rm{dec}}_{\rm{eff}}}  -   \int_{t_*}^{t_{dec}}dt' \left[ \frac{1 + z^{*}_{\rm{obs}}}{ 1+z_{\rm{obs}}(t') } \frac{\dot S_{\rm{obs}}(r_{\rm{obs}}(t'),t')}{S_{\rm{obs}}(r_{\rm{obs}}(t'),t')}  - \frac{1 + z^{*}_{\rm{eff}}}{ 1+z_{\rm{eff}}(t') } \frac{\dot S_{\rm{eff}}(t')}{S_{\rm{eff}}(t')}  \right].\\
\end{align}
Therefore	
\begin{align}
\epsilon =  \int_{t_*}^{t_{dec}}dt' \left[ \frac{1 + z^{*}_{\rm{obs}}}{ 1+z_{\rm{obs}}(t') } \frac{\dot S_{\rm{obs}}(r_{\rm{obs}}(t'),t')}{S_{\rm{obs}}(r_{\rm{obs}}(t'),t')}  - \frac{1 + z^{*}_{\rm{eff}}}{ 1+z_{\rm{eff}}(t') } \frac{\dot S_{\rm{eff}}(t')}{S_{\rm{eff}}(t')}  \right],\\
\end{align}
which obviously goes to zero for $t_* \rightarrow t_{dec}$. Since the void by construction converges to FLRW at $r\rightarrow \infty$, the function $ \frac{1 + z^{*}_{\rm{obs}}}{ 1+z_{\rm{obs}}(t') } \frac{\dot S_{\rm{obs}}(r_{\rm{obs}}(t'),t')}{S_{\rm{obs}}(r_{\rm{obs}}(t'),t')}  - \frac{1 + z^{*}_{\rm{eff}}}{ 1+z_{\rm{eff}}(t') } \frac{\dot S_{\rm{eff}}(t')}{S_{\rm{eff}}(t')} $ must be monotonically decreasing with time, and the integral can be estimated as
\begin{align}
\left|\epsilon \right| \le &  \left| \int_{t_*}^{t_{dec}}dt' \left[ \frac{1 + z^{*}_{\rm{obs}}}{ 1+z_{\rm{obs}}(t') } \frac{\dot S_{\rm{obs}}(r_{\rm{obs}}(t'),t')}{S_{\rm{obs}}(r_{\rm{obs}}(t'),t')}  - \frac{1 + z^{*}_{\rm{eff}}}{ 1+z_{\rm{eff}}(t') } \frac{\dot S_{\rm{eff}}(t')}{S_{\rm{eff}}(t')}  \right] \right|,\nonumber\\
 = & \left| t_* \times \left[ \frac{1 + z^{*}_{\rm{obs}}}{ 1+z_{\rm{obs}}(t') } \frac{\dot S_{\rm{obs}}(r_{\rm{obs}}(t'),t')}{S_{\rm{obs}}(r_{\rm{obs}}(t'),t')}  - \frac{1 + z^{*}_{\rm{eff}}}{ 1+z_{\rm{eff}}(t') } \frac{\dot S_{\rm{eff}}(t')}{S_{\rm{eff}}(t')}  \right]_{t' = t_*}\right|,\nonumber\\
 = & \left| t_* \times \left[  \frac{\dot S_{\rm{obs}}(r^V_*,t_*)}{S_{\rm{obs}}(r^V_*,t_*)}  -\frac{\dot S_{\rm{eff}}(t_*)}{S_{\rm{eff}}(t_*)}  \right]\right|.\label{eq:app_tcmb_epsilon}
\end{align}
We dropped $r$ in $\frac{\dot S_{\rm{eff}}(t_*)}{S_{\rm{eff}}(t_*)} $, as this ratio is independent of $r$.
With equation~(\ref{eq:app_tcmb_epsilon}) we can integrate from the observer backwards in time until a time $t_*$ for which $\epsilon$ is a few orders of magnitude smaller than $1 -  \frac{1 + z^{*}_{\rm{obs}}}{ 1 + z^{*}_{\rm{eff}}} $, such that up to high accuracy $T_{\rm{eff}}  =  T_{\rm{obs}} \frac{1 + z^{*}_{\rm{obs}}}{ 1 + z^{*}_{\rm{eff}}} $. It is trivial to see that $\epsilon =0$ for voids embedded inside FLRW, at a time $t_*$ for which the photon is outside the void.

It is also at this coordinate $\{ r^*_{\rm{obs}}, t^*_{\rm{obs}}\}$ that we can pretend the effective FLRW to start, in order to determine the cosmological parameters $\Omega^{\rm eff}_i$ describing the FLRW cosmology of the CMB observer, as explained in section~\ref{sec:technicalresults}.


\section{The goodness of fit of a void model to a supernova observation}\label{app:snchi2}
\subsection{Likelihood of a multivalued function}

The quantities that are measured from each supernova in a supernova survey, are its redshift $z$ and its angular diameter distance $d_{\rm A}$, or equivalently the luminosity distance $d_{\rm L}=(1+z)^2 d_{\rm A}$. The theoretical function that is fit to the data is hence $d_{\rm A}(z)$. In an FLRW-universe, both redshift and time vary monotonically along a photon geodesic from the supernova to the observer, so there is no ambiguity when exchanging $t$, $r$ and $z$ as time parameters. For each value of $z$, there is a unique combination of $\{t,r\}$ which determines the angular diameter distance to the observer as $d_{\rm A}=R(r,t)$.

Inside the mass-compensated void however, the photon usually experiences a blueshift when crossing the dense shell that surrounds the void, as shown in the left panel in Fig.~\ref{fig:da3p_ex}. This implies that even though coordinate distance $r$ and cosmic time $t$ vary monotonically along the geodesic, $z$ no longer does so. The same value $z$ is met up to three times along the same geodesic. As a consequence, for one value of redshift $z$, the theory predicts three possible outcomes for $d_{\rm A}(z)$, as shown in the right panel of Fig.~\ref{fig:da3p_ex}.

\FIGURE{ 
\includegraphics[width=0.499\textwidth]{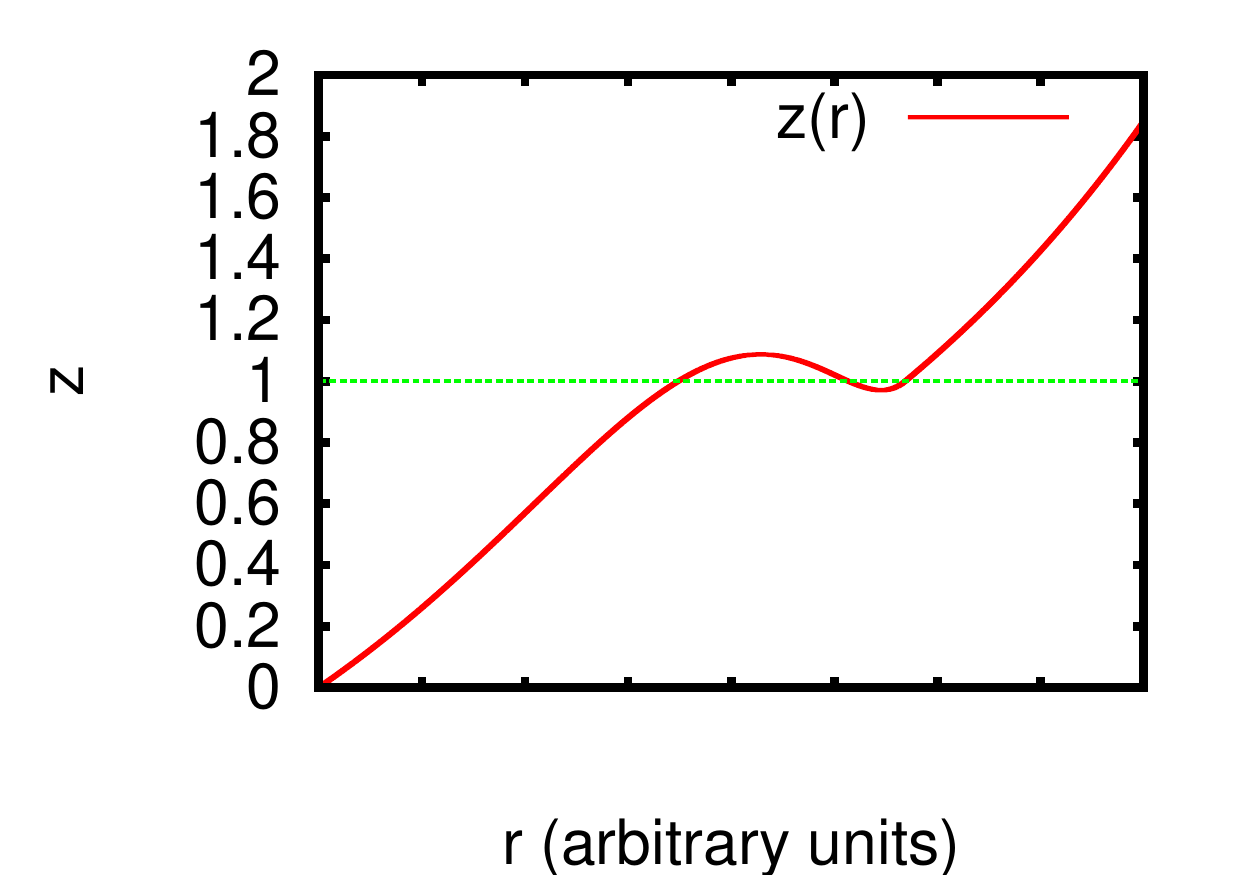}
\includegraphics[width=0.499\textwidth]{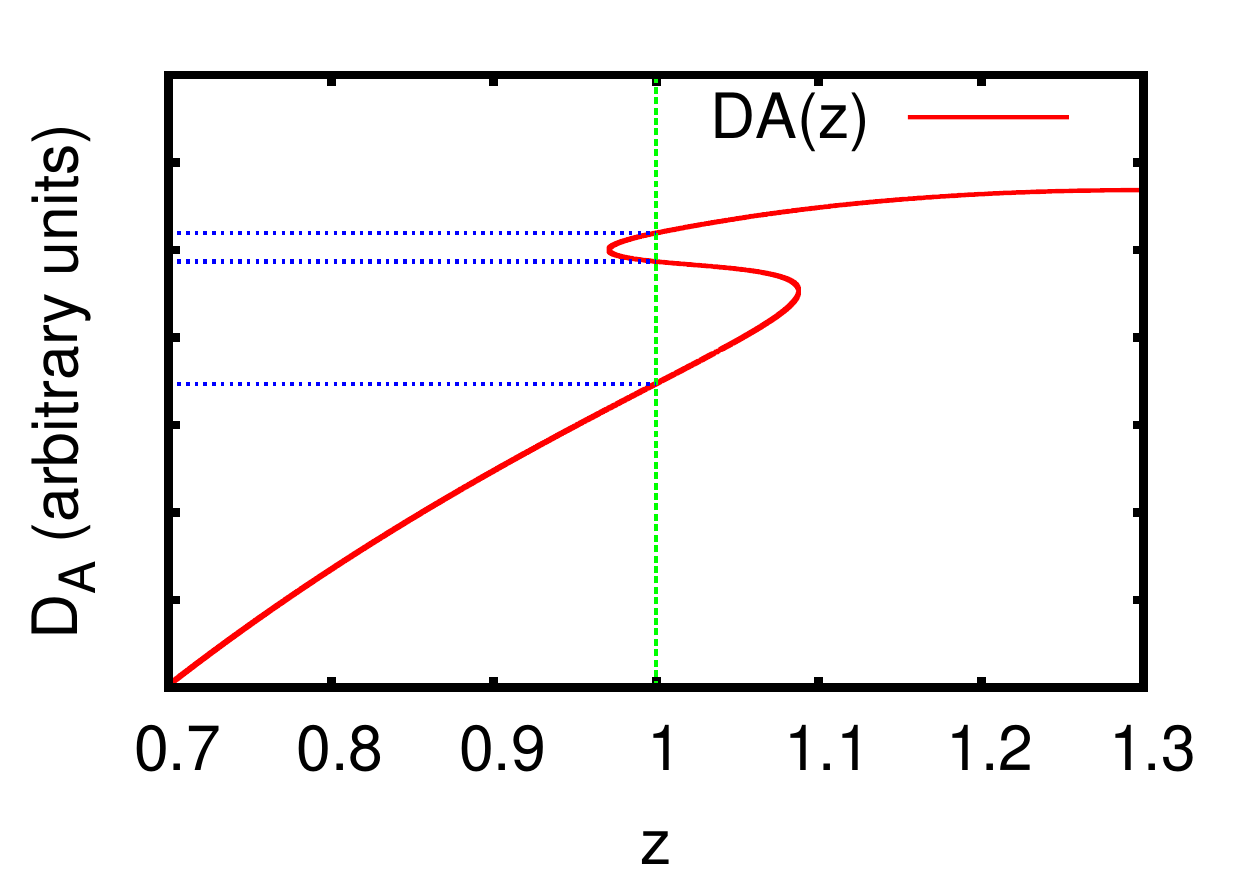}
\caption{{\em Left:} Redshift $z$ as a function of coordinate distance $r$ along a photon geodesic pointing from an observer ($r=0$) to a supernova, for a void with $\delta_0=-0.8$ and $z_b$=1. The photon experiences a blueshift when crossing the shell around $z=1$. The definition of the coordinate size $L$ of the void is such that the highest value of $r$ for which $z(r)=z_b$ is equal to $r=L$, hence the rightmost $r$ in this figure for which $z$ crosses the value of $z_b$. {\em Right:} The angular diameter distance $d_{\rm A}(z)$ for the same model, exemplifying the multiple theoretical predictions for the value of $d_{\rm A}$ corresponding to one value of $z$.  }\label{fig:da3p_ex}
} 

Let us for consider a single data point for the moment. To construct the likelihood of an observation given a model $M$, we need to calculate the probability of measuring a value $d_i$ when $M$ is the underlying model, $p(d_i|M)$. For a gaussian measurement error at measurement $i$, the probability of measuring a certain value $d_i$ given a certain theoretical prediction $x_i$
\begin{align}
	p(d_i|x_i) =  \frac{1}{\sqrt{2\pi} \sigma_i} \exp \left[ -\frac{1}{2} \frac{(x_i - d_i)^2}{\sigma_i^2} \right].\label{eq:chi2_pdx}
\end{align}
Taking into account that the theory may predict a certain spread in $x_i$, such that for each $x_i$ there is a probabiliy $p(x_i|M)$, the probability $p(d_i|M)$ becomes
\begin{align}
	p(d_i|M) = \int dx\,\, p(x_i|M) \frac{1}{\sqrt{2\pi} \sigma_i} \exp \left[ -\frac{1}{2} \frac{(x_i - d_i)^2}{\sigma_i^2} \right].\label{eq:chi2_pdM}
\end{align}

If the theory predicts only one possible outcome $x_i$, we simply have $p(x_i|M)=\delta(x_i-x_{\rm predicted})$, such that Eq.~(\ref{eq:chi2_pdx}) and Eq.~(\ref{eq:chi2_pdM}) are identical, exchanging $x_i$ and $x_{\rm predicted}$. In our scenario however, for certain data points the theory predicts $N$ values, hence
\begin{align}
	p(x_i|M) = \sum_{j=1}^N c_{ij} \delta(x_i-x_{ij})\, ,
\end{align}
normalized such that $\sum_{j=1}^N c_{ij} =1$. The indices $ij$ denote the $j$th prediction at position $i$. For example, in our scenario, $x_{ij}$ is the $j$th angular diameter distance predicted by the theory for a redshift $z_i$, and $c_{ij}$ is its corresponding weight.
The likelihood of the observation $d_i$ then becomes
\begin{align}
	p(d_i|M) = \sum_{j=1}^N \frac{c_{ij}}{\sqrt{2\pi} \sigma_i} \exp \left[ -\frac{1}{2} \frac{(x_{ij} - d_i)^2}{\sigma_i^2} \right].\label{eq:chi2_pdM2}
\end{align}

For multiple data points, the likelihood $\mathcal{L}$ of the data give the model is given by
\begin{align}
	\mathcal{L}=\prod_{i} p(d_i|M)\, .
\end{align}
For single valued predictions, carrying a subscript $s$ for clarity, this reduces to the well known $\chi^2$ statistic,
\begin{align}
	\mathcal{L}_s &=  \frac{1}{({2\pi})^{\frac{N}{2}}\prod_i^N\sigma_i } \exp \left[ -\frac{1}{2} \sum_i^N \frac{(x_i - d_i)^2}{\sigma_i^2} \right],\nonumber\\
		&= C_1 e^{ -\frac{1}{2} \chi^2},\label{eq:chi2_simpleLike}
\end{align}
where $C_1=\frac{1}{({2\pi})^{\frac{N}{2}}\prod_i^N\sigma_i } $. In our scenario, the likelihood becomes,
\begin{align}
\mathcal{L} &=  \frac{1}{({2\pi})^{\frac{N}{2}}} \prod_i^N\frac{1}{\sigma_i } 	\left( \sum_{j=1}^N {c_{ij}} \exp \left[ -\frac{1}{2} \frac{(x_{ij} - d_i)^2}{\sigma_i^2} \right] \right)\label{eq:chi2_like} \\
&\equiv  C_1 e^{ -\frac{1}{2} \chi_{\rm eff}^2} \, ,
\end{align}
where the last line defines the effective $\chi_{\rm eff}^2 \equiv -2 \ln \mathcal{L} + 2 \ln C_1$.

\subsection{Marginalization over  \texorpdfstring{$H_0$}{H0}}
In the results of supernova observations, the normalization is unknown~\cite{Kessler:2009ys}. The normalization is set by choosing a value for $H_0$, such that a dimensionless $H_0 d_{\rm A}$ is obtained. In a Markov Chain Monte Carlo analysis (MCMC), the normalization of the supernovae would influence the location of the best fit point in parameter space, thereby falsely interfering with the preferred values of $H_0$ of other datasets in the same analysis, and falsely constraining $H_0$ around a certain value. Therefore we cannot ignore this effect during the analysis and marginalize over $H_0$ after the MCMC has finished, but we have to marginalize over $H_0$ already for the evaluation of the likelihood of each individual choice of parameters. This is well known ~\cite{Bridle:2001zv, Lewis:2002ah, Gong:2007wx}, and is implemented in {\sc cosmomc}~\cite{Lewis:2002ah} by means of analytical marginalization. As in our scenario the likelihood function is different from the simple $\chi^2$ function, also the marginalization changes.

The analytical marginalization for the single valued case, with a Jeffreys prior on $H_0$, for gaussian errors in the distance modulus $\mu \equiv 5 \log_{10}\left[ (1+z)^2 H_0 d_{\rm A}(z) \right] + 25$ leads to
\begin{align}
 \mathcal{L}_{s, {\rm marg}} =& \int  dx \, \mathcal{L}_{s} \\
 =& \exp \left[ -\frac{1}{2}\sum_i \frac{(\mu_{i,{\rm{obs}}} - \mu_{i, {\rm th}})^2}{\sigma_i^2} + \frac{1}{2}\left(\sum_i \frac{\mu_{i,{\rm{obs}}} - \mu_{i, {\rm th}}}{\sigma_i^2}\right)^2\left( \sum_i \frac{1}{\sigma_i^2}\right)^{-1} \right.\\
   &\left.+ \ln 2\pi   - \frac{1}{2} \ln \sum_i \frac{1}{\sigma_i^2} \right],\label{eq:chi2_marg}
\end{align}
with $x = 5 \log_{10} H_0$.

In the multivalued scenario, the product of the sum in the likelihood in Eq.~(\ref{eq:chi2_like}) can be expanded, such that each term is exactly of the form of Eq.~(\ref{eq:chi2_simpleLike}) and the marginalization integral can be performed for each term individually. However, if the theory predicts for example $N$ values for $M$ data points, this implies evaluating the relevant terms in Eq~(\ref{eq:chi2_marg}) $N^M$ times, which becomes already a huge number for few datapoints. In practice this is impossible to do, and we resorted to a numerical evaluation of the marginalization integral. Taking into account that for the supernovae alone the likelihood is of the order $-\ln\mathcal{L} \sim 120$, a rescaling had to be performed in the numerical integration. As it turns out, the numerical marginalization of the likelihood takes only $\mathcal{O}(10^{-3})$ s, where evaluating the $N^M$ terms could take forever.


\clearpage
\section{Figures and Tables}
\label{figurestables}

\FIGURE{ 
\includegraphics[width=0.95\textwidth]{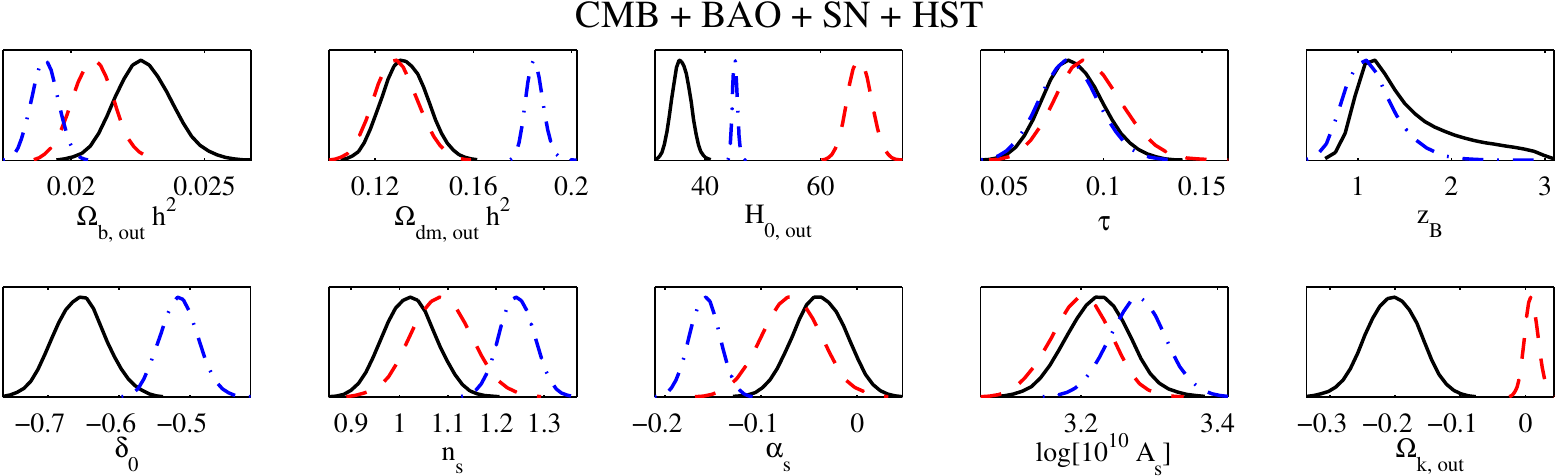}
\caption{The marginalized 1D posterior probabilities of the free parameters, on all of which we had a flat prior, when fitting to CMB+BAO+SN+HST. $\Lambda$CDM in dashed, red, EdS Void in dashed-dotted, blue, and Curved Void in solid, black. Only $\Omega_{k, {\rm out}}$ is a derived parameter in the Void scenario, but since it is a free parameter for $\Lambda$CDM, we do plot it here for comparison. Note that our choice $k_{\rm pivot}=0.05$ Mpc/$h$ affects the central values of the scalar tilt and its running, which has no consequences.}\label{fig:marge_prim_CBHS}
} 

\FIGURE{ 
\includegraphics[width=0.6\textwidth]{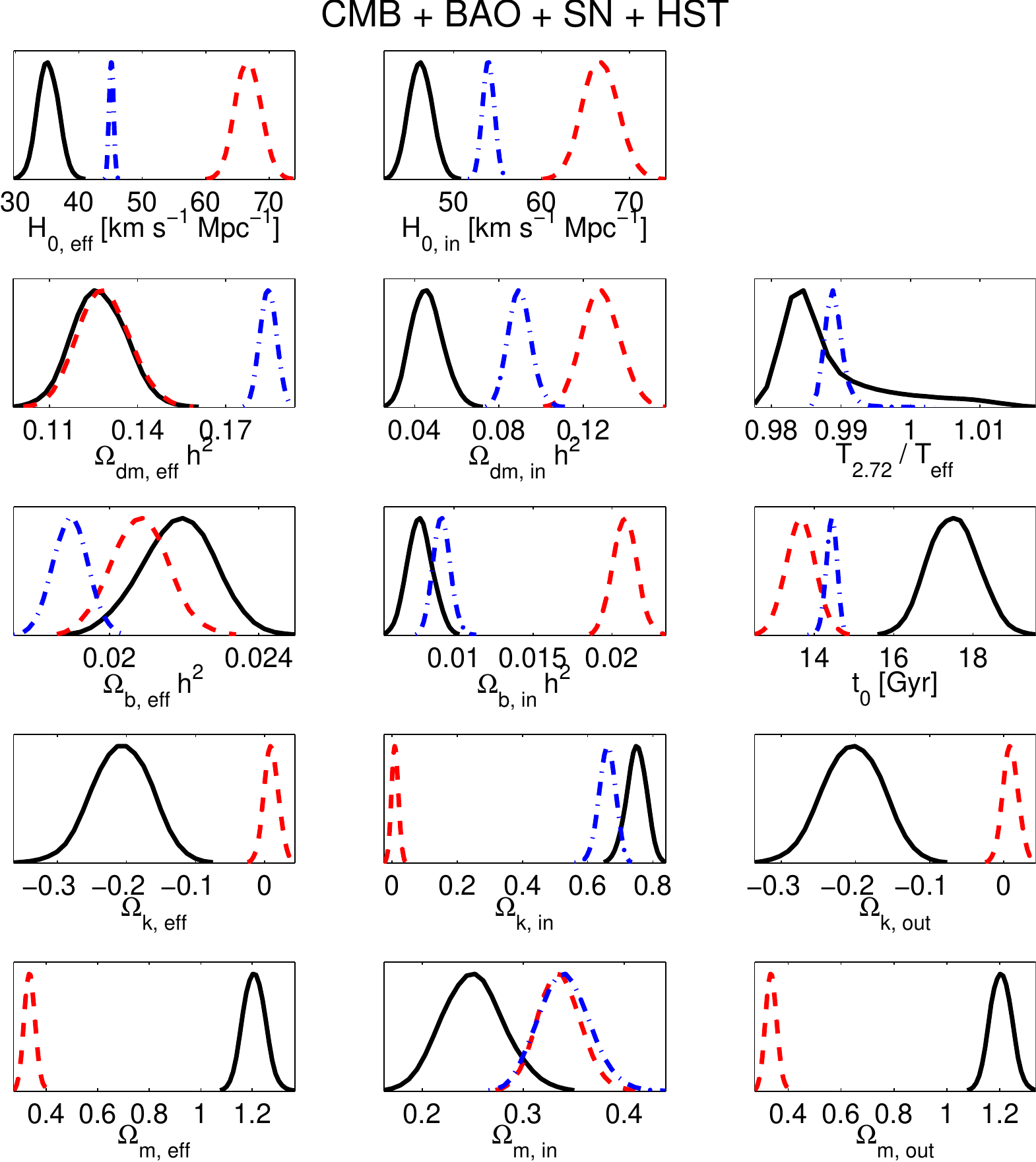}
\caption{The marginalized 1D posterior probabilities of derived parameters, when fitting to CMB+BAO+SN+HST. $\Lambda$CDM in dashed, red, EdS Void in dashed-dotted, blue, and Curved Void in solid, black. In $\Lambda$CDM we have $H_{0, {\rm in}}=H_{0, {\rm out}}$.  The variable $H_{0, {\rm in}}$ is fit to the HST value. }\label{fig:marge_sec_CBHS}
} 

\renewcommand\arraystretch{\myarraystretch}
\TABLE{ 
\resizebox{\textwidth}{!}{
\begin{tabular}{l || l | l || l | l || l | l}
\multicolumn{7}{c}{CMB+BAO+SN+HST}\\ \hline
Parameter & \multicolumn{2}{c||}{$\Lambda$CDM} &  \multicolumn{2}{c||}{Curved Void } &  \multicolumn{2}{c}{EdS Void}\\
\hline
$ \Omega_{b, {\rm out}} h^2 $ & $ 2.06 \times 10^{-2} $  &  $ 2.10^{+0.15}_{-0.14} \times 10^{-2} $ & $ 2.22 \times 10^{-2} $  &  $ 2.38^{+0.28}_{-0.25} \times 10^{-2} $ & $ 1.94 \times 10^{-2} $  &  $ 1.95^{+0.10}_{-0.10} \times 10^{-2} $ \\
$ \Omega_{dm, {\rm out}} h^2 $ & $ 0.126 $  &  $ 0.128^{+0.018}_{-0.017} $ & $ 0.138 $  &  $ 0.143^{+0.020}_{-0.020} $ & $ 0.185 $  &  $ 0.188^{+0.011}_{-0.007} $ \\
$ H_{0, {\rm out}} $ {\tiny [km s$^{-1}$ Mpc$^{-1}$]}$ $ & $ 66.3 $  &  $ 66.7^{+4.2}_{-4.0} $ & $ 37.2 $  &  $ 37.4^{+3.1}_{-3.1} $ & $ 45.3 $  &  $ 45.5^{+1.3}_{-0.9} $ \\
$ \tau $ & $ 8.69 \times 10^{-2} $  &  $ 9.24^{+3.81}_{-3.15} \times 10^{-2} $ & $ 8.32 \times 10^{-2} $  &  $ 8.55^{+3.55}_{-2.88} \times 10^{-2} $ & $ 7.93 \times 10^{-2} $  &  $ 8.25^{+3.11}_{-2.68} \times 10^{-2} $ \\
$ \Omega_{k, {\rm out}} $ & $ 7.27 \times 10^{-3} $  &  $ 7.95^{+20.44}_{-20.42} \times 10^{-3} $ & $ -0.162 $  &  $ -0.194^{+0.075}_{-0.081} $ &  -  &  - \\
$ n_{{\rm S}} $ & $ 1.07 $  &  $ 1.08^{+0.12}_{-0.11} $ & $ 1.06 $  &  $ 1.04^{+0.10}_{-0.11} $ & $ 1.22 $  &  $ 1.24^{+0.07}_{-0.07} $ \\
$ \alpha_{{\rm S}} $ & $ -6.27 \times 10^{-2} $  &  $ -6.38^{+5.19}_{-5.79} \times 10^{-2} $ & $ -5.84 \times 10^{-2} $  &  $ -4.75^{+5.59}_{-4.95} \times 10^{-2} $ & $ -0.146 $  &  $ -0.157^{+0.029}_{-0.030} $ \\
$ \log[10^{10} A_{{\rm S}}] $ & $ 3.21 $  &  $ 3.20^{+0.08}_{-0.08} $ & $ 3.22 $  &  $ 3.23^{+0.09}_{-0.09} $ & $ 3.29 $  &  $ 3.28^{+0.08}_{-0.08} $ \\
$ z_B $ &  -  &  - & $ 1.02 $  &  $ 1.98^{+1.02}_{-1.01} $ & $ 1.10 $  &  $ 1.30^{+0.74}_{-0.42} $ \\
$ \delta_0 $ &  -  &  - & $ -0.743 $  &  $ -0.757^{+0.078}_{-0.055} $ & $ -0.636 $  &  $ -0.630^{+0.063}_{-0.049} $ \\
\hline$ \Omega_\Lambda $ & $ 0.659 $  &  $ 0.658^{+0.046}_{-0.051} $ &  -  &  - &  -  &  - \\
$ \Omega_{m, {\rm out}} $ & $ 0.333 $  &  $ 0.334^{+0.044}_{-0.038} $ & $ 1.16 $  &  $ 1.19^{+0.08}_{-0.08} $ & $ 1.00 $  &  $ 1.00 $ \\
$ H_{0, {\rm in}} $ {\tiny [km s$^{-1}$ Mpc$^{-1}$]}$ $ & $ 66.3 $  &  $ 66.7^{+4.2}_{-4.0} $ & $ 49.5 $  &  $ 50.3^{+3.2}_{-3.6} $ & $ 56.5 $  &  $ 56.7^{+1.7}_{-1.8} $ \\
$ T_{2.72} / T_{{\rm eff}} $ & $ 1.00 $  &  $ 1.00 $ & $ 0.980 $  &  $ 0.997^{+0.027}_{-0.019} $ & $ 0.985 $  &  $ 0.986^{+0.007}_{-0.003} $ \\
$ t_0 $ [Gyr]  & $ 13.8 $  &  $ 13.7^{+0.7}_{-0.7} $ & $ 17.0 $  &  $ 16.8^{+1.3}_{-1.1} $ & $ 14.4 $  &  $ 14.3^{+0.3}_{-0.4} $ \\
$ H_{0, {\rm eff}} $ {\tiny [km s$^{-1}$ Mpc$^{-1}$]}$ $ & $ 66.3 $  &  $ 66.7^{+4.2}_{-4.0} $ & $ 37.2 $  &  $ 35.1^{+3.6}_{-3.6} $ & $ 45.3 $  &  $ 45.2^{+0.7}_{-0.7} $ \\
$ \Omega_{b, {\rm eff}} h^2 $ & $ 2.06 \times 10^{-2} $  &  $ 2.10^{+0.15}_{-0.14} \times 10^{-2} $ & $ 2.22 \times 10^{-2} $  &  $ 2.11^{+0.25}_{-0.22} \times 10^{-2} $ & $ 1.94 \times 10^{-2} $  &  $ 1.92^{+0.09}_{-0.10} \times 10^{-2} $ \\
$ \Omega_{dm, {\rm eff}} h^2 $ & $ 0.126 $  &  $ 0.128^{+0.018}_{-0.017} $ & $ 0.138 $  &  $ 0.127^{+0.020}_{-0.020} $ & $ 0.185 $  &  $ 0.185^{+0.005}_{-0.005} $ \\
$ \Omega_{k, {\rm eff}} $ & $ 7.27 \times 10^{-3} $  &  $ 7.95^{+20.44}_{-20.42} \times 10^{-3} $ & $ -0.162 $  &  $ -0.204^{+0.083}_{-0.094} $ &  -  &  - \\
$ \Omega_{m, {\rm eff}} $ & $ 0.333 $  &  $ 0.334^{+0.044}_{-0.038} $ & $ 1.16 $  &  $ 1.20^{+0.09}_{-0.08} $ & $ 1.00 $  &  $ 1.00 $ \\
$ \Omega_{b, {\rm in}} h^2 $ & $ 2.06 \times 10^{-2} $  &  $ 2.10^{+0.15}_{-0.14} \times 10^{-2} $ & $ 1.01 \times 10^{-2} $  &  $ 1.04^{+0.23}_{-0.17} \times 10^{-2} $ & $ 1.10 \times 10^{-2} $  &  $ 1.12^{+0.14}_{-0.12} \times 10^{-2} $ \\
$ \Omega_{dm, {\rm in}} h^2 $ & $ 0.126 $  &  $ 0.128^{+0.018}_{-0.017} $ & $ 6.30 \times 10^{-2} $  &  $ 6.29^{+1.80}_{-1.57} \times 10^{-2} $ & $ 0.105 $  &  $ 0.108^{+0.015}_{-0.012} $ \\
$ \Omega_{k, {\rm in}} $ & $ 7.27 \times 10^{-3} $  &  $ 7.95^{+20.44}_{-20.42} \times 10^{-3} $ & $ 0.702 $  &  $ 0.710^{+0.055}_{-0.095} $ & $ 0.636 $  &  $ 0.630^{+0.049}_{-0.063} $ \\
$ \Omega_{m, {\rm in}} $ & $ 0.333 $  &  $ 0.334^{+0.044}_{-0.038} $ & $ 0.298 $  &  $ 0.290^{+0.095}_{-0.055} $ & $ 0.364 $  &  $ 0.370^{+0.063}_{-0.049} $
\end{tabular}}
\caption{The best fit parameters (each left column) and marginalized posterior probabilities (each right column) with $95\%$ C.L. errors  of all free parameters (top ten parameters) and derived parameters (from the eleventh parameter down), for the three models at stake when simultaneously fitting CMB + BAO + SN + HST. In the Curved Void-model, the parameters $\Omega_k$ is actually a derived parameter, as we take a flat prior on $\Omega_{dm, {\rm out}}$ in that case (see Table~\ref{tab:priors}). Let us again emphasize that our pivot scale $k_{\rm pivot}=0.05$ Mpc / $h$, which leads to the lower value of $n_{\rm s}$, since we allow for a running.}\label{tab:CBHS}
} 
\renewcommand\arraystretch{1}

\FIGURE{ 
\includegraphics[width=0.95\textwidth]{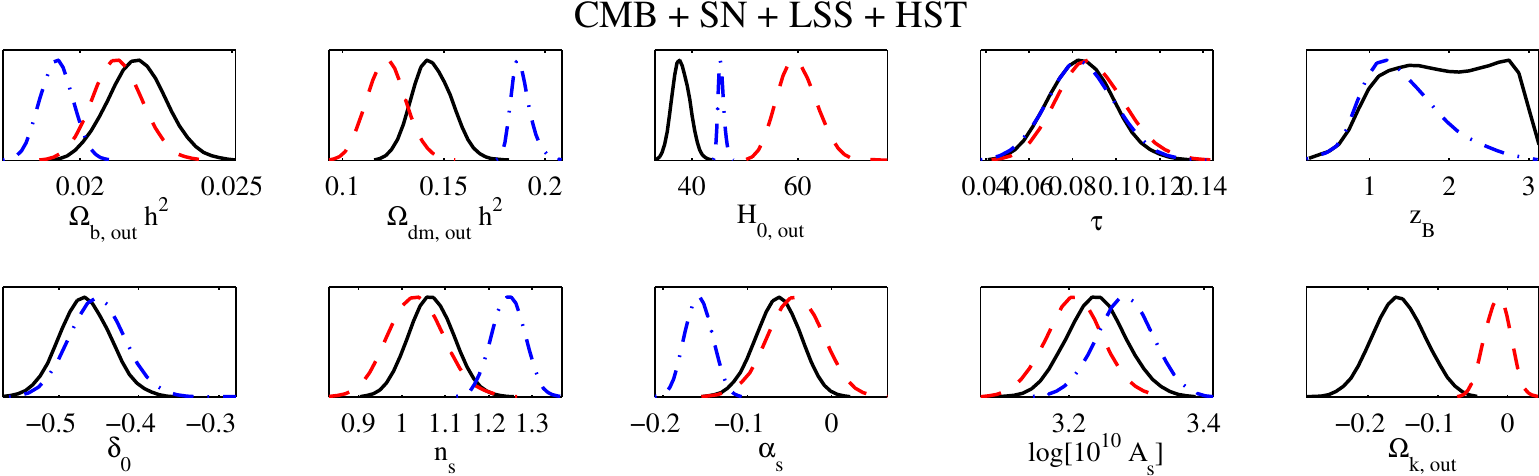}
\caption{The marginalized 1D posterior probabilities of the free parameters, on all of which we had a flat prior, when fitting to CMB+SN+SDSS+HST. $\Lambda$CDM in dashed, red, EdS Void in dashed-dotted, blue, and Curved Void in solid, black. Only $\Omega_{k, {\rm out}}$ is a derived parameter in the Void scenario, but since it is a free parameter for $\Lambda$CDM, we do plot it here for comparison. Note that our choice $k_{\rm pivot}=0.05$ Mpc / $h$ affects the central values of the scalar tilt and its running, which has no consequences.}\label{fig:marge_prim_CHSS}
} 

\FIGURE{ 
\includegraphics[width=0.6\textwidth]{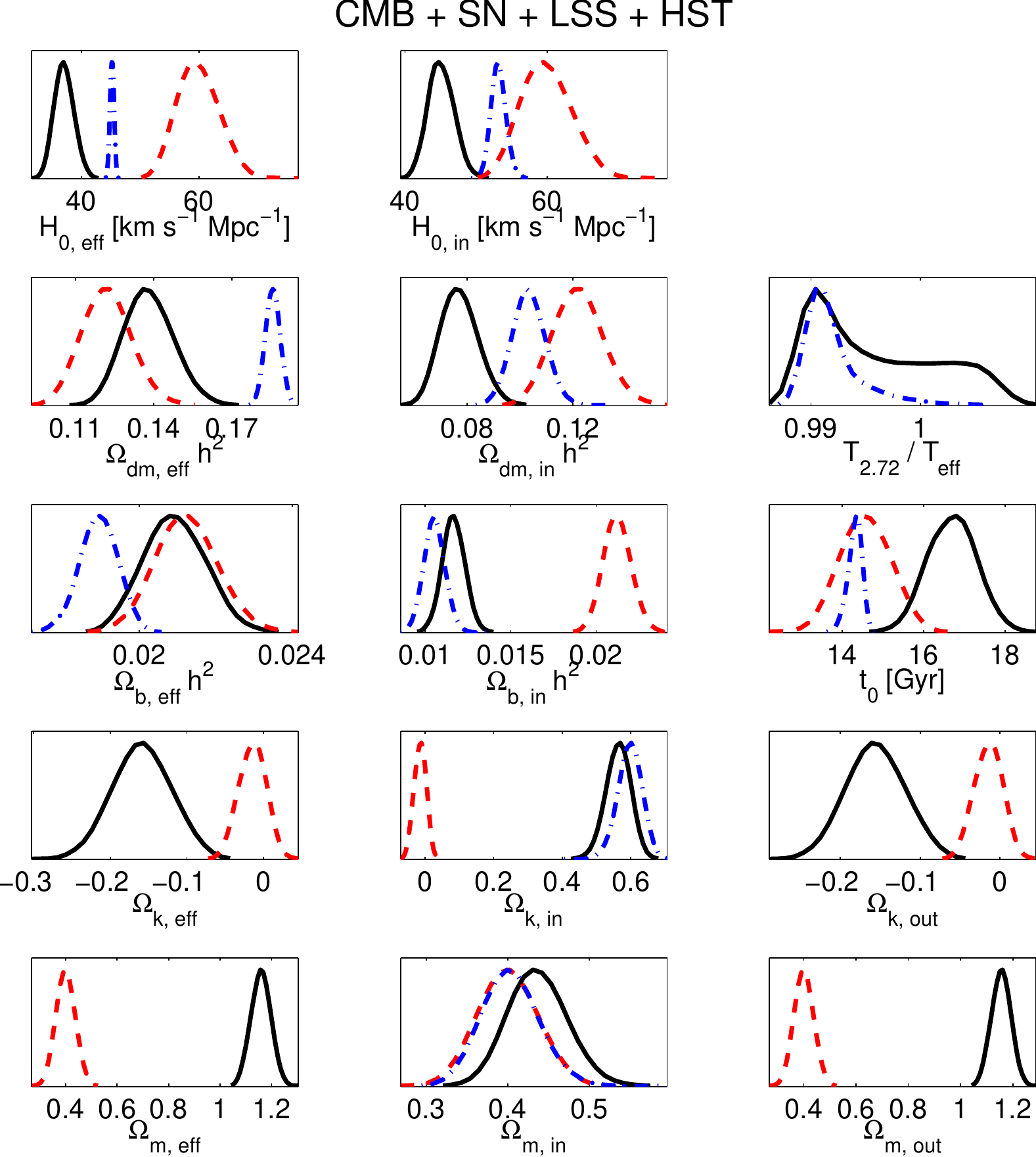}
\caption{The marginalized 1D posterior probabilities of derived parameters, when fitting to CMB+SN+SDSS+HST. $\Lambda$CDM in dashed, red, EdS Void in dashed-dotted, blue, and Curved Void in solid, black. In $\Lambda$CDM we have $H_{0, {\rm in}}=H_{0, {\rm out}}$.  The variable $H_{0, {\rm in}}$ is fit to the HST value. }\label{fig:marge_der_CHSS}
} 

\renewcommand\arraystretch{\myarraystretch}
\TABLE{ 
\resizebox{\textwidth}{!}{
\begin{tabular}{l || l | l || l | l || l | l}
\multicolumn{7}{c}{CMB+SN+SDSS+HST}\\ \hline
Parameter & \multicolumn{2}{c||}{$\Lambda$CDM} &  \multicolumn{2}{c||}{Curved Void } &  \multicolumn{2}{c}{EdS Void}\\
\hline
$ \Omega_{b, {\rm out}} h^2 $ & $ 2.21 \times 10^{-2} $  &  $ 2.12^{+0.16}_{-0.15} \times 10^{-2} $ & $ 2.24 \times 10^{-2} $  &  $ 2.19^{+0.19}_{-0.18} \times 10^{-2} $ & $ 1.90 \times 10^{-2} $  &  $ 1.92^{+0.10}_{-0.10} \times 10^{-2} $ \\
$ \Omega_{dm, {\rm out}} h^2 $ & $ 0.111 $  &  $ 0.122^{+0.019}_{-0.017} $ & $ 0.148 $  &  $ 0.145^{+0.020}_{-0.017} $ & $ 0.186 $  &  $ 0.188^{+0.011}_{-0.008} $ \\
$ H_{0, {\rm out}} $ {\tiny [km s$^{-1}$ Mpc$^{-1}$]}$ $ & $ 56.7 $  &  $ 60.0^{+7.2}_{-6.1} $ & $ 38.3 $  &  $ 38.0^{+3.2}_{-2.9} $ & $ 45.2 $  &  $ 45.6^{+1.3}_{-0.9} $ \\
$ \tau $ & $ 8.44 \times 10^{-2} $  &  $ 8.86^{+3.16}_{-2.76} \times 10^{-2} $ & $ 8.13 \times 10^{-2} $  &  $ 8.40^{+2.90}_{-2.64} \times 10^{-2} $ & $ 7.57 \times 10^{-2} $  &  $ 8.43^{+3.10}_{-2.89} \times 10^{-2} $ \\
$ \Omega_{k, {\rm out}} $ & $ -3.50 \times 10^{-2} $  &  $ -1.45^{+3.30}_{-3.42} \times 10^{-2} $ & $ -0.162 $  &  $ -0.156^{+0.070}_{-0.071} $ &  -  &  - \\
$ n_{\rm s} $ & $ 0.970 $  &  $ 1.03^{+0.13}_{-0.12} $ & $ 1.06 $  &  $ 1.07^{+0.10}_{-0.10} $ & $ 1.25 $  &  $ 1.25^{+0.07}_{-0.07} $ \\
$ \alpha_{\rm s} $ & $ -1.08 \times 10^{-2} $  &  $ -4.30^{+6.05}_{-6.44} \times 10^{-2} $ & $ -6.09 \times 10^{-2} $  &  $ -6.36^{+4.91}_{-5.15} \times 10^{-2} $ & $ -0.158 $  &  $ -0.158^{+0.031}_{-0.030} $ \\
$ \log[10^{10} A_{\rm s}] $ & $ 3.20 $  &  $ 3.21^{+0.08}_{-0.08} $ & $ 3.25 $  &  $ 3.24^{+0.08}_{-0.08} $ & $ 3.26 $  &  $ 3.28^{+0.08}_{-0.08} $ \\
$ z_B $ &  -  &  - & $ 2.90 $  &  $ 1.91^{+1.09}_{-1.15} $ & $ 1.16 $  &  $ 1.42^{+1.11}_{-0.76} $ \\
$ \delta_0 $ &  -  &  - & $ -0.465 $  &  $ -0.467^{+0.065}_{-0.061} $ & $ -0.449 $  &  $ -0.451^{+0.068}_{-0.064} $ \\ \hline
$ \Omega_\Lambda $ & $ 0.622 $  &  $ 0.614^{+0.057}_{-0.061} $ &  -  &  - &  -  &  - \\
$ \Omega_{m, {\rm out}} $ & $ 0.413 $  &  $ 0.400^{+0.071}_{-0.070} $ & $ 1.16 $  &  $ 1.16^{+0.07}_{-0.07} $ & $ 1.00 $  &  $ 1.00 $ \\
$ H_{0, {\rm in}} $ {\tiny [km s$^{-1}$ Mpc$^{-1}$]}$ $ & $ 56.7 $  &  $ 60.0^{+7.2}_{-6.1} $ & $ 45.6 $  &  $ 45.1^{+3.7}_{-3.3} $ & $ 52.7 $  &  $ 53.2^{+2.2}_{-2.0} $ \\
$ T_{2.72} / T_{{\rm{eff}}} $ & $ 1.00 $  &  $ 1.00 $ & $ 1.01 $  &  $ 0.996^{+0.011}_{-0.007} $ & $ 0.990 $  &  $ 0.992^{+0.008}_{-0.003} $ \\
$ t_0 $ [Gyr] & $ 15.3 $  &  $ 14.6^{+1.2}_{-1.2} $ & $ 16.5 $  &  $ 16.7^{+1.2}_{-1.1} $ & $ 14.4 $  &  $ 14.3^{+0.3}_{-0.4} $ \\
$ H_{0, {\rm{eff}}} $ {\tiny [km s$^{-1}$ Mpc$^{-1}$]}$ $ & $ 56.7 $  &  $ 60.0^{+7.2}_{-6.1} $ & $ 36.4 $  &  $ 37.1^{+3.3}_{-3.1} $ & $ 45.2 $  &  $ 45.2^{+0.7}_{-0.7} $ \\
$ \Omega_{b, {\rm{eff}}} h^2 $ & $ 2.21 \times 10^{-2} $  &  $ 2.12^{+0.16}_{-0.15} \times 10^{-2} $ & $ 2.03 \times 10^{-2} $  &  $ 2.09^{+0.16}_{-0.15} \times 10^{-2} $ & $ 1.90 \times 10^{-2} $  &  $ 1.89^{+0.09}_{-0.10} \times 10^{-2} $ \\
$ \Omega_{dm, {\rm{eff}}} h^2 $ & $ 0.111 $  &  $ 0.122^{+0.019}_{-0.017} $ & $ 0.134 $  &  $ 0.138^{+0.019}_{-0.017} $ & $ 0.186 $  &  $ 0.186^{+0.005}_{-0.005} $ \\
$ \Omega_{k, {\rm{eff}}} $ & $ -3.50 \times 10^{-2} $  &  $ -1.45^{+3.30}_{-3.42} \times 10^{-2} $ & $ -0.168 $  &  $ -0.158^{+0.072}_{-0.075} $ &  -  &  - \\
$ \Omega_{m, {\rm{eff}}} $ & $ 0.413 $  &  $ 0.400^{+0.071}_{-0.070} $ & $ 1.17 $  &  $ 1.16^{+0.07}_{-0.07} $ & $ 1.00 $  &  $ 1.00 $ \\
$ \Omega_{b, {\rm in}} h^2 $ & $ 2.21 \times 10^{-2} $  &  $ 2.12^{+0.16}_{-0.15} \times 10^{-2} $ & $ 1.20 \times 10^{-2} $  &  $ 1.17^{+0.13}_{-0.13} \times 10^{-2} $ & $ 1.05 \times 10^{-2} $  &  $ 1.05^{+0.12}_{-0.11} \times 10^{-2} $ \\
$ \Omega_{dm, {\rm in}} h^2 $ & $ 0.111 $  &  $ 0.122^{+0.019}_{-0.017} $ & $ 7.93 \times 10^{-2} $  &  $ 7.70^{+1.42}_{-1.25} \times 10^{-2} $ & $ 0.102 $  &  $ 0.103^{+0.012}_{-0.012} $ \\
$ \Omega_{k, {\rm in}} $ & $ -3.50 \times 10^{-2} $  &  $ -1.45^{+3.30}_{-3.42} \times 10^{-2} $ & $ 0.560 $  &  $ 0.563^{+0.067}_{-0.073} $ & $ 0.594 $  &  $ 0.596^{+0.064}_{-0.072} $ \\
$ \Omega_{m, {\rm in}} $ & $ 0.413 $  &  $ 0.400^{+0.071}_{-0.070} $ & $ 0.440 $  &  $ 0.437^{+0.073}_{-0.067} $ & $ 0.406 $  &  $ 0.404^{+0.072}_{-0.064} $ \\
\end{tabular}
}
\caption{The best fit parameters (each left column) and marginalized posterior probabilities (each right column) with $95\%$ C.L. errors of all free parameters (top ten parameters) and derived parameters (from the eleventh parameter down), for the three models at stake when simultaneously fitting CMB + SN + SDSS + HST. In the C-model, the parameters $\Omega_k$ is actually a derived parameter, as we take a flat prior on $\Omega_{dm, {\rm out}}$ in that case. Let us again emphasize that our pivot scale $k_{\rm pivot}=0.05$ Mpc / $h$, which leads to the lower value of $n_{\rm s}$, since we allow for a running. The most notable difference with Table~\ref{tab:CBHS} is the smaller value of the density contrast $\delta_0$. Since we use FLRW perturbation theory for the LSS inside the void, a fit to the matter power spectrum has considerable weight on the value of $\Omega_{dm}$ inside the void. Note that this approximation is not justified.} \label{tab:CHSS}
} 
\renewcommand\arraystretch{1}

\renewcommand\arraystretch{\myarraystretch}
\TABLE{ 
\resizebox{\textwidth}{!}{
\begin{tabular}{l | l | l | l | l | l | l}
\multicolumn{7}{c}{CMB+BAO+SN+HST}\\ \hline
Parameter &$\Lambda$CDM&{Profile A}&{Profile B}& {Profile C } & {Profile D}&{Profile E}\\
\hline
$ \Omega_{b, {\rm out}} h^2 $ & $ 2.12 \times 10^{-2} $ & $ 2.22 \times 10^{-2} $ & $ 2.44 \times 10^{-2} $ & $ 2.48 \times 10^{-2} $ & $ 2.23 \times 10^{-2} $ & $ 2.65 \times 10^{-2} $ \\
$ \Omega_{dm, {\rm out}} h^2 $ & $ 0.126 $ & $ 0.138 $ & $ 0.152 $ & $ 0.151 $ & $ 0.136 $ & $ 0.194 $ \\
$ H_{0, {\rm out}} $ {\tiny [km s$^{-1}$ Mpc$^{-1}$]}$ $ & $ 66.4 $ & $ 37.2 $ & $ 38.4 $ & $ 38.2 $ & $ 36.9 $ & $ 40.7 $ \\
$ \tau $ & $ 8.68 \times 10^{-2} $ & $ 8.32 \times 10^{-2} $ & $ 8.22 \times 10^{-2} $ & $ 8.95 \times 10^{-2} $ & $ 7.59 \times 10^{-2} $ & $ 8.93 \times 10^{-2} $ \\
$ \Omega_{k, {\rm out}} $ & $ 5.63 \times 10^{-3} $ & $ -0.162 $ & $ -0.194 $ & $ -0.203 $ & $ -0.168 $ & $ -0.332 $ \\
$ n_{{\rm S}} $ & $ 1.06 $ & $ 1.06 $ & $ 1.06 $ & $ 1.04 $ & $ 1.02 $ & $ 1.17 $ \\
$ \alpha_{{\rm S}} $ & $ -5.43 \times 10^{-2} $ & $ -5.84 \times 10^{-2} $ & $ -5.99 \times 10^{-2} $ & $ -5.12 \times 10^{-2} $ & $ -4.22 \times 10^{-2} $ & $ -0.130 $ \\
$ \log[10^{10} A_{{\rm S}}] $ & $ 3.19 $ & $ 3.22 $ & $ 3.22 $ & $ 3.25 $ & $ 3.23 $ & $ 3.20 $ \\
$ z_B $ &  - & $ 1.02 $ & $ 2.78 $ & $ 2.91 $ & $ 1.01 $ & $ 2.13 $ \\
$ \delta_0(k_{\mt{max}}+k_{\mt{max,2}}, t_0) $ &  - &  $-0.743$ & $ -0.806 $ & $ -0.837 $ & $ -0.964 $ & $ -0.794 $ \\
$ L_{2} / L $ &  - &  - & $ 0.129 $ & $ 8.18 \times 10^{-2} $ & $ 6.74 \times 10^{-3} $ & $ 1.64 $ ($\alpha$   in Eq.~(\ref{eq:profE}) \\
$ \delta(k_{max}, t_0) $ &  - & - & $ -0.728 $ & $ -0.741 $ & $ -0.751 $ & $ -0.627 $ \\ \hline
$ \Omega_\Lambda $ & $ 0.660 $ &  - &  - &  - &  - &  - \\
$ \Omega_{m, {\rm out}} $ & $ 0.334 $ & $ 1.16 $ & $ 1.19 $ & $ 1.20 $ & $ 1.17 $ & $ 1.33 $ \\
$ H_{0, {\rm in}} $ {\tiny [km s$^{-1}$ Mpc$^{-1}$]}$ $ & $ 66.4 $ & $ 49.5 $ & $ 52.9 $ & $ 53.5 $ & $ 55.2 $ & $ 56.6 $ \\
$ T_{2.72} / T_{{\rm eff}} $ & $ 1.00 $ & $ 0.980 $ & $ 1.01 $ & $ 1.02 $ & $ 0.980 $ & $ 1.37 $ \\
$ t_0 $ [Gyr] & $ 13.7 $ & $ 17.0 $ & $ 16.4 $ & $ 16.4 $ & $ 17.1 $ & $ 15.1 $ \\
$ H_{0, {\rm eff}} $ {\tiny [km s$^{-1}$ Mpc$^{-1}$]}$ $ & $ 66.4 $ & $ 37.2 $ & $ 34.4 $ & $ 33.8 $ & $ 36.9 $ & $ 19.9 $ \\
$ \Omega_{b, {\rm eff}} h^2 $ & $ 2.12 \times 10^{-2} $ & $ 2.22 \times 10^{-2} $ & $ 1.99 \times 10^{-2} $ & $ 1.96 \times 10^{-2} $ & $ 2.23 \times 10^{-2} $ & $ 7.65 \times 10^{-3} $ \\
$ \Omega_{dm, {\rm eff}} h^2 $ & $ 0.126 $ & $ 0.138 $ & $ 0.124 $ & $ 0.120 $ & $ 0.136 $ & $ 5.59 \times 10^{-2} $ \\
$ \Omega_{k, {\rm eff}} $ & $ 5.63 \times 10^{-3} $ & $ -0.162 $ & $ -0.211 $ & $ -0.223 $ & $ -0.168 $ & $ -0.605 $ \\
$ \Omega_{m, {\rm eff}} $ & $ 0.334 $ & $ 1.16 $ & $ 1.21 $ & $ 1.22 $ & $ 1.17 $ & $ 1.60 $ \\
$ \Omega_{b, {\rm in}} h^2 $ & $ 2.12 \times 10^{-2} $ & $ 1.01 \times 10^{-2} $ & $ 9.00 \times 10^{-3} $ & $ 7.89 \times 10^{-3} $ & $ 4.84 \times 10^{-3} $ & $ 1.10 \times 10^{-2} $ \\
$ \Omega_{dm, {\rm in}} h^2 $ & $ 0.126 $ & $ 6.30 \times 10^{-2} $ & $ 5.59 \times 10^{-2} $ & $ 4.82 \times 10^{-2} $ & $ 2.96 \times 10^{-2} $ & $ 8.04 \times 10^{-2} $ \\
$ \Omega_{k, {\rm in}} $ & $ 5.63 \times 10^{-3} $ & $ 0.702 $ & $ 0.768 $ & $ 0.804 $ & $ 0.887 $ & $ 0.715 $ \\
$ \Omega_{m, {\rm in}} $ & $ 0.334 $ & $ 0.298 $ & $ 0.232 $ & $ 0.196 $ & $ 0.113 $ & $ 0.285 $ \\
\end{tabular}
}
\caption{The best fit parameters of all free parameters (top twelve parameters) and derived parameters (from the thirteenth parameter down), for the $\Lambda$CDM and all considered curvature profiles when simultaneously fitting CMB + BAO + SN + HST. The parameter $ \delta_0(k_{\mt{max}}+k_{\mt{max,2}}, t_0) $ is the actual density contrast between the centre of the void and the outer FLRW, where in the case of Profile E the outer FLRW is never reached. The parameter $ \delta(k_{\mt{max}}, t_0) $ gives the density contrast that would be present if $k_{\mt{max, 2}}=0$, hence defining both $k_{\mt{max}}$ and $k_{\mt{max, 2}}$ for a given $ \delta_0(k_{\mt{max}}+k_{\mt{max,2}}, t_0) $.} \label{tab:Exotic}
} 
\renewcommand\arraystretch{1}

\setlength{\totfigwidth}{0.8\textwidth}
\FIGURE{ 
\includegraphics[width=\totfigwidth]{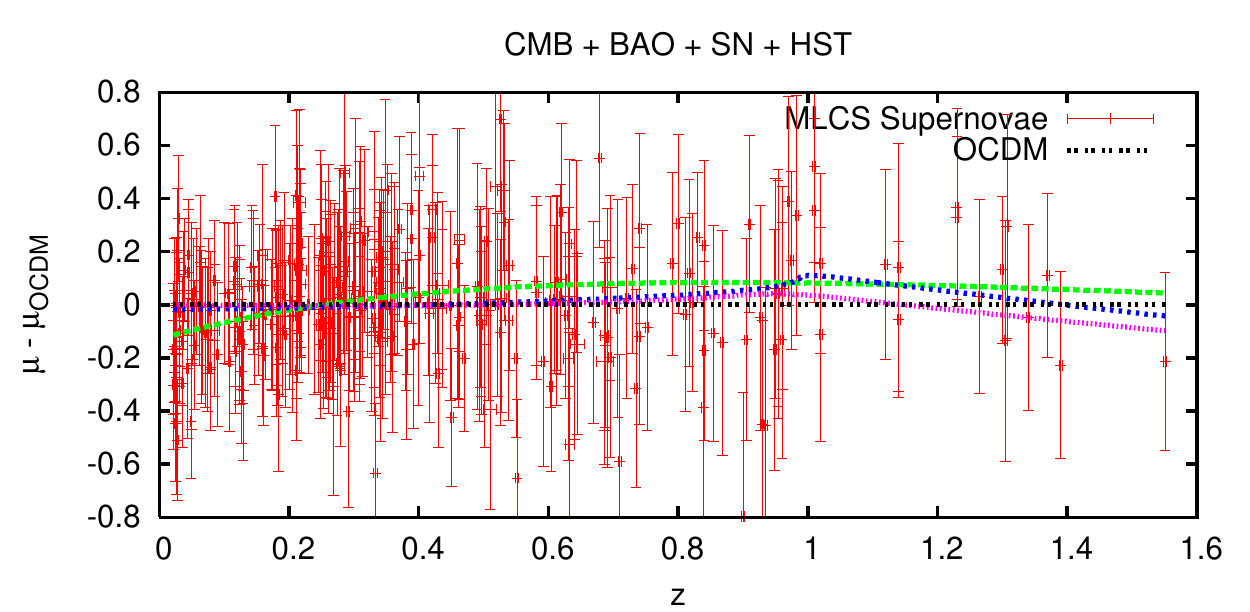}\vspace{-0.05\totfigwidth}\\
\hspace{-0.02\totfigwidth}\vspace{-0.01\totfigwidth}
\includegraphics[width=1.02\totfigwidth]{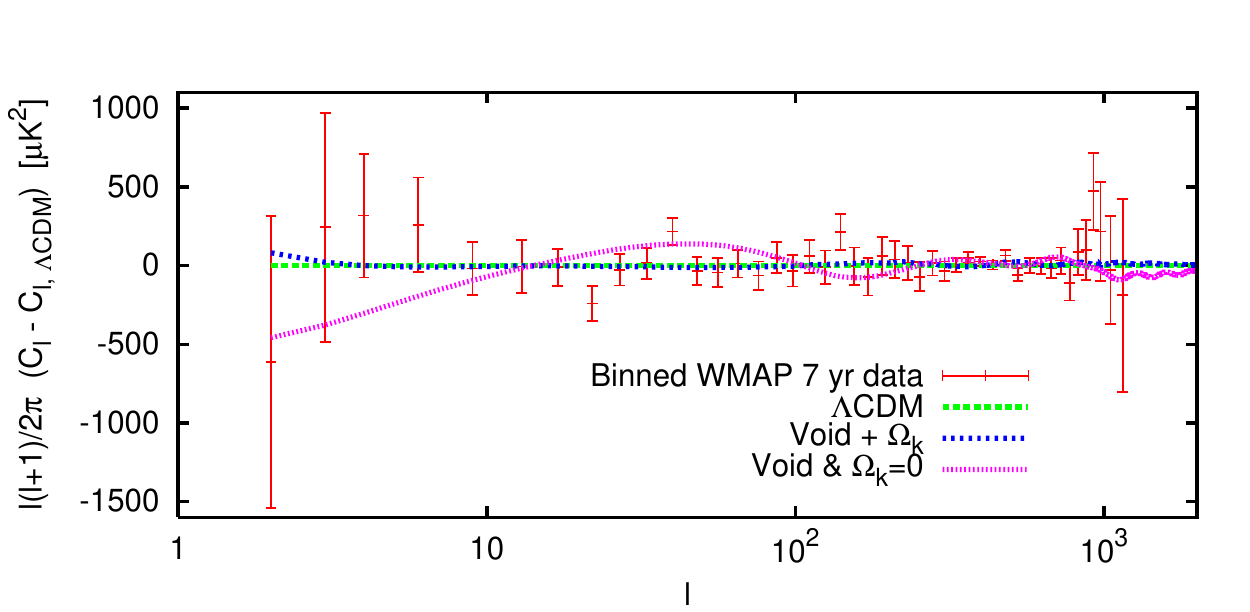}\\
\includegraphics[width=0.6\totfigwidth]{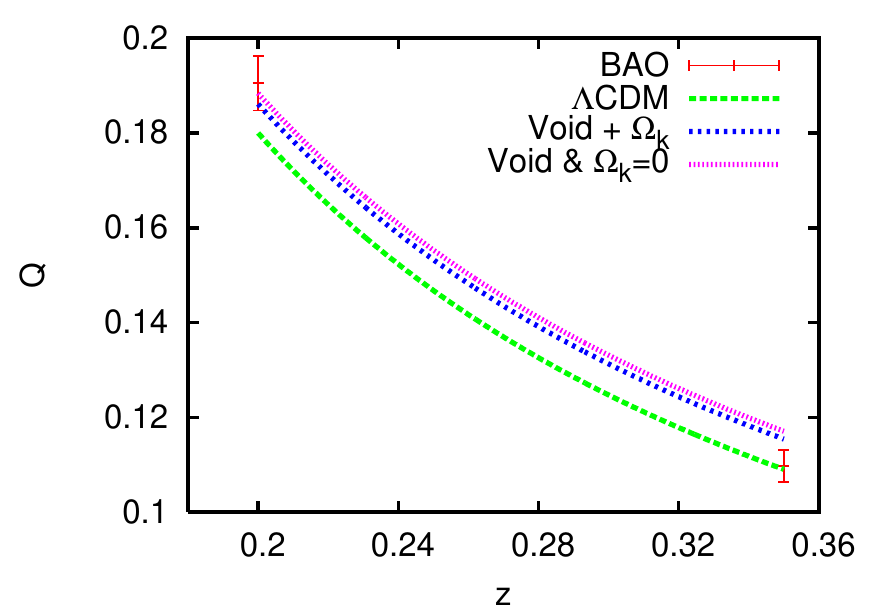}\hspace{0.04\totfigwidth}
\raisebox{0.035\totfigwidth}{\includegraphics[width=0.355\totfigwidth]{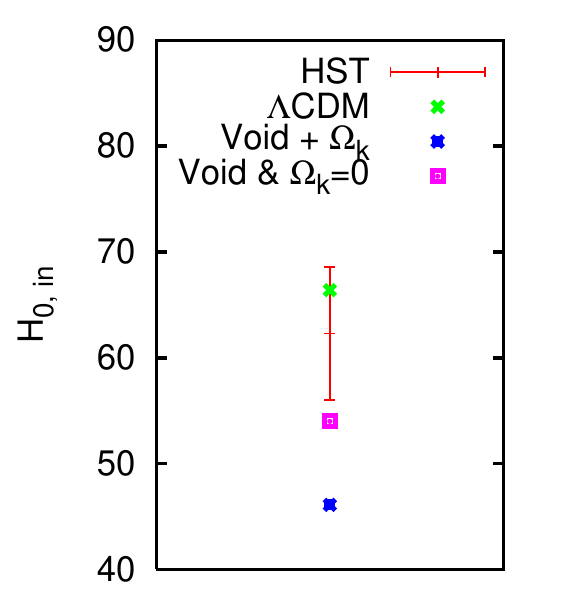}}
\caption{Plots of all the datasets and the best fit models when fitting to CMB + BAO  + SN + HST. The top panel shows SN data and predictions normalized to a reference model with $\Omega_m = 0.3$ and $\Omega_k = 0.7$ (OCDM). The second panel from the top shows CMB data and predicitions normlized to the $\Lambda$CDM best fit model for this combination of datasets. The bottom left panel shows the BAO, the bottom right panel $H_{0, {\rm in}}$. In all plots $\Lambda$CDM is dashed green, the Curved Void-model dotted blue, and the EdS Void-model fine dotted magenta.}\label{fig:CBHS_all}
} 

\FIGURE{ 
\includegraphics[width=\totfigwidth]{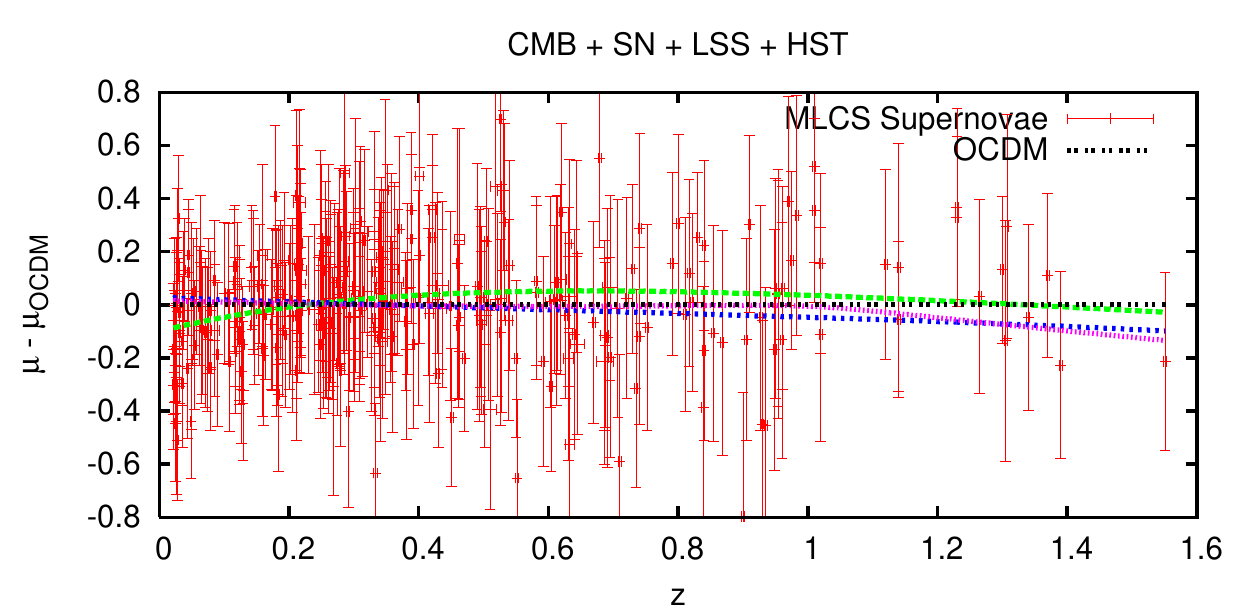}\vspace{-0.05\totfigwidth} \\
\hspace{-0.02\totfigwidth}\vspace{-0.01\totfigwidth}\includegraphics[width=0.7\totfigwidth]{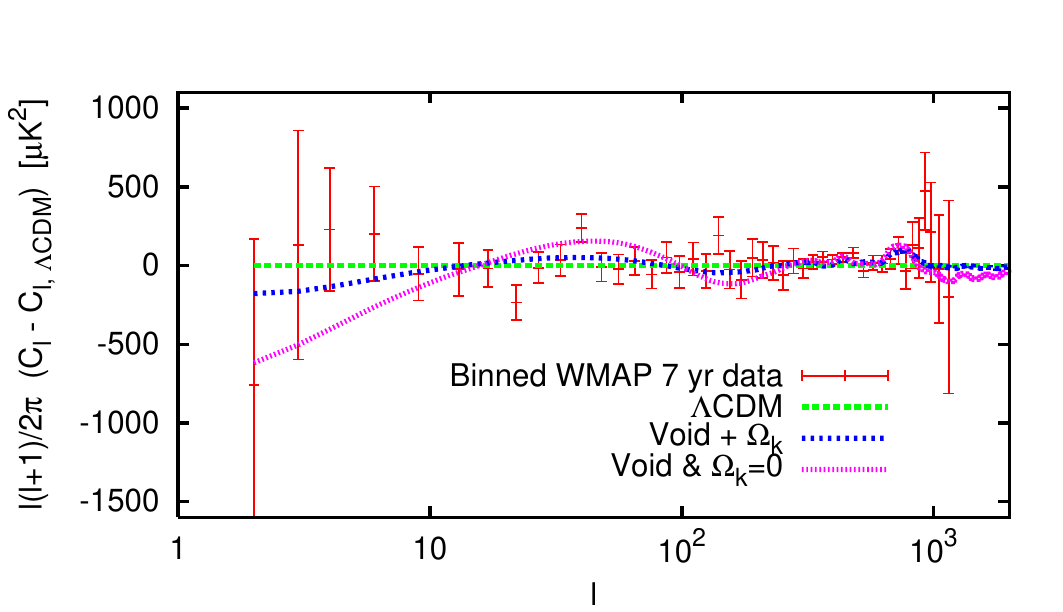}
\raisebox{0.035\totfigwidth}{\hspace{-0.02\totfigwidth}\vspace{0pt}\includegraphics[width=0.305\totfigwidth]{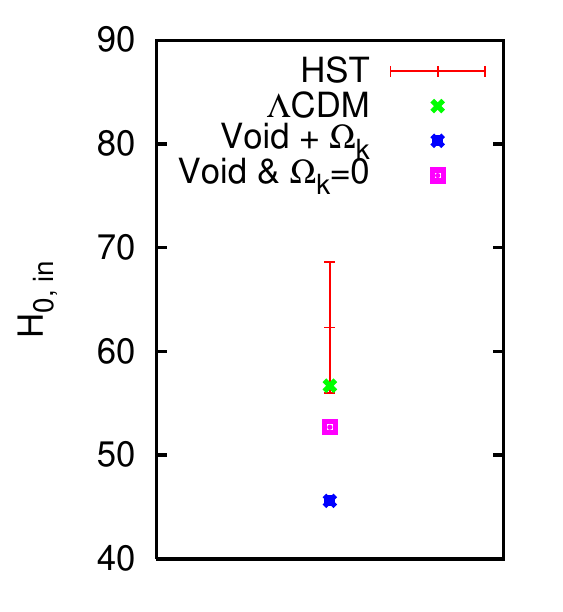}}\\
\includegraphics[width=\totfigwidth]{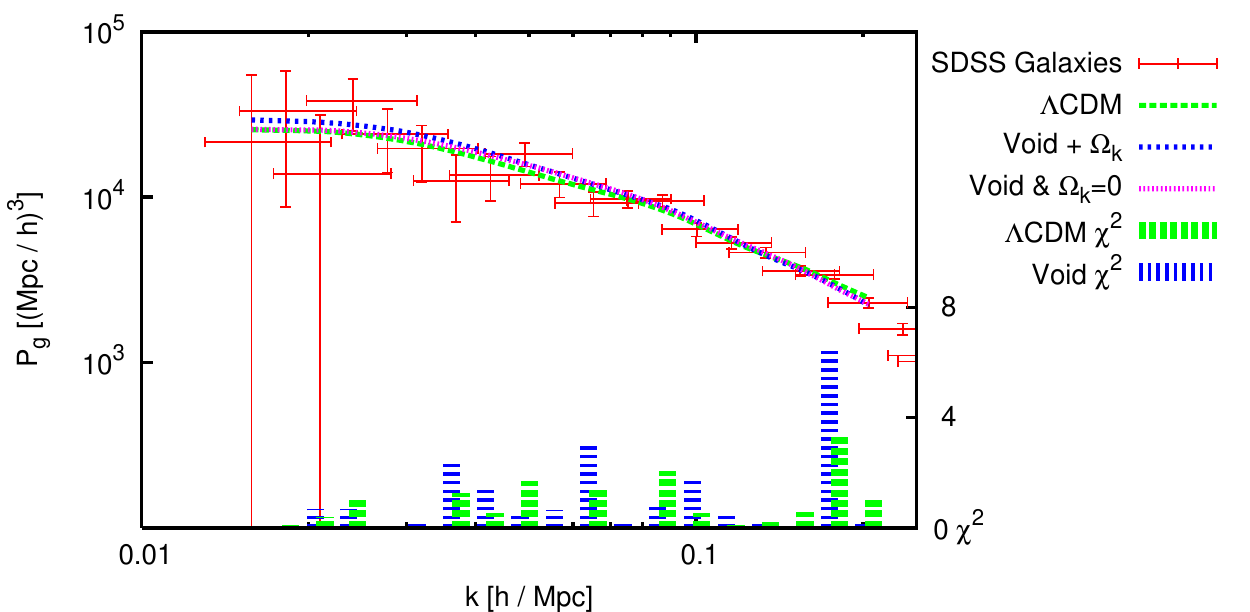}\vspace{-0.01\totfigwidth}
\caption{Plots of all the datasets and the best fit models when fitting to CMB + SDSS  + SN + HST. The top panel shows SN data and predictions normalized to a reference model with $\Omega_m = 0.3$ and $\Omega_k = 0.7$ (OCDM). The second panel from the top shows CMB data and predicitions normlized to the $\Lambda$CDM best fit model for this combination of datasets. The bottom left panel shows $H_{0, {\rm in}}$, and the bottom right panel shows the galaxy power spectrum, as well as the contribution to the total $\chi^2$ for each separate data point, in a bar chart. In all plots $\Lambda$CDM is dashed green, the Curved Void-model dotted blue, and the EdS Void-model fine dotted magenta.}\label{fig:CHSS_all}
} 

\clearpage 

\bibliographystyle{JHEP}
\bibliography{refs}

\end{document}